\documentclass[rmp,twocolumn,nofootinbib]{revtex4}

\usepackage{graphics}
\usepackage{rotating}
\usepackage{epsfig}
\usepackage{color}

\def\inbar{\,\vrule height1.5ex width.4pt depth0pt}
\def\IR{\relax{\rm I\kern-.18em R}}
\def\IC{\relax\hbox{$\inbar\kern-.3em{X\rm C}$}}

\begin{document}
\title{Feshbach Resonances in Ultracold Gases}

\author{Cheng Chin}
\affiliation{Department of Physics and James Franck Institute,
University of Chicago, Chicago, Illinois 60637, USA}

\author{Rudolf Grimm}
\affiliation{Institute of Experimental Physics and Center for
Quantum Physics, University of Innsbruck, Technikerstra\ss{}e 25,
6020 Innsbruck, Austria} \affiliation{Institute for Quantum Optics
and Quantum Information, Austrian Academy of Sciences,
Otto-Hittmair-Platz 1, 6020 Innsbruck, Austria}

\author{Paul Julienne and Eite Tiesinga}
\affiliation{Joint Quantum Institute, National Institute of
Standards and Technology and University of Maryland, 100 Bureau
Drive, Gaithersburg, Maryland 20899-8423, USA}

\begin{abstract}
Feshbach resonances are the essential tool to control the
interaction between atoms in ultracold quantum gases. They have
found numerous experimental applications, opening up the way to
important breakthroughs. This Review broadly covers the phenomenon
of Feshbach resonances in ultracold gases and their main
applications. This includes the theoretical background and models
for the description of Feshbach resonances, the experimental methods
to find and characterize the resonances, a discussion of the main
properties of resonances in various atomic species and
mixed {atomic} species systems, and an overview of key
experiments with atomic Bose-Einstein condensates, degenerate Fermi
gases, and ultracold molecules.
\end{abstract}

\date{\today}
\maketitle
\tableofcontents

\hyphenation{Fesh-bach}

\section{Introduction}
\label{sec:intro}

\subsection{Ultracold gases and Feshbach resonances: Scope of the Review}

The great impact of ultracold atomic and molecular quantum gases on present-day physics is linked to the extraordinary degree of control that such systems offer to investigate the fundamental behavior of quantum matter under various conditions. The interest goes beyond atomic and molecular physics, reaching far into other fields, like condensed matter, few- and many-body physics. In all these applications, Feshbach resonances represent the essential tool to control the interaction between the atoms, which has been the key to many great breakthroughs.

Ultracold gases are generally produced by laser cooling \cite{Metcalf1999} and
subsequent evaporative cooling \cite{Ketterle1997}. At temperatures
in the nanokelvin range and typical number densities somewhere
between $10^{12}$\,cm$^{-3}$ and $10^{15}$\,cm$^{-3}$,
quantum-degenerate states of matter are formed when the atomic
de-Broglie wavelength exceeds the typical interparticle distance and
quantum statistics governs the behavior of the system. The
attainment of Bose-Einstein condensation (BEC) in dilute ultracold
gases marked the starting point of a new era in physics
\cite{Anderson1995, Bradley1995, Davis1995}, and degenerate atomic Fermi
gases entered the stage a few years later \cite{DeMarco1999,
Schreck2001, Truscott2001}. The developments of the techniques to
cool and trap atoms by laser light were recognized with the 1997
Nobel prize in physics \cite{Cohentannoudji1998, Chu1998,
Phillips1998}. Only four years later, the achievement of BEC in
dilute gases of alkali atoms and early fundamental studies of the
properties of the condensates led to the 2001 Nobel prize
\cite{Cornell2002,Ketterle2002} \footnote{For overviews on laser
cooling and trapping, BEC, and ultracold Fermi gases see the
proceedings of the Varenna summer schools in 1991, 1998, and 2006
\cite{Varenna1991, Varenna1998, Varenna2006}. For reviews on the
theory of degenerate quantum gases of bosons and fermions see
\cite{Dalfovo1999} and \cite{Giorgini2007}, respectively, and the
textbooks by \cite{stringaribook} and \cite{pethickbook}.}.

In this Review, we give a broad coverage of Feshbach resonances in view of the manifold applications they have found in ultracold gases. Regarding theory, we focus on the underlying two-body physics and on models to describe Feshbach resonances. In the experimental part we include applications to few- and many-body physics; we discuss typical or representative results, instead of the impossible attempt to exhaustively review all developments in this rapidly growing field. Several aspects of Feshbach resonances and related topics have already been reviewed elsewhere. An early review on Feshbach resonance theory was given by \cite{Timmermans1999}. In another theoretical review, \cite{Duine2004} focussed on atom-molecule coherence. \cite{Kohler2006} and \cite{Hutson2006a} reviewed the formation of ultracold molecules near Feshbach resonances. The closely related topic of photoassociation was reviewed by \cite{Jones2006}.

In Sec.~\ref{sec:theory}, we start with a presentation of the
theoretical background. Then, in Sec.~\ref{sec:species}, we present
the various experimental methods to identify and characterize
Feshbach resonances. There we also discuss the specific interaction
properties of different atomic species, which can exhibit vastly
different behavior. In Sec.~\ref{sec:qugas}, we present important
applications of interaction control in experiments on atomic Bose
and Fermi gases. In Sec.~\ref{sec:molecules}, we discuss properties
and applications of ultracold molecules created via Feshbach
association. Finally, in Sec.~\ref{sec:related}, we discuss some
related topics, like optical Feshbach resonances, interaction
control in optical lattices, few-body physics, and the relation to
molecular scattering resonances and cold chemistry.

\subsection{Basic physics of a Feshbach resonance}
\label{ssec:basic}

The physical origin and the elementary properties of a Feshbach
resonance can be understood from a simple picture. Here we outline the
basic ideas, referring the reader to Sec.~\ref{sec:theory} for a
more detailed theoretical discussion.

\begin{figure}
\includegraphics[width=2.8in]{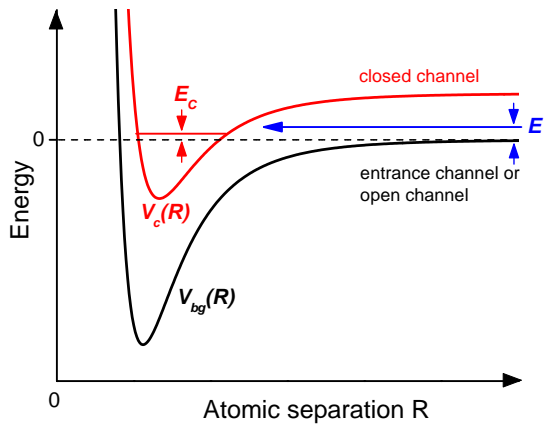}
\caption{Basic two-channel model for a Feshbach resonance. The
phenomenon occurs when two atoms colliding at energy $E$ in the entrance channel resonantly couple to a molecular bound state with energy
$E_c$ supported by the closed channel potential. In the ultracold domain,
collisions take place near zero-energy, $E \rightarrow
0$.  Resonant coupling
is then conveniently realized by magnetically tuning $E_c$ near 0, if the
magnetic moments of the closed and open channel differ.} \label{fig:intro_potential}
\end{figure}

We consider two molecular potential curves $V_\mathrm{bg}(R)$ and
$V_\mathrm{c}(R)$, as illustrated in Fig.~\ref{fig:intro_potential}.
For large internuclear distances $R$, the background potential
$V_\mathrm{bg}(R)$ asymptotically connects to two free atoms in the
ultracold gas. For a collision process, having the very small energy
$E$, this potential represents the energetically open channel, in
the following referred to as the {\em entrance channel}. The other
potential, $V_\mathrm{c}(R)$, representing the {\em closed channel},
is important as it can support bound molecular states near the
threshold of the open channel.

A Feshbach resonance occurs when the bound molecular state in the
closed channel energetically approaches the scattering state in
the open channel. Then even weak coupling can lead to strong mixing
between the two channels. The energy difference can be controlled
via a magnetic field when the corresponding magnetic moments are
different. This leads to a {\em magnetically tuned Feshbach
resonance}. The magnetic tuning method is the common way to achieve
resonant coupling and it has found numerous applications, as will be
extensively discussed in this Review. Alternatively, resonant
coupling can be achieved by optical methods, leading to {\em optical
Feshbach resonances} with many conceptual similarities to the
magnetically tuned case; see Sec.~\ref{ssec:optfesh}.
Such resonances are promising for cases where magnetically tunable resonances are absent.

\begin{figure}
\includegraphics[width=3.2in]{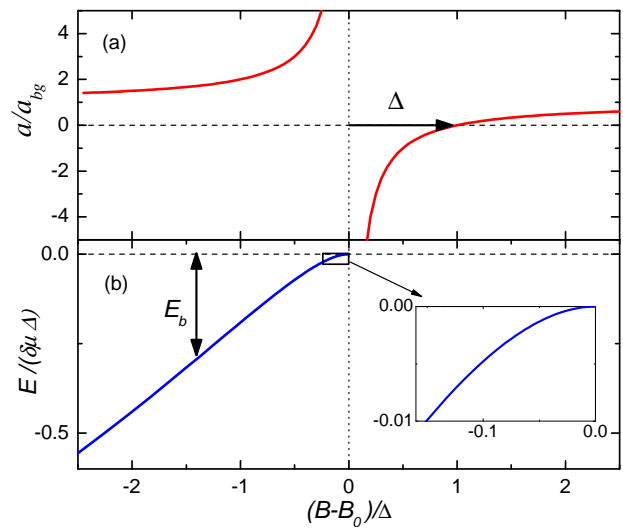}
\caption{Scattering length $a$ (Panel (a)) and molecular state energy $E$ (Panel (b)) near a
magnetically tuned Feshbach resonance.  The binding energy is defined to be positive, $E_b=-E$.  The inset shows the universal regime near the point of resonance where $a$ is very large and positive.} \label{fig:intro_e&a}
\end{figure}

A magnetically tuned Feshbach resonance can be described by a simple
expression\footnote{This simple expression applies to resonances
without { inelastic} two-body channels. Some Feshbach resonances, especially the optical ones, feature two-body decay. A more general discussion including { inelastic decay} is given in Sec.~\ref{sssec:resscatt}}, introduced by
\cite{Moerdijk1995}, for the $s$-wave scattering length $a$ as a
function of the magnetic field $B$,
\begin{equation}
 a(B) = a_\mathrm{bg} \left( 1 - \frac{\Delta}{B-B_0} \right) \,.
      \label{II.A.21}
\end{equation}
Figure~\ref{fig:intro_e&a}(a) illustrates this resonance expression.
The {\em background scattering length}
$a_\mathrm{bg}$, which is the scattering length associated with $V_\mathrm{bg}(R)$, represents the off-resonant value.  It is directly related to the energy of the last-bound vibrational level of $V_\mathrm{bg}(R)$.
The parameter $B_0$ denotes the {\em resonance
position}, where the scattering length diverges
($a\rightarrow\pm\infty$), and the parameter $\Delta$ is
the {\em resonance width}. Note that both $a_\mathrm{bg}$ and
$\Delta$ can be positive or negative. An important point is the {\em
zero crossing} of the scattering length associated with a Feshbach
resonance; it occurs at a magnetic field $B=B_0 + \Delta$.
Note also that we will use G as the magnetic field unit in this Review, because of its near-universal usage among groups working in this field; 1 G $=$ $10^{-4}$ T.

The energy of the weakly bound molecular state near the resonance
position $B_0$  is shown in
Fig.~\ref{fig:intro_e&a}(b), relative to the threshold of two free
atoms with zero kinetic energy. The energy approaches threshold at $E=0$ on
the side of the resonance where $a$ is large and positive.   Away from resonance, the
energy varies linearly with $B$ with a slope given by $\delta \mu$, the difference in magnetic moments of the open and closed channels.  Near
resonance the coupling between the two channels mixes in entrance-channel contributions and strongly bends the molecular state.

In the vicinity of the resonance position at $B_0$, where the two
channels are strongly coupled, the scattering length is very
large. For large positive values of $a$, a ``dressed'' molecular state exists  with a binding energy given by
\begin{equation}
 E_b = \frac{\hbar^2}{2\mu a^2}\,,
 \label{II.A.4}
\end{equation}
where $\mu$ is the reduced mass of the atom pair.  In this limit
$E_b$ depends quadratically on the magnetic detuning $B-B_0$ and
results in the bend seen in the inset to Fig.~\ref{fig:intro_e&a}.
This region is of particular interest because of its {\em universal}
properties; { here} the state can be described in terms of
a single effective molecular potential having scattering length $a$.
In this case, the wavefunction for the relative atomic motion is a
quantum halo state which extends to a very large size on the order
of $a$; the molecule is then called a {\em halo dimer}; see
Sec.~\ref{sssec:halo}.

A very useful distinction can be made between resonances that exist
in various systems; see Sec.~\ref{sssec:resstrength}. For narrow
resonances with a width $\Delta$ typically well below 1\,G (see
Appendix) the universal range persist only for a very small fraction
of the width. In contrast, broad resonances with a width typically
much larger than 1\,G tend to have a large universal range extending
over a considerable fraction of the width. The first class of
resonances is referred to as {\em closed-channel dominated
resonances}, whereas the second class is called {\em
entrance-channel dominated resonances}. For the distinction between
both classes, the width $\Delta$ is not the only relevant parameter.
Also, the background scattering length $a_{\rm bg}$ and the
differential magnetic moment $\delta \mu$ need to be taken into
account. Section~\ref{sssec:resstrength} discusses this important
distinction in terms of a dimensionless {\em resonance strength}.

\begin{figure}
\epsfxsize=3in \epsfbox{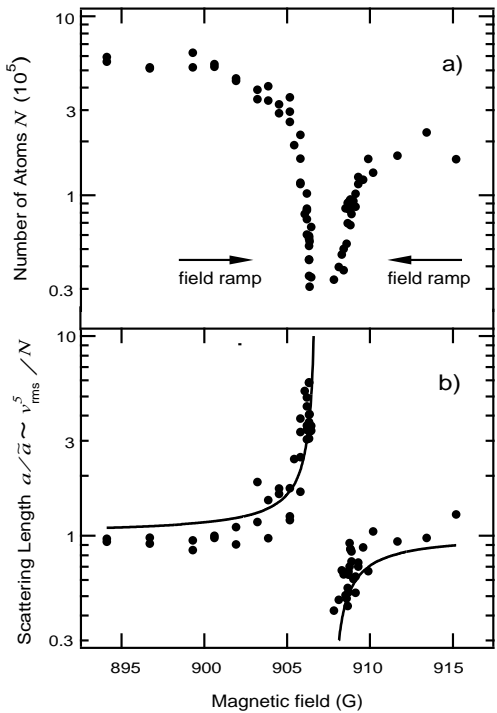}
\caption{Observation of a magnetically tuned Feshbach resonance in
an optically trapped BEC of Na atoms. The upper panel shows a strong
loss of atoms near the resonance, which is due to enhanced
three-body recombination. The lower panel shows the dispersive shape
of the scattering length { $a$ near the resonance, as determined}
from measurements of the mean-field interaction by expansion of the
condensate after release from the trap; { here $a$ is normalized to the background value $a_{\rm bg}$}. The magnetic field is given in G,
where 1\,G = $10^{-4}$\,T.  Reprinted by permission from Macmillan Publishers Ltd: Nature \cite{Inouye1998}, copyright 1998.}
\label{fig:firstobsNa}
\end{figure}

Figure \ref{fig:firstobsNa} shows the observation of a
Feshbach resonance as reported in 1998
by \cite{Inouye1998} for an optically trapped BEC of Na atoms. This
early example highlights the two most striking features of a
Feshbach resonance, the tunability of the scattering length
according to Eq.~\ref{II.A.21} and the fast loss of atoms in the
resonance region. The latter can be attributed to strongly enhanced
three-body recombination and molecule formation near a Feshbach
resonance; see Sec.~\ref{sssec:lossspect}.

A Feshbach resonance in an ultracold atomic gas can serve as a gateway into the molecular world and is thus strongly connected to the field of ultracold molecules; see Sec.~\ref{sec:molecules}. Various techniques have been developed to associate molecules near Feshbach resonances. Ultracold molecules
produced in this way are commonly referred to as {\em Feshbach
molecules}. The meaning of this term is not precisely defined, as
Feshbach molecules can be transferred to many other states near
threshold or to much more deeply bound states, thus being converted
to more conventional molecules. We will use the term Feshbach
molecule for any molecule that exists near the threshold in an
energy range set by the quantum of energy for near-threshold vibrations.
The universal halo state is a special, very weakly bound case of a Feshbach
molecule.

\subsection{Historical remarks}

Early investigations on phenomena arising from the coupling of a bound state to the continuum go back to the 1930s. \cite{Rice1933} considered how a bound state predissociates into a continuum, \cite{Fano1935, Fano2005} described asymmetric line shapes occurring in such a situation as a result of quantum interference, and \cite{Beutler1935} reported on the observation of highly asymmetric lineshapes in rare gas photoionization spectra. Nuclear physicists considered basically the same situation, having nuclear scattering experiments in mind instead of atomic physics. \cite{Breit1936} considered the situation in the limit when the bound state plays a dominant role and the asymmetry disappears. Later, interference and line-shape asymmetry were taken into account by several authors \cite{Blatt1952}.

Herman Feshbach (1917-2000) and Ugo Fano (1912-2001) developed their thorough treatments of the resonance phenomena that arise from the coupling of a discrete state to the continuum. Their work was carried out independently, using different theoretical approaches. While Feshbach's work originated in the context of nuclear physics \cite{Feshbach1958,Feshbach1962}, Fano approached the problem on the background of atomic physics \cite{Fano1961}, reformulating and extending his earlier work \cite{Fano1935,Fano2005}. Nowadays, the term ``Feshbach resonance'' is most widely used in the literature for the resonance phenomenon itself, but sometimes also the term ``Fano-Feshbach resonance'' appears. As a curiosity Feshbach himself considered his name being attached to a well-known resonance phenomenon as a mere atomic physics jargon \cite{Kleppner2004,Rau2005}. Fano's name is usually associated with the asymmetric lineshape of such a resonance, well known in atomic physics as a ``Fano profile''.

A prominent example for the observation of a Feshbach resonance in atomic physics is the experiment of \cite{Bryant1977} on photodetachment by the negative ion of hydrogen. Near a photon energy of 11\,eV two prominent resonances were seen, one of them being a Feshbach resonance and the other one a ``shape resonance''; see Sec.~\ref{sssec:resscatt}. Many more situations where Feshbach resonances play an important role can be found in atomic, molecular, and chemical physics; see \cite{Spence1975, Gauyacq1982, Macarthur1985, Nieh1990, Weber1999} for a few examples. In such experiments, the resonances occur when the scattering energy is varied. This is in contrast to the experiments on ultracold gases, where scattering takes place in the zero-energy limit and the resonances occur when an external field tunes bound states near threshold.

In the context of quantum gases, Feshbach resonance were first considered by \cite{Stwalley1976}, who suggested the existence of magnetically
induced Feshbach resonances in the scattering of spin-polarized hydrogen and deuterium atoms (H+D and D+D). He pointed to enhanced inelastic decay near these resonances and suggested that { they} should be avoided to maintain stable spin-polarized hydrogen gases. { A related loss resonance in hydrogen was observed by \cite{Reynolds1986}.} The positive aspect of such resonances was first pointed out by \cite{Tiesinga1993}, who showed that they can be used to change the sign and strength of the interaction between ultracold atoms. In 1998, the possibility of interaction tuning via Feshbach resonances was demonstrated by \cite{Inouye1998} for a $^{23}$Na BEC, as already discussed in the preceding Section. In the same year, \cite{Courteille1998} demonstrated a Feshbach resonance in a trapped sample of $^{85}$Rb atoms through the enhancement of photoassociative loss induced by a probe laser.

\begin{figure}
\includegraphics[width=3.2in]{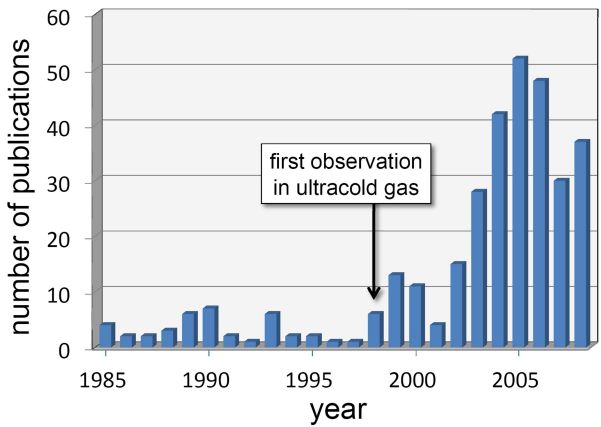}
\caption{Number of publications per year {(from 1985 to 2008)} with ``Feshbach resonances'' appearing in the title. Data from ISI Web of Science.} \label{fig:publications}
\end{figure}

The very important role of Feshbach resonances in present-day quantum gas experiments can be highlighted by looking at the number of publications per year with ``Feshbach resonances'' in the title; see Fig.~\ref{fig:publications}. Before 1998, one finds just a few publications with the great majority not related to ultracold atoms. Then, after 1998, a substantial increase is observed as a result of the first successful experiments with Feshbach resonances in ultracold gases. It then took a few years until Feshbach resonances had become a fully established tool and opened up many new applications in the field. This is reflected in the steep increase of the publication rate in the period from 2002 to 2004.

\section{Theoretical background}\label{sec:theory}

This review primarily concentrates on magnetically tunable
resonances, described in detail in the next Sections, while
Section~\ref{ssec:optfesh} discusses optical changes in scattering
lengths. Here we describe the two-body physics of collision
resonances, not the few-body or  many-body aspects.  Properties of a number of
magnetic Feshbach resonances are tabulated in  the Appendix; see
Table~\ref{tab:resonances}.

\subsection{Basic collision physics}\label{ssec:basiccoll}

The theory for describing  2-body collisions is described in a
number of textbooks \cite{Mott1965,Messiah1966,Taylor1972}.  Let us
first consider the collision of two structureless atoms, labeled $1$
and $2$ with masses $m_1$ and $m_2$ interacting under the influence
of the potential $V({\bf R})$, where $\bf R$ is the vector between
the positions of the two atoms with magnitude $R$. The separated atoms are
prepared in a plane wave with relative kinetic energy
$E=\hbar^2k^2/(2\mu)$ and relative momentum $\hbar {\bf k}$, where
$\mu= m_1 m_2/(m_1+m_2)$ is the reduced mass of the pair.  The plane
wave in turn is expanded in a standard sum over spherical harmonic
functions $Y_{\ell m_\ell}({\bf \hat{R}})$, where $\ell$ is the
relative angular momentum, $m_\ell$ is its projection along a
space-fixed $z$ axis, and ${\bf \hat{R}}={\bf R}/R$ is the direction
vector  on the unit sphere \cite{Messiah1966}.  This expansion is
called the partial wave expansion, and the various partial waves
$\ell = 0,1,2,\ldots$ are designated $s$-, $p$-, $d$-, $\ldots$
waves.

If the potential $V({\bf R})$ is isotropic, depending only on the
magnitude of ${\bf R}$, there is no coupling among partial waves,
each of which is described by the solution
$\psi_\ell(R)=\phi_\ell(R)/R$ to the Schr{\"o}dinger equation
\begin{equation}
- \frac{\hbar^2}{2\mu}\frac{d^2 \phi_\ell(R)}{dR^2} +
V_\ell(R)\phi_\ell(R) = E \phi_\ell(R)  \,,
 \label{II.A.1}
\end{equation}
where $V_\ell(R)=V(R) + \hbar^2 \ell(\ell+1)/(2\mu R^2)$ includes
the centrifugal potential, which is repulsive for $\ell >0$ and
vanishes for the $s$-wave.  We assume $V(R)\to0$ as
$R \to \infty$, so that $E$ represents the energy of the separated
particles.  This equation has a spectrum of $N_\ell$ bound state
solutions at discrete energies $E_{n\ell}$ for $E<0$ and a
continuous spectrum of scattering states with $E>0$.  While bound
states are normally labeled by vibrational quantum number
$v=0,\ldots,N_\ell-1$ counting up from the bottom of the potential, we
prefer to label threshold bound states  by quantum number $n=-1,
-2,\ldots$ counting down from the top of the potential for the last,
next to last, etc., bound states.   The bound state solutions
$|n\ell\rangle$ are normalized to unity, $|\langle
{n\ell}|{n\ell}\rangle |^2=1$, and $\phi_{n\ell}(R)=\langle R
|n\ell\rangle \to 0$ as $R\to\infty$.  The scattering solutions,
representing the incident plane wave plus a scattered wave, approach
\begin{equation}
  \phi_\ell(R,E) \to c \frac{\sin(kR -\pi \ell/2 + \eta_\ell(E))}{\sqrt{k}}e^{i\eta_\ell(E)}
  \label{II.A.2}
\end{equation}
as $R\to\infty$, where $\eta_\ell(E)$ is the scattering phase shift
and $c=\sqrt{2 \mu/(\pi \hbar^2)}$ is a constant that ensures the
wave function $|E \ell \rangle$ is normalized per unit energy,
$\langle E \ell | E' \ell\rangle=\int_0^\infty
\phi^*_\ell(R,E)\phi_\ell(R,E')dR = \delta(E-E')$.  The scattering
phase shift is the key parameter that incorporates the effect of the
whole potential on the collision event.

\cite{Sadeghpour2000} reviews the special properties of scattering
phase shift near a collision threshold when $k\to0$.  If $V(R)$
varies as $1/R^s$ at large $R$, then $\tan\eta_\ell \propto
k^{2\ell+1}$ if $2\ell+1 \le s-2$ and $\tan\eta_\ell \propto
k^{s-2}$ if $2\ell+1 \ge s-2$.  While Levinson's theorem shows that
$\eta_\ell \to N_\ell \pi$ as $k \to 0$, we need not
consider the $N_\ell \pi$ part of the phase shift in this review.
For van der Waals potentials with $s=6$, the threshold $\tan
\eta_\ell$ varies as $k$ and $k^3$ for $s$- and $p$-waves, and as
$k^4$ for all other partial waves.  The properties of $s$-wave
collisions are of primary interest for cold neutral atom collisions,
where near threshold, a more precise statement of the variation of
$\tan \eta_0$ with $k$ is given by the effective range expansion,
\begin{equation}
  k \cot \eta_0(E) = -\frac{1}{a} + \frac{1}{2} r_0 k^2  \,,
  \label{II.A.3}
\end{equation}
where $a$ is called the $s$-wave scattering length and $r_0$ the
effective range.  For practical purposes, it often suffices to
retain only the scattering length term and use $\tan \eta_0(E) =
-ka$.   Depending on the potential, the scattering length can have
any value, $-\infty < a <+\infty$.

When the scattering length is positive and sufficiently large, that
is, large compared to the characteristic length scale of the
molecular potential (see Section~\ref{sssec:vdw}), the last $s$-wave
bound state of the potential, labeled by index $n=-1$ and $\ell=0$,
is just below threshold with a binding energy $E_b=-E_{-1,0}$ given
by Eq.~(\ref{II.A.4}) in the Introduction. The domain of
universality, where scattering and bound state properties are solely
characterized by the scattering length and mass, are discussed at
length in recent reviews \cite{Braaten2006,Kohler2006}. The
universal bound state wave function takes on the form
$\phi_{-1,0}(R)=\sqrt{2/a} \exp{(-R/a)}$ at large $R$.  Such a state
exists almost entirely at long range beyond the outer classical
turning point of the potential.  Such a bound state is known as a
``halo state,'' also studied in nuclear physics~\cite{Riisager1994}
and discussed in Sec.~\ref{sssec:halo}.

\subsubsection{Collision channels}\label{sssec:channels}

The atoms used in cold collision experiments generally have spin
structure.  For each atom $i=1$ or $2$ in a collision the electronic
orbital angular momentum ${\bf L}_i$ is coupled to the total
electronic spin angular momentum ${\bf S}_i$ to give a resultant
${\bf j}_i$, which in turn is coupled to the nuclear spin ${\bf
I}_i$ to give the total angular momentum ${\bf f}_i$.   The
eigenstates of each atom are designated by the composite labels
$q_i$.  At zero magnetic field these labels are $f_im_i$, where
$m_i$ is the projection of ${\bf f}_i$.  For example,  alkali-metal
atoms that are commonly used in Feshbach resonance experiments, have
$^2$S$_{1/2}$ electronic ground states with quantum numbers $L_i=0$
and $S_i=1/2$, for which there are only two values of $f_i=$ $I_i -
1/2$ and $I_i + 1/2$ when $I_i \ne 0$.   Whether $f_i$ is an integer
or half an odd integer determines whether the atom is a composite
boson or fermion.

A magnetic field ${\bf B}$ splits these levels into a manifold of
Zeeman sublevels.  Only the projection $m_i$ along the field remains
a good quantum number, and $B=0$ levels with the same $m_i$ but
different $f_i$ can be mixed by the field.  Even at high field,
where the individual $f_i$ values no longer represent good quantum
numbers, the $f_i$ value still can be retained as a label,
indicating the value at $B=0$ with which the level adiabatically
correlates.

\begin{figure}
\includegraphics[width=2.8in,angle=270]{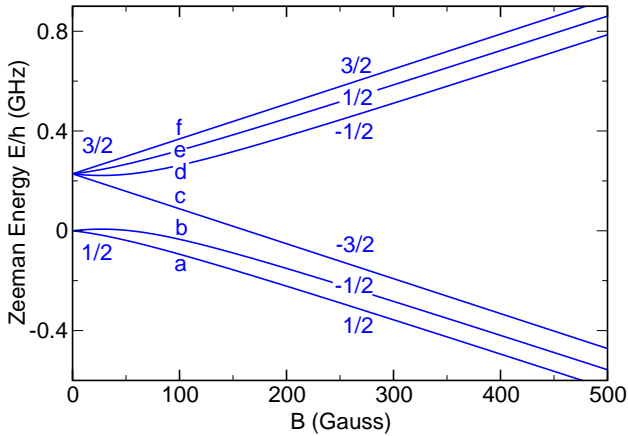}
\caption{Atomic energy levels of the ${^6}$Li atom, which has
$S=1/2$, $I=1$, and $f=1/2$ and $3/2$.  The figure shows both the
projection $m$ of $f$ and the alphabetic shorthand notation $q_i$
$=$ $a$, $b$, $c$, $d$, $e$, and $f$ used to label the levels in
order of increasing energy. } \label{fig:Li_Zeeman}
\end{figure}

Figure~\ref{fig:Li_Zeeman} indicates the Zeeman energy levels versus
$B$ for the $^6$Li atom, a fermion, according to the classic
Breit-Rabi formula~\cite{Breit1931}.  The two $f_i$ levels are split
at $B=0$ by the hyperfine energy, $E_\mathrm{hf}/h=228$ MHz.  At
large fields the lower group of three levels are associated with the
quantum numbers $m_S=-1/2$, while the upper group has $m_S=+1/2$.
The figure also shows our standard notation for atomic Zeeman levels
for any species and any field strength.  We label states by lower
case Roman letters $a$, $b$, $c$, $\ldots$ in order of increasing
energy.  Some authors prefer to label the levels in order
numerically as 1, 2, 3, $\ldots$  The notation $q_i$ can
symbolically refer to the $f_im_i$, alphabetical, or numerical
choice of labeling

The collision event between two atoms is defined by preparing the
atoms in states $q_1$ and $q_2$ while they are separated by a large
distance $R$, then allowing them to come together, interact, and
afterwards separate to two atoms in states $q'_1$ and $q'_2$.  If
the two final states are the same as the initial ones, ${q_1,q_2}
={q'_1,q'_2}$, the collision is said to be elastic, and the atoms
have the same relative kinetic energy $E$ before and after the
collision. If one of the final states is different from an initial
state, the collision is said to be inelastic.  This often results in
an energy release that causes a loss of cold atoms when the
energetic atoms escape from the shallow trapping potential.  We will
concentrate primarily on collisions where the two-body inelastic
collision rate is zero or else very small in comparison to the
elastic rate, since this corresponds in practice to most cases of
practical experimental interest.  This condition is necessary for
efficient evaporative cooling or to prevent rapid decay of the cold
gas.  Section \ref{sssec:lossspect} discusses how atom loss due to
3-body collisions can be used to detect the presence of 2-body
resonances.

In setting up the theory for the collision of two atoms, the
scattering channels are defined by the internal states of the two
atoms $1$ and $2$ and the partial wave,
$|\alpha\rangle=|q_1q_2\rangle|\ell m_\ell\rangle$, where
$\langle{\hat R}|\ell m_\ell\rangle=Y_{\ell m_\ell}({\bf{\hat R}})$.
Since for collisions in a magnetic field the quantum number
$M=m_1+m_2+m_\ell$ is strictly conserved, a scattering channel can
be conveniently labeled by specifying the set of quantum numbers
$\{q_1q_2\ell M\}$.  For $s$-waves, where $\ell=m_\ell=0$ and
$M=m_1+m_2$, it is only necessary to specify the quantum numbers
$\{q_1q_2\}$ to label a channel.

When the two atoms are of the same isotopic species, the wave
function must be symmetric (antisymmetric) with respect to exchange
of identical bosons (fermions).  We assume such symmetrized and
normalized functions, as described by \cite{Stoof1988}.  Exchange
symmetry ensures that identical atoms in identical spin states can
only  collide in $s$-, $d$-, $\ldots$ waves for the case of bosons
and in $p$-, $f$- $\ldots$ waves in the case of fermions; in all
other cases, collisions in all partial waves are allowed.

The channel energy $E_\alpha=E(q_1) + E(q_2)$ is the internal energy
of the separated atoms. Let us assume that the atoms are prepared in
channel $\alpha$ with relative kinetic energy $E$ so that the total
energy is $E_\mathrm{tot}=E_\alpha+E$.  Any channel $\beta$ with
$E_\beta \le E_\mathrm{tot}$ is called an open channel, and any
channel with $E_\beta> E_\mathrm{tot}$ is called a closed channel.
A collision can produce atoms in an open channel after the
collision, but not in a closed channel, since the atoms do not have
enough energy to separate to the product atoms.

\subsubsection{Collision rates}\label{sssec:collrate}

The partial collision cross section for starting in open channel $\alpha$
with relative kinetic energy $E$
and ending in open channel $\beta$ can be expressed in terms of the
$S_{\alpha,\beta}(E)$ element of the multichannel unitary scattering matrix $\bf S$.
The cross section for elastic scattering at energy $E$ in channel $\alpha$ is
\begin{equation}
 \sigma_{\mathrm{el},\alpha}(E) = g_\alpha \frac{\pi}{k^2} |1-S_{\alpha,\alpha}(E)|^2 \,,
 \label{II.A.6}
\end{equation}
whereas the unitarity property of ${\bf S}$ allows us to express the
cross section for loss of atoms from channel $\alpha$ as
\begin{equation}
  \sigma_{\mathrm{loss},\alpha}(E) = g_\alpha \frac{\pi}{k^2} \left (1-|S_{\alpha,\alpha}(E)|^2 \right ) \,.
   \label{II.A.7}
\end{equation}
The corresponding partial elastic and inelastic rate coefficients
$K_{\mathrm{el},\alpha}(E)$ and $K_{\mathrm{loss},\alpha}(E)$ are found
by multiplying these partial cross sections by the relative collision velocity
$v=\hbar k/\mu$.
The factor $g_\alpha=1$ except for certain special cases
involving identical particles.  The factor $g_\alpha=2$ for describing
thermalization or inelastic collisions in a normal Maxwellian gas of two
atoms of the same species in identical spin states.  Inelastic decay of a
pure Bose-Einstein condensate has $g_\alpha=1$~\cite{Stoof1989,Kagan1985}.

If only one open channel $\alpha$ is present, collisions are purely
elastic and  $S_{\alpha,\alpha}(E)=\exp(2i\eta_\alpha(E))$.
For $s$-waves the real-valued $\tan \eta_\alpha(E) \to -ka_\alpha$ as $k
\to 0$ and $a_\alpha$ is the scattering length for channel $\alpha$.
When other open channels are present, the amplitude
$|S_{\alpha,\alpha}(E)|$ is no longer unity, and for $s$-wave we can represent
the complex phase $\eta_\alpha(E) \to -k\tilde{a}_\alpha$ for $k \to
0$ in terms of a complex scattering length
\cite{Bohn1996,Balakrishnan1997}
\begin{equation}
 \tilde{a}_\alpha=a_\alpha-ib_\alpha \,,
  \label{II.A.8}
\end{equation}
where $a$ and $b$ are real, and $1-|S_{\alpha,\alpha}(E)|^2  \to
4kb_\alpha \ge 0$ as $k \to 0$.  The threshold behavior is
\begin{equation}
 \sigma_{\mathrm{el},\alpha}(E) = 4 \pi g_\alpha (a_\alpha^2+b_\alpha^2) \,,
  \label{II.A.9}
\end{equation}
for the $s$-wave elastic collision cross section and
\begin{equation}
 K_{\mathrm{loss},\alpha}(E) = \frac{2h}{\mu}g_\alpha b_\alpha \,.
  \label{II.A.10}
\end{equation}
for inelastic collisions that remove atoms from channel $\alpha$.
Both $\sigma_{\mathrm{el},\alpha}$ and $K_{\mathrm{loss},\alpha}$
approach constant values when $E$ is sufficiently small.

The unitarity property of the $S$-matrix also sets an upper bound on
the cross sections.  Since there is a rigorous upper bound
of $|S_{\alpha,\alpha}(E)| \le 1$, we find that the elastic
scattering cross section is maximum
\begin{equation}
 \sigma_{\mathrm{el},\alpha}(E) = \frac{4\pi}{k^2}g_\alpha  \,,
 \label{II.A.6b}
 \end{equation}
for any channel $\alpha$ (and thus any partial
wave $\ell$) when $S_{\alpha,\alpha}(E)=-1$. Furthermore,
$\sigma_{\mathrm{loss},\alpha}(E)$, if nonvanishing, has a maximum
value of $\sigma_{\mathrm{loss},\alpha}(E)=g_\alpha \pi /k^2$ when
$S_{\alpha,\alpha}(E)=0$.  These limits are called the unitarity limits
of the cross sections.  For $s$-wave collisions this limit is approached
at quite low energy approximately equal to $E \approx \hbar^2/(2\mu
a^2_\alpha)$, where $ka_\alpha \approx 1$.

In order to compare with experimental data the partial rate coefficients
must be summed over partial waves and thermally averaged over the
distribution of relative collision velocities at temperature $T$.
This defines the total rate coefficients $K_{\mathrm{el},q_1q_2}(T)$
and $K_{\mathrm{loss},q_1q_2}(T)$ when the atoms are prepared in states
$q_1$ and $q_2$, respectively.  Often the temperatures are sufficiently
small that only the $s$-wave entrance channel contributes.

\subsubsection{Resonance scattering}\label{sssec:resscatt}

The idea of resonance scattering in atomic and molecular systems has
been around since the earliest days of quantum physics, as described
in the introduction.   A conventional ``resonance'' occurs when the
phase shift changes rapidly by $\approx \pi$ over a relatively
narrow range of energy, due to the presence of a quasibound level of
the system that is coupled to the scatttering state of the colliding
atoms.  Such a resonance may be due to a quasibound level trapped
behind a repulsive barrier of a single potential, or may be due to
some approximate bound state which has a different symmetry and
potential from that of the colliding atoms.  The former is commonly
known as a ``shape resonance'', whereas the latter is often called a
``Feshbach resonance'', in honor of Herman Feshbach, who developed a
theory and a classification scheme for resonance scattering
phenomena in the context of nuclear
physics~\cite{Feshbach1958,Feshbach1962}.  We will follow here
Fano's configuration interaction treatment of resonant scattering
\cite{Fano1961}, which is common in atomic physics.  A variety of
treatments of the two-body physics of resonances in the context of
ultracold Bose gases have been given, for example,
\cite{Timmermans1999,Duine2004,Goral2004,Marcelis2004,Raoult2004}.

We first consider the standard scattering picture away from any
collision threshold defined by a two-channel Hamiltonian $H$. Assume
that we can describe our system to a good approximation by two
uncoupled ``bare'' channels, as schematically shown in
Fig.~\ref{fig:intro_potential}.  One is the open background
scattering channel $|bg\rangle$ with scattering states $|E\rangle =
\phi_\mathrm{bg}(R,E) |bg\rangle$ labeled by their collision energy
$E$.  The other is the closed channel $|c\rangle$ supporting a bound
state $|C\rangle = \phi_c(R) |c\rangle$ with eigenenergy $E_c$.  The
functions $\phi_c(R)$ and $ \phi_\mathrm{bg}(R,E)$ are the solutions
to Eq.~(\ref{II.A.1}) for the background potential
$V_\mathrm{bg}(R)$ and the closed channel potential $V_c(R)$
respectively.  Here $\phi_c(R)$ is  normalized to unity.  The
scattering in the open channel is characterized by a background
phase shift $\eta_\mathrm{bg}(E)$. When the Hamiltonian coupling
$W(R)$ between the two channels is taken into account, then the two
states become mixed, or ``dressed'', by the interaction, and the
scattering phase picks up a resonant part due to the bound state
embedded in the scattering continuum,
\begin{equation}
  \eta(E) = \eta_\mathrm{bg}(E) + \eta_\mathrm{res}(E) \,,
  \label{II.A.11}
\end{equation}
where $\eta_\mathrm{res}(E)$ takes on the standard Breit-Wigner form
(Mott and Massey, 1965; Taylor, 1972):
\begin{equation}
   \eta_\mathrm{res}(E) = -\tan^{-1} \left (\frac{\frac{1}{2}\Gamma(E_c)}{E - E_c -\delta E(E_c)} \right )\,.
    \label{II.A.12}
\end{equation}
The interaction $W(R)$, which vanishes at large $R$, determines two key features of the resonance,
namely, its width
\begin{equation}
  \Gamma(E) = 2 \pi |\langle C |  W(R) | E\rangle|^2 \,,
      \label{II.A.13}
\end{equation}
and its shift $\delta E$ to a new position at $E_c+\delta E(E)$,
\begin{equation}
 \delta E(E)= {\cal{P}}\int_{-\infty}^\infty \frac{|\langle C |  W(R) | E' \rangle|^2}{E-E'} dE'
     \label{II.A.14}
\end{equation}
where $\cal{P}$ implies a principal part integral, which includes a
sum over the contribution from any discrete bound states in the
spectrum of the background channel.  When the resonance energy is
not near the channel threshold, it is normally an excellent
approximation to take the width and shift as energy-independent
constants, $\Gamma(E_c)$ and $\delta E(E_c)$,  evaluated at the
resonance energy $E_c$, as in Eq.~(\ref{II.A.12}). The resonance
phase changes by $\approx \pi$ when $E$ varies over a range on the
order of $\Gamma$ from below to above resonance.

The essential difference between conventional and threshold
resonance scattering is that if $E_c$ is close to the open channel
threshold at $E=0$, the explicit energy-dependence of the width and
shift become crucial \cite{Bohn1999,Julienne2006,Marcelis2004}:
\begin{equation}
   \eta_\mathrm{res}(E) = -\tan^{-1} \left (\frac{\frac{1}{2}\Gamma(E)}{E - E_c -\delta E(E)} \right )\,.
       \label{II.A.15}
\end{equation}
The threshold laws for the $s$-wave width and shift as $k \to 0$ are
\begin{eqnarray}
  \frac{1}{2} \Gamma(E) & \to & (ka_\mathrm{bg})\Gamma_0 \,.  \label{II.A.16} \\
   E_c + \delta E(E) & \to & E_0 \,.  \label{II.A.17}
\end{eqnarray}
where $\Gamma_0$ and $E_0$ are $E$-independent constants.
Since $\Gamma(E)$ is positive definite, $\Gamma_0$ has the same sign
as $a_\mathrm{bg}$. Combining these limits with the background phase
property, $ \eta_\mathrm{bg}(E) \to -k a_\mathrm{bg}$, and for the
sake of generality, adding a decay rate $\gamma/\hbar$ for the decay
of the bound state into all available loss channels, gives in the limit
of $k \to 0$,
\begin{equation}
 \tilde{a} = a-ib= a_\mathrm{bg} + \frac{a_\mathrm{bg}\Gamma_0}{-E_0+i(\gamma/2)} \,.
      \label{II.A.18}
\end{equation}
{ The unique role of scattering resonances in the ultracold domain comes from the
ability to tune the threshold resonance position $E_0$ through zero by varying either an
external magnetic field with strength $B$ or optical field with frequency $\nu$.

Both magnetically and optically
tunable resonances are treated by the same theoretical formalism
given above, although the physical mechanisms determining the
coupling and tuning are quite different. }  In the case of a magnetically tunable resonance,
the channel can often be chosen so that $\gamma$ is zero or small
enough to be ignored, whereas optical resonances are always
accompanied by decay processes $\gamma$ due to decay of the excited
state. The resonance strength $\Gamma_0$ is fixed for magnetic
resonances, but $\Gamma_0(I)$ for optical resonances can be turned
off and on by varying the laser intensity $I$. It may also be possible to gain some control over $\Gamma_0$ by using a combination of electric and magnetic fields~\cite{Marcelis2008}.

In the case of a magnetically tunable resonance, there is a
difference $\delta \mu = \mu_\mathrm{atoms} - \mu_c$ between the
magnetic moment $\mu_\mathrm{atoms}$  of the separated atoms and the
magnetic moment $\mu_c$ of the bare bound state $|C\rangle$. {  Thus, the
energy $E_c$ of the state $|C\rangle$ relative to the channel energy of the separated atoms,
\begin{equation}
E_c = \delta \mu (B-B_c) \,,
\label{II.A.20}
\end{equation}
can be tuned by varying the magnetic field, and $E_c$ is zero at
a magnetic field equal to $B_c$. } Then, given that
$\gamma=0$, the scattering length takes on the simple form given in
Eq.~(\ref{II.A.21}),
\begin{equation}
 a(B) = a_\mathrm{bg} - a_\mathrm{bg}\frac{\Delta}{B-B_0} \,,
\end{equation}
where
\begin{equation}
 \Delta = \frac{\Gamma_0}{\delta \mu} \quad {\rm and}\quad
 B_0=B_c+\delta B
      \label{II.A.21b}
\end{equation}
are the width and the position of the singularity in the scattering length,
shifted due to the interaction between the closed and open channels
by an amount $\delta B = -\delta E/\delta \mu$.   Note that $\Delta$
has the same sign as $\delta \mu/a_\mathrm{bg}$.
Figure~\ref{fig:intro_e&a} schematically illustrates the scattering length near
the point of resonance $B_0$.

The complex scattering length of an optically tunable resonance at
laser frequency $\nu$ includes the collisional loss due to excited
state decay~\cite{Fedichev1996,Bohn1999},
\begin{equation}
 \tilde{a}(\nu,I) = a_\mathrm{bg} +\frac{a_\mathrm{bg}\Gamma_0(I)}{h(\nu-\nu_c-\delta \nu(I))+i(\gamma/2)} \,.
      \label{II.A.22}
\end{equation}
where the optically induced width $\Gamma_0(I)$ and shift $\delta
\nu(I)$ are linear in $I$, and $\nu_c$ represents the frequency of
the unshifted optical transition between the excited bound state and
the collisional state of the two atoms at $E=0$.

\begin{figure}
\includegraphics[scale=0.3]{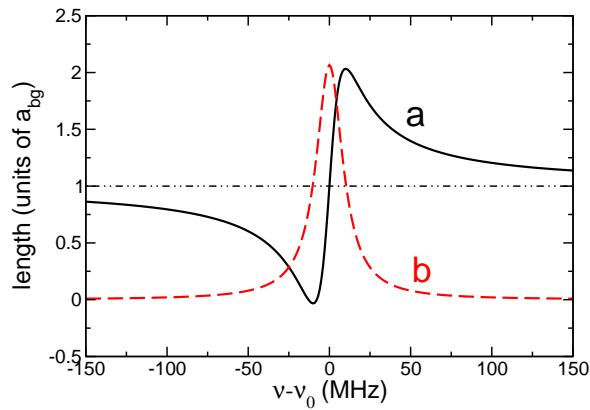}
\caption{Scattering length for an optically tunable Feshbach
resonance as a function of laser tuning $\nu-\nu_0$. The lengths $a$
and $b$ are defined in Eqs.~(\ref{II.A.23}) and (\ref{II.A.24}).
Here $a_{\rm bg}=5.29$ nm, $\Gamma_0/h=21$ MHz at $I=500$
W$/$cm$^2$, $a_{\rm res}=5.47$ nm, and $\gamma/h=20$ MHz. Numerical
values for the strength and spontaneous linewidth of the resonance
are typical for $^{87}$Rb and are taken from Fig.~1 of
\cite{Theis2004}.} \label{fig:opticalscat}
\end{figure}

Whenever bound state decay is present, whether for magnetically or
optically tunable resonances, Eq.~(\ref{II.A.18}) shows that
resonant control of the scattering length,
\begin{equation}
 a= a_\mathrm{bg} - a_\mathrm{res} \frac{\gamma E_0}{E_0^2 + \left( \gamma/2 \right)^2}   \,,
     \label{II.A.23}
\end{equation}
is accompanied by collisional loss given by
\begin{equation}
 b = \frac{1}{2} a_\mathrm{res}  \frac{ \gamma^2}{E_0^2 + \left( \gamma/2 \right)^2} \,.
     \label{II.A.24}
\end{equation}
The resonant length parameter
\begin{equation}
  a_\mathrm{res} = a_\mathrm{bg}\frac{\Gamma_0}{\gamma}
  \label{II.A.25}
\end{equation}
is useful for defining the strength of an optical resonance
\cite{Bohn1997,Ciurylo2005} or any other resonance with strong decay
\cite{Hutson2006}.   Figure~\ref{fig:opticalscat} gives an example
of such a resonance.  The scattering length has its
maximum variation of $a_\mathrm{bg} \pm a_\mathrm{res}$ at $E_0=\pm
\gamma/2$, where $b=a_\mathrm{res}$. Resonances with $a_\mathrm{res}\leq
|a_\mathrm{bg}|$ only allow relatively small changes in scattering
length{, yet $b$ remains large enough that they}
are typically accompanied by large inelastic rate coefficients.
On the other hand, if $a_\mathrm{res} \gg |a_\mathrm{bg}|$, losses can be
overcome by using large detuning, since the change in scattering length
is $a-a_\mathrm{bg}=-a_\mathrm{res} (\gamma/E_0)$ when $|E_0|\gg \gamma$,
whereas $b/|a-a_\mathrm{bg}|=\frac{1}{2}|\gamma/E_0| \ll 1$.

The resonance length formalism is quite powerful.  By introducing
the idea of an energy-dependent scattering length
\cite{Blume2002,Bolda2002} it can be extended to Feshbach resonances
in reduced dimensional systems such as pancake or cigar shaped
optical lattice cells \cite{Naidon2006}.

While this discussion has concentrated on resonant scattering
properties for $E>0$, the near-threshold resonant properties of
bound Feshbach molecules for energy $E<0$ are very important aspects
of Feshbach physics; see Fig.~\ref{fig:intro_e&a} and
~\cite{Kohler2006}.   In particular, as the bound state becomes more
deeply bound, the closed channel character of the bound state
increases and the binding energy $E_b$ is no longer described by the
universal expression in Eq.~(\ref{II.A.4}).    The ''dressed'' or
true molecular bound state of the system with energy $-E_b$ is a
mixture of closed and background channel components,
\begin{equation}
 | \psi_b(R)\rangle = \sqrt{Z} \phi_c(R) |c\rangle +\chi_\mathrm{bg}(R) |bg\rangle \,,
 \label{eq:Zpsi}
\end{equation}
where $0 \le Z \le 1$ represents the fraction of the eigenstate $|
\psi_b(R)\rangle$ in the closed channel component \cite{Duine2003}.
Unit normalization of $ | \psi_b(R)\rangle$ ensures that $\int
|\chi_\mathrm{bg}(R)|^2dR = 1-Z$.  Since the variation of the energy
$-E_b$ with a parameter $x$ of the Hamiltonian satisfies
the Hellman-Feynmann theorem
 $\partial(-E_b)/\partial x =  \langle  \psi_b |\partial H/\partial  x |  \psi_b \rangle$,
it follows from Eq.~(\ref{eq:Zpsi}) that
\begin{equation}
Z =\frac{ \partial(-E_b)}{\partial E_c} = \frac{\delta \mu_b}{\delta \mu} \,.
\label{eq:Zeq}
\end{equation}
Here $\delta \mu_b = \partial E_b/\partial B =  \mu_\mathrm{atoms} -
\mu_b$ is the difference between the magnetic moment of the
separated atoms and the  magnetic moment $\mu_b$ of the ''dressed''
molecular eigenstate.  Since $\delta \mu_b$ vanishes in the limit $B
\to B_0$, where $E_b \to 0$ according to the universality condition
in Eq.~(\ref{II.A.4}), then $Z$ vanishes in this limit also.
Section \ref{sssec:AnalMol} develops more specific  properties and
conditions for $E_b$ and $Z$ in this limit.

\subsection{Basic molecular physics}\label{ssec:basicmol}

\begin{figure}
\includegraphics[width=2.8in,clip,,angle=270]{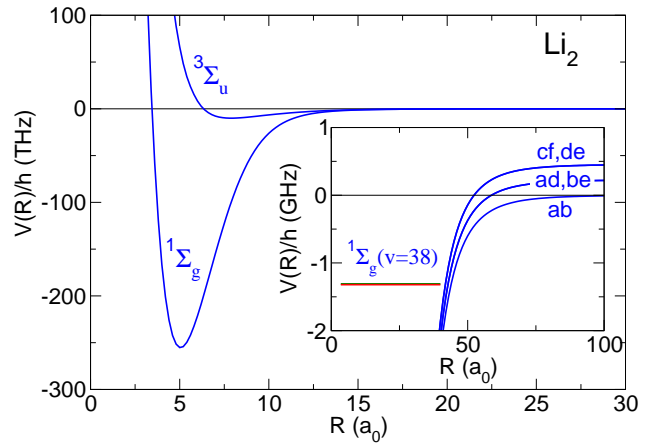}
\caption{Molecular potentials $V(R)/h$ versus $R$ of the two
electronic states of Li$_2$ that correlate with two separated $^2$S
atoms.  The inset shows an expanded view of the long-range $s$-wave
potentials of $^6$Li at $B=0$, indicating the five hyperfine states
of the separated atoms (see Fig.~\ref{fig:Li_Zeeman}) for which the
total angular momentum has projection $M=0$.  The inset also shows
the last two nearly degenerate bound states (unresolved on the
figure) of the $^6$Li$_2$ molecule from a coupled channels
calculation.   It is a good approximations to label these nearly
degenerate levels as the $I=0$ and $2$ components of the total
nuclear spin ${\bf I}={\bf I}_1+{\bf I}_2$ of the last $v=38$
vibrational level of the $^1\Sigma_g^+$ potential.}
\label{fig:Li_2_potentials}
\end{figure}

Most atoms that can be trapped at ultracold temperatures have ground
S-states with zero electronic orbital angular momentum ($L=0$), as
for alkali-metal or alkaline-earth-metal atoms.  The collision
between two atoms is controlled by the electronic Born-Oppenheimer
interaction potential(s) between them.  All potentials are isotropic
for the interaction of two $S$-state atoms.  We restrict our
discussion of molecular physics to such cases.  Figure
~\ref{fig:Li_2_potentials} shows as an example the $^1\Sigma_g^+$
and $^3\Sigma_u^+$ potentials for two ground state $^2$S Li atoms,
which are analogous to the similar  potentials for the H$_2$
molecule or other alkali-metal atoms.  The superscripts $1$ and $3$
refer to singlet and triplet coupling of the spins of the unpaired
electrons from each atom, i.e., the total electron spin ${\bf
S}={\bf S}_1+{\bf S}_2$ has quantum numbers $S=$ 0 and 1. The
$\Sigma$ refers to zero projection of electronic angular momentum on
the interatomic axis for the $S$-state atoms, and $g/u$ refers to
{\it gerade/ungerade} electronic inversion symmetry with respect to
the center of mass of the molecule.  The $g/u$ symmetry is absent
when the two atoms are not of the same species.

The Born-Oppenheimer potentials are often available from {\it ab
initio} or semi-empirical sources.  When $R$ is sufficiently small,
typically less than $R_\mathrm{ex} \approx 20$ a$_0$ $\approx$ 1 nm
for alkali-metal atoms, electron exchange and chemical bonding
effects determine the shape of the potentials.  For $R \gg
R_\mathrm{ex}$, the potentials are determined by the long-range
dispersion interaction represented by a sum of second-order
multipolar interaction terms.

\subsubsection{Van der Waals bound states and scattering}\label{sssec:vdw}

Many aspects of ultracold neutral atom interactions and of Feshbach
resonances, in particular,  can be understood qualitatively and even
quantitatively from the scattering and bound state properties of the
long range van der Waals potential.  The properties of this
potential relevant for ultracold photoassociation spectroscopy have
been reviewed in \cite{Jones2006}.  Its analytic properties are
discussed by \cite{Mott1965,Gribakin1993,Gao1998,Gao2000}.

In the case of $S$-state atoms, the lead term in the long-range part
of all Born-Oppenheimer potentials for a given atom pair
has the {\it same} van
der Waals potential characterized by a single $C_6$ coefficient
for the pair.
Consequently, all $q_1 q_2$ spin combinations have the long-range
potential
\begin{equation}
   V_\ell(R) =
     -\frac{C_6}{R^6} + \frac{\hbar^2}{2 \mu} \frac{ \ell(\ell+1)}{R^2} \,.
   \label{II.B.1.1}
\end{equation}

A straightforward consideration of the units in Eq.~(\ref{II.B.1.1})
suggests that it is useful to define length and energy scales
\begin{equation}
  R_\mathrm{vdw} = \frac{1}{2} \left ( \frac{2 \mu C_6}{\hbar^2} \right )^{1/4}\,{\rm and} \,\,\,\,\,
  E_\mathrm{vdw}= \frac{\hbar^2}{2 \mu} \frac{1}{R_\mathrm{vdw}^2}\,.
  \label{II.B.2}
\end{equation}
\cite{Gribakin1993} defined an alternative van der Waals length
scale which they called the mean scattering length:
\begin{equation}
\bar{a}=4\pi/\Gamma(1/4)^2\,R_\mathrm{vdw}=0.955978\dots\,R_\mathrm{vdw}\,,
\label{II.B.2b}
\end{equation}
 where $\Gamma(x)$ is the Gamma function.   A corresponding energy scale is $\bar{E}=\hbar^2/(2\mu \bar{a}^2)= 1.09422\ldots E_\mathrm{vdw}$.  The parameter $\bar{a}$ occurs frequently in formulas based on the van der Waals potential.   Table~\ref{tab:vdw} gives the values of $R_\mathrm{vdw}$ and $E_\mathrm{vdw}$ for several cases.  Values
of $C_6$ for other systems are tabulated in
\cite{Derevianko1999,Porsev2006,Tang1976}.

The van der Waals energy and length scales permit a simple physical
interpretation~\cite{Julienne1989}.   A key property for ultracold
collisions is that $C_6/R^6$ becomes large compared to the collision
energy $E$ when $R < R_\mathrm{vdw}$.  Thus, the wave function for
any partial wave oscillates rapidly with $R$ when
$R<R_\mathrm{vdw}$, since the local momentum $\hbar
k(R)=\sqrt{2\mu(E-V(R))}$ becomes  large compared to the asymptotic
$\hbar k$.   On the other hand, when $R>R_\mathrm{vdw}$, the
wavefunction approaches its asymptotic form with oscillations on the
scale determined by the long de Broglie wavelength of the ultracold
collision.  The energy scale $E_\mathrm{vdw}$ determines the nature
of the connection between the long- and short-range forms of the
wavefunction.  The de Broglie wavelength $\lambda=2 \pi
(R_\mathrm{vdw})$ for $E=E_\mathrm{vdw}$.  When $E\ll
E_\mathrm{vdw}$ so that $\lambda \gg R_\mathrm{vdw}$,  a WKB
connection can not be made near $R_\mathrm{vdw}$ between the
asymptotic $s$-wave and the short-range wavefunction (see Fig.~15 of
\cite{Jones2006}).  Consequently, the quantum properties of the
collision are manifest for $E < E_\mathrm{vdw}$.

The van der Waals length also characterizes the extent of
vibrational motion for near-threshold bound state.   The outer
turning point for classical motion for all low $\ell$ bound states
is on the order of $R_\mathrm{vdw}$.  The wave function for $\ell=0$
oscillates rapidly for  $R < R_\mathrm{vdw}$ and decays
exponentially as $e^{-k_b R}$ for $R \gg R_\mathrm{vdw}$, where
$\hbar^2 k_b^2/(2\mu)$ is the binding energy.   The only case where
the wave function extends far beyond $R_\mathrm{vdw}$ is that of the
last $s$ wave bound state for the case of the universal halo
molecule, where $a \gg R_\mathrm{vdw}$; see
Secs.~\ref{ssec:basiccoll} and \ref{sssec:halo}.

The van der Waals potential determines the interaction over a wide
zone between $R_\mathrm{vdw}$ and the much smaller $R_\mathrm{ex}$
where chemical forces become important.  Thus, near-threshold bound
and scattering state properties are determined to a large extent by
the long-range van der Waals potential.  The effect of  short-range
is then contained within the {\em phase} of the wave function, or
equivalently, the log derivative~\cite{Moerdijk1994,Vogels2000}.   More
precisely, for any $R_z$ satisfying $R_\mathrm{ex} <R_z \ll
R_\mathrm{vdw}$ so that $k(R_z) \gg k$, the wave function phase is
nearly independent of $E$ and almost the same for all near-threshold
bound or scattering states.  In fact, the phase is nearly
independent of partial wave $\ell$ as well, since the centrifugal
potential is typically small compared to the van der Waals
potential for such an  $R_z$.  Using this phase as a boundary
condition for propagating the wavefunction to large $R$ in the
asymptotic domain determines the energy-dependent scattering phase
$\eta_\ell(E)$ and bound state energies.  In fact the phase of the
wave function in the zone $R_\mathrm{ex} <R_z \ll R_\mathrm{vdw}$ is
uniquely related to the $s$-wave scattering length~\cite{Gao1998b}.
Consequently, to a good approximation the near-threshold bound
states and scattering properties for all low partial waves are
determined by the $s$-wave scattering length, the $C_6$ constant,
and the reduced mass~\cite{Gao2001}.

\begin{figure}
\vspace*{0.2cm}
\includegraphics[width=3.0in]{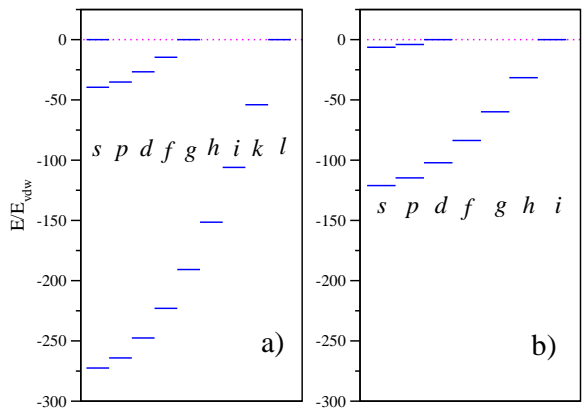}
\vspace*{0.2cm} \caption{Bound-state energies of the last
vibrational levels of two atoms interacting via a van der Waals
potential as a function of partial wave $\ell$. The zero of energy
is at two free atoms with zero collision energy (``at threshold'').
The lowest partial waves are shown. Panel a) shows the bound state
structure up to $\ell=8$ when the scattering length of the colliding
atoms is infinite or equivalently that there is an $s$-wave bound
state with zero binding energy. Panel b) shows the bound state
structure up to $\ell=6$ when the scattering length is
$a=0.956\ldots R_\mathrm{vdw}=\bar{a}$, or equivalently there is a
$d$-wave bound state with zero binding energy. The length
$R_\mathrm{vdw}$ and energy $E_\mathrm{vdw}$ are defined in the
text. Adapted from \cite{Gao2000}. } \label{fig:C6boundstates}
\end{figure}

\cite{Gao2000} has worked out the energies $E_{n,\ell}$ of the bound
states of all partial waves for a van der Waals potential as a
function of the $s$-wave scattering length, where $n=-1,-2,\ldots$
is the vibrational quantum number and $\ell$ is the rotational
quantum number of the bound state. He showed that the energies of
weakly bound states have a $\Delta\ell=4$ periodicity.
Figure~\ref{fig:C6boundstates} shows bound state energies as
function of $\ell$ for two values of $a$. In the left panel
$a=\pm\infty$ so there is a $s$-wave bound state with $E=0$. The
figure shows that for $\ell=4$ there also is bound state with
$E/E_\mathrm{vdw}=0$. In fact \cite{Gao2000} showed for
$\ell=8,12,\dots$ there will be a bound state at zero energy as
well.  The right panel in Fig.~\ref{fig:C6boundstates} shows that
when $a=\bar{a}$ there is a bound state at zero energy for $\ell=2$.
There also will be a bound state at zero energy for
$\ell=6,10,\dots$.

Figure~\ref{fig:C6boundstates} can also be used to define the
concept of ``energy bins'' in which, regardless of the value of $a$,
there must be a bound state.  Bins are most easily defined by
starting from a case with a bound state at zero binding energy. By
changing the short-range log derivative its binding energy can be
increased, or its energy lowered, and at some point the binding
energy is so large that a {\it new} bound state appears at zero
binding energy. This is exactly the situation depicted in
Fig.~\ref{fig:C6boundstates}a for $s$ and $g$ waves. In other words,
for $s$ waves there must be a $n=-1$ bound state between $-39.5\,
E_\mathrm{vdw}$ and $0\, E_\mathrm{vdw}$, while for $g$ waves there
must be a $n=-1$ bound state between $-191\, E_\mathrm{vdw}$ and
$0\, E_\mathrm{vdw}$. The $n=-2$ $s$-wave bound state appears
between $-272\, E_\mathrm{vdw}$ and $-39.5\, E_\mathrm{vdw}$.
Figure~\ref{fig:C6boundstates}b can similarly be used to define the
bins for other waves.

When the scattering length is large compared to $\bar{a}$ and
positive, a simple expression for the van der Waals correction to
the binding energy of the last $s$-wave bound state can be worked
out \cite{Gribakin1993}:
\begin{equation}
  E_{-1,0} = -\frac{\hbar^2}{2\mu (a-\bar{a})^2} \,.
  \label{II.B.3}
\end{equation}
The universal formula in Eq.~(\ref{II.A.4}) only applies in the
limit that $a \gg \bar{a}$ and $|E_{-1,0}| \ll E_\mathrm{vdw}$.
\cite{Gao2004} has worked out higher order corrections to the
binding energy due to the van der Waals potential, which can be
recast as
\begin{equation}
  E_{-1,0} = -\frac{\hbar^2}{2\mu (a-\bar{a})^2}[1+\frac{g_1\bar{a}}{a-\bar{a}}+\frac{g_2\bar{a}^2}{(a-\bar{a})^2}+...]\,.
  \label{II.B.3a}
\end{equation}
Here $g_1=\Gamma(1/4)^4/6\pi^2-2=0.9179...$, $g_2=(5/4)
g_1^2-2=-0.9468...$ are constants.

Similarly, the effective range of the potential in
Eq.~(\ref{II.A.3}) is also determined from the van der Waals
potential, given the $s$-wave scattering length
\cite{Flambaum1999,Gao1998b}:
\begin{equation}
 r_0=\frac{\Gamma(1/4)^4}{6\pi} \bar{a} \left ( 1 -2\frac{\bar{a}}{a}+2\left (\frac{\bar{a}}{a} \right )^2 \right)
 \label{II.B.4}
\end{equation}
where $\Gamma(1/4)^4/(6\pi)\approx2.9179$.  When $a \gg \bar{a}$,
this simplifies to $r_0=2.9179\bar{a}$.  Note that $r_0$ diverges as
$a \to 0$.

\begin{table}
\caption{ Characteristic van der Waals scales $R_\mathrm{vdw}$ and $
E_\mathrm{vdw}$ for several atomic species. (1 amu = 1/12 mass of a
$^{12}$C atom, 1 au= 1 $E_h a_0^6$ where $E_h$ is a hartree and 1
$a_0$= 0.0529177... nm) } \label{tab:vdw}
\begin{tabular}{ccccccl}
\hline\hline
Species & mass & C$_6$  & $R_\mathrm{vdw}$ & $E_\mathrm{vdw}/k_B$ & $E_\mathrm{vdw}/h$\\
     & (amu) & (au) & ($a_0$) & (mK) & (MHz) \\
\hline
${^6}$Li & 6.0151223 & 1393.39$^a$ & 31.26 & 29.47  & 614.1 \\
${^{23}}$Na & 22.9897680  & 1556$^b$  & 44.93  & 3.732   &   77.77 \\
${^{40}}$K & 39.9639987  & 3897$^b$  & 64.90  & 1.029   & 21.44  \\
$^{40}$Ca &  39.962591 & 2221$^c$ & 56.39 & 1.363  & 28.40 \\
${^{87}}$Rb & 86.909187  & 4698$^d$  & 82.58  & 0.2922  & 6.089\\
$^{88}$Sr &  87.905616 & 3170$^c$ & 75.06 & 0.3497 & 7.287\\
${^{133}}$Cs & 132.905429  & 6860$^e$  & 101.0  & 0.1279  & 2.666\\
\hline\hline
\end{tabular}\\
a. \cite{Yan1996}
b. \cite{Derevianko1999}\\
c. \cite{Porsev2002}\\
d. \cite{vanKempen2002} e. \cite{Chin2004}
\end{table}

The energy levels of the van der Waals potential are not exact due
to the slight influence from the actual short-range potential and
extremely long-range retardation corrections.  They are nevertheless
relatively accurate guides to the expected energy spectrum for real
molecules. For example when the scattering length is slightly larger
than $\bar{a}$, which corresponds to Fig.~\ref{fig:C6boundstates}b
with all bound states shifted to slightly more positive energies, the
$d$-wave bound state becomes a shape resonance, that is, a decaying
quasibound state with $E>0$ trapped behind the $d$-wave centrifugal
barrier. For $^{23}$Na and $^{87}$Rb the experimentally observed
scattering length is 10\%-20\% larger than  $\bar{a}$ and, indeed, in
both cases a $d$-wave shape resonance has been observed under various
circumstances \cite{Samuelis2000,Boesten1997,Thomas2004,Buggle2004}.
Similarly, a $p$-wave shape resonance occurs when $a$ is slightly
larger than $2\bar{a}$, as for $^{40}$K~\cite{DeMarco1999} and
$^{171}$Yb~\cite{Kitagawa2007}.  In addition~\cite{Kitagawa2007} show how
the scattering length and binding energies of the last few bound states
for the single potential of the Yb $+$ Yb  interaction are related as
the reduced mass is changed by using different isotopic combinations of
Yb atoms. The scattering length and binding energies can be ''tuned''
over a wide range by choosing different pairs of atoms among the seven
stable isotopes of Yb.

The $\delta\ell=4$ characteristic of van der Waals potentials also
has practical consequences for ultra-cold scattering.  For $^{85}$Rb
the scattering length has been found to be large compared to
$R_\mathrm{vdw}$ and a $g$-wave shape resonance has been observed
\cite{Boesten1996}.  For $^{133}$Cs the scattering length is large
compared to $R_\mathrm{vdw}$, and numerous $g$-wave bound states
with binding energies much smaller than $E_\mathrm{vdw}$ were
observed by \cite{Chin2004} at low magnetic field. In fact, some of
these bound states appear as magnetic Feshbach resonances in the
collision of two Cs atoms. Recently, a weakly-bound $l=8$ or
$l$-wave state has been observed as well~\cite{Mark2007a}.

\subsubsection{Entrance- and closed-channel dominated resonances:
Resonance strength}\label{sssec:resstrength}

The van der Waals theory is very useful for characterizing and
classifying the basic properties of the resonances discussed in
Section~\ref{sssec:resscatt} by expressing lengths in units of
$\bar{a}$ and energies in units of $\bar{E}$; see
Eq.~(\ref{II.B.2b}). The numerator of the resonant term in
Eq.~(\ref{II.A.18}) defines a resonance strength parameter to be
$a_\mathrm{bg}\Gamma_0$, { where $\Gamma_0=\delta \mu \Delta$;
see Eq.~(\ref{II.A.21b}).   It is helpful to define a dimensionless
resonance strength parameter $s_\mathrm{res}$ to be
\begin{equation}
  s_\mathrm{res} = r_\mathrm{bg} \frac{\Gamma_0}{\bar{E}}
  = \frac{a_\mathrm{bg} }{\bar{a}} \frac{\delta \mu \Delta}{\bar{E}} \,.
  \label{II.B.5}
\end{equation} }
where $r_\mathrm{bg}=a_\mathrm{bg}/\bar{a}$ is the dimensionless
background scattering length.  The sign of $s_\mathrm{res}$ is
always positive.  The resonance phase in Eq.~(\ref{II.A.15}) is
determined by the tunable resonance position and the resonance width
and shift.  In the limit $E \to 0$, both the width
 \begin{equation}
 \frac{1}{2}  \Gamma(E)= (k\bar{a})(\bar{E}s_\mathrm{res})  \label{II.B.6}
 \end{equation}
 and the shift~\cite{Goral2004,Julienne2006}
 \begin{equation}
 \delta E =\frac{1-r_\mathrm{bg}}{1+\left ( 1-r_\mathrm{bg} \right )^2} (\bar{E}s_\mathrm{res}) \,,
 \label{II.B.7}
 \end{equation}
are proportional to $\bar{E} s_\mathrm{res}$.
Section~\ref{sssec:Li6res} describes widths and shifts for some
typical resonances.  Sections~\ref{sssec:anvdwmod},
\ref{sssec:SquareWell}, and \ref{sssec:AnalMol}   give additional
analytic properties of threshold scattering and bound states
associated with Feshbach resonances and show how Eq.~(\ref{II.B.7})
can be derived.

{ The strength parameter $s_\mathrm{res}$ allows us to classify
Feshbach resonances into two limiting cases. \cite{Stoll2005,Kohler2006}
used $\eta=1/s_\mathrm{res}$ to do this. When $s_\mathrm{res}\gg 1$,
the resonance is called an \emph{entrance channel dominated
resonance}. Here, the near-threshold scattering and bound states
have the spin character of
the entrance channel for detuning $E_0$ over a large fraction of the
width $\Gamma_0$ and thus for $B-B_0$ over a large fraction of the resonance width
$\Delta$.  In this regime, the resonance can be well modeled by the $B$-dependent
scattering length of Eq.~(\ref{II.A.21}). The bound state is universal with
$Z\ll 1$ (see Eq.~(\ref{eq:Zpsi})) over this large detuning range
and with a binding energy well-approximated by
Eq.~(\ref{II.A.4}).   Resonances of this
type have the largest resonance width $\Delta$ and are conventionally
called ``broad resonances''.}

{ Resonances with $s_\mathrm{res}\ll 1$ are called
\emph{closed channel dominated resonances}.  Here, the near-threshold
scattering and bound states have the spin character of
the entrance channel only over a small fraction of the
width $\Gamma_0$ near $E_0=0$ and thus over a
small fraction of the resonance width $\Delta$ near $B=B_0$.
A universal bound state only exists over this small detuning range.
Thus, the closed channel fraction $Z$ is only small near $B=B_0$ and is
near unity over a wide detuning range away from $B=B_0$.  Such
resonances need to be modeled by a coupled channels description.
Resonances of this type often have a small
width $\Delta$ and are conventionally called ``narrow resonances.''

It should be emphasized that the conventional use of ``broad'' or
``narrow'' resonances referring to those that can or cannot be
modelled by single channel model is not rigorously defined.  Exceptions
exist where resonances with apparently
broad widths are actually closed channel dominated. The terms introduced here, entrance
and closed channel dominance, better reflect the nature of
the near threshold states over a detuning range on the order of the width
$\Delta$ and can be unambiguously assigned to a
resonance by evaluating $s_\mathrm{res}$.}

{ Section \ref{sssec:Li6res} illustrates the differences
between entrance and closed channel dominated resonances by giving
specific examples of such resonances.
Section~\ref{sssec:AnalMol} develops a simple model for the bound
states for any type of resonance, and shows that the norm $Z$ of the
closed channel part vanishes in the limit that $E_0 \to 0$ near the point of
resonance $B_0$, even for closed channel dominated resonances.
~\cite{Szymanska2005} discuss in detail the implication of the distinction between
open and closed channel dominance for the modeling of many-body systems, a topic
that is beyond the scope of this review.}

Figure~\ref{fig:ResStrength} in the Appendix illustrates the wide
range of resonance strengths $s_\mathrm{res}$ and widths $\Delta$
observed for various alkali atom resonances.   Broad resonances with
$\Delta$ larger than $\sim$ 1 G tend to { have $s_\mathrm{res} >1$ and
thus be } entrance channel
dominated ones.  Narrow resonances with  $\Delta$ smaller than
$\sim$ 1 G tend to { have $s_\mathrm{res} <1$ and
thus be } closed channel dominated ones.
{ A notable exceptions is the $^7$Li 737 G resonance with
$s_\mathrm{res} <1$ that is very broad yet tends towards being closed
channel dominated; see Section~\ref{sssec:Li6res}.}

Equation~(\ref{II.B.6}) allows us to address the question whether a
sharp resonance feature appears at small but finite collision energy
above threshold.  A condition for having a sharp resonance is that
the width $\frac{1}{2}\Gamma(E)$ should be smaller than the
collision energy $E$.  It is convenient to rewrite
Eq.~(\ref{II.B.6}) as $\frac{1}{2}\Gamma(E) =
(s_\mathrm{res}/(k\bar{a})) E$.  For an entrance channel dominated
resonance with $s_\mathrm{res} \gg 1$ and $k\bar{a} <1$ or
$E<\bar{E}$, it follows that $\frac{1}{2}\Gamma(E)
>E$ .  Thus, there can be no sharp resonance features evident in the
above-threshold phase $\eta(E,B)$ of an entrance channel dominated
resonance when $E <\bar{E}$.  A sharp resonance feature can only
appear when $E \gg \bar{E}$.   \cite{Nygaard2006} illustrate this
case for a resonance involving $^{40}$K  atoms.   On the other hand,
for a closed channel dominated resonance with $s_\mathrm{res} \ll 1$
a sharp resonance feature in $\eta(E,B)$ with $\frac{1}{2}\Gamma(E)
\ll E$ can appear immediately above threshold.

\subsubsection{Coupled channels picture of molecular interactions}\label{sssec:coupchan}

While many insights can be gained from the properties of the long
range van der Waals potential, actual calculations require taking
into account the full molecular Hamiltonian, including not only the
full range of the Born-Oppenheimer potentials but also the various
spin-dependent couplings among them.  In general, the potential
should be viewed as a spin-dependent potential matrix, the elements
of which account for the interaction among the various spin states
of the atoms.  The wave function for atoms prepared in channel
$\alpha$ can be written as a coupled channels expansion in the
separated atom spin basis described in Section ~\ref{sssec:channels}
\cite{Mott1965,Stoof1988, Gao1996, Mies2000a,Hutson2008}:
\begin{equation}
 | \psi_\alpha(R,E)\rangle = \sum_\beta |\beta \rangle \phi_{\beta\alpha}(R,E)/R \,.
  \label{eq:psi}
\end{equation}
The allowed states $|\beta\rangle$ in this expansion are those that
have the same projection quantum number $M=m_1+m_2+m_\ell$ (see
Section~\ref{sssec:channels}).  In addition parity conservation
implies that all partial waves $\ell_\beta$ in the expansion are
even if $\ell_\alpha$ is even and odd if $\ell_\alpha$ is odd.

Subsituting Eq.~(\ref{eq:psi}) in the Schr{\"o}dinger equation gives
the coupled channels equations for the vector $\mathbf{\phi}(R,E)$:
\begin{equation}
- \frac{\hbar^2}{2\mu}\frac{d^2\mathbf{\phi}(R,E)}{dR^2} +  {\bf
V}(R) \mathbf{\phi}(E,R) = E \mathbf{\phi}(E,R) \,.
 \label{eq:SE}
\end{equation}
The solutions to these coupled equations give the bound and
scattering states of the interacting atoms.  The potential matrix
${\bf V}(R)$ gives the matrix elements of the Hamiltonian between
the channel basis sets.  It takes on the following form in the
asymptotic spin basis:
\begin{equation}
V_{\alpha\beta} = \left (E_\alpha + \frac{\hbar^2 \ell_\alpha(\ell_\alpha+1)}{2\mu
r^2} \right ) \delta_{\alpha\beta} +V_\mathrm{int,\alpha \beta}(R)
\label{eq:V}
\end{equation}
The interaction matrix $V_\mathrm{int}(R)$ contains the electronic
Born-Oppenheimer potentials discussed in Section~\ref{ssec:basicmol}
and the relativistic electron spin-dependent interactions:
\begin{equation}
  {\bf V}_\mathrm{int}(R) = {\bf V}_\mathrm{el}(R) + {\bf V}_\mathrm{ss}(R)
  \label{eq:Vint}
\end{equation}
For collisions of $S$ state atoms the term ${\bf V}_\mathrm{el}(R)$
represents the strong isotropic electronic interaction that is
diagonal in $\ell$ and $m_\ell$ but off-diagonal in the atomic
channel quantum numbers $q_1 q_2$.  The diagonal elements
$V_{\mathrm{el},\alpha\alpha}$ vary at long range as the van der
Waals potential (see Fig.~\ref{fig:Li_2_potentials}).
It has normally been unnecessary to include small retardation or
nonadiabatic corrections to long range molecular potentials in order
to fit experimental data on ground state collisions  within their
experimental error; see for example~\cite{Kitagawa2007}.  The
off-diagonal elements $V_{\mathrm{el},\alpha\beta}$, where $\beta
\ne \alpha$, decrease exponentially at large $R$ as the exchange
potential and become small compared to the atomic hyperfine
splitting for $R>R_\mathrm{ex}$.   The ${\bf V}_\mathrm{el}$
coupling is responsible for elastic scattering and inelastic
spin-exchange collisions and gives rise to the largest resonance
strengths.

The term ${\bf V}_\mathrm{ss}(r)$ in Eq.~(\ref{eq:Vint}) represents
 weak relativistic spin-dependent interactions.  These include the spin-spin dipole interaction~\cite{Stoof1988,Moerdijk1995} and the second-order spin-orbit interaction~\cite{Kotochigova2000}, important for heavy atoms~\cite{Leo2000}.  The two contributions are both anisotropic and are
off-diagonal in both $q_1 q_2$ and $\ell$.  Thus, ${\bf
V}_\mathrm{ss}(r)$ couples different partial waves.  At long range
${\bf V}_\mathrm{ss}(r)$ is proportional to $\alpha^2/R^3$, where $\alpha=1/137.0426$ is
the fine structure constant.  This anisotropic potential only
contributes diagonal terms for partial waves $\ell \ge 1$, and does
not contribute to the potential $V_\mathrm{int,\alpha \alpha}(R)$
when $\alpha$ represents an $s$-wave channel.   The ${\bf
V}_\mathrm{ss}(r)$ coupling is responsible for weak inelastic
relaxation and normally gives rise to small resonance
strengths.

The Born-Oppenheimer potentials are normally never known with
sufficient accuracy to permit accurate calculations of threshold
scattering properties.  Consequently, it is usually necessary to
vary the short range potentials over some range of $R <
R_\mathrm{ex}$ to calibrate theoretical models so they reproduce
measured threshold bound state or scattering data.  In some cases
the van der Waals coefficients are accurately known, whereas in
other cases they need to be varied to fit the data as well.  Once
this is done, coupled channels theoretical models typically are
robust and predictive of near-threshold collision and bound state
properties.  Some examples of high quality theoretical models based
on fitting Feshbach resonance data are given by
\cite{vanAbeelen1999} for $^{23}$Na,  \cite{Chin2000} and
\cite{Leo2000} for $^{133}$Cs, \cite{Marte2002} for $^{87}$Rb,
\cite{Bartenstein2005} for $^6$Li, \cite{Werner2005} for $^{52}$Cr,
and \cite{Ferlaino2006,Pashov2007} for $^{40}$K$^{87}$Rb.
Information on other models can be found in the references listed
in Sections \ref{ssec:species_B} and \ref{ssec:species_C}.

\subsubsection{Classification and molecular physics of Feshbach resonances}\label{sssec:feshcoupl}

The previous Sections have laid the groundwork for classifying and
understanding the properties of Feshbach resonance states in
ultracold collisions of ground $S$-state atoms.  This classification
can be made according to the quantum numbers  $\{q_1q_2\ell
M|q_\mathrm{c}\ell_\mathrm{c}\}$, where $\{q_1q_2\ell M\}$
characterize the entrance channel (see Section~\ref{sssec:channels})
and $\{q_\mathrm{c}\ell_\mathrm{c}\}$ characterize the "bare" closed
channel bound state that gives rise to the resonance.  Such a bound
state has the same $M$ as the entrance channel.   Some possible
choices for quantum numbers comprising the composite $q_\mathrm{c}$
are given below.

It is important to note that $\ell_c$ need not be the same as the
entrance channel partial wave $\ell$.  Parity conservation ensures
that $|\ell -\ell_\mathrm{c}|$ is even.  In the case of two ${\bf
L}=0$ atoms  the ${\bf V}_\mathrm{el}$ term is isotropic and only gives
rise to nonzero matrix elements when $\ell_\mathrm{c} =\ell$.
On the other hand, $\ell_\mathrm{c}$ can be different from $\ell$ for the
anisotropic ${\bf V}_\mathrm{ss}$ term.  We
are primarily concerned with entrance channel $s$-waves, although
some resonances with $p$-wave (e.g., $^6$Li, $^{40}$K, $^{133}$Cs)
or $d$-wave ($^{52}$Cr) entrance channels are known in the $\mu$K
domain.

We find it is convenient to  designate resonances according to the
value of the closed channel bound state quantum number
$\ell_\mathrm{c}$, as shown in Table~\ref{tab:resclass}.  If
$\ell_\mathrm{c}$ is even (odd), we assume an $s$- ($p$-) wave
entrance channel unless otherwise stated.  The strongest resonances
with the largest widths $\Delta$ are $s$-wave resonances with
$\ell=\ell_\mathrm{c}=0$, and are due to the ${\bf V}_\mathrm{el}$
term in the Hamiltonian.   A number of weak resonances with small
$\Delta$ are known where the $s$-wave entrance channel is coupled
through the ${\bf V}_\mathrm{ss}$ term to bound states with even
$\ell_\mathrm{c}$ such as 2 or 4.   Following
Table~\ref{tab:resclass} the latter are designated as $d$-wave or
$g$-wave resonances, respectively.  For example,  $d$-wave resonances
are known for $^{87}$Rb \cite{Marte2002} and $g$-wave resonances for Cs
\cite{Chin2004}.  For $s$-wave entrance channels the $g$-wave resonances
are only possible due to second-order coupling in ${\bf V}_\mathrm{ss}$.
Entrance channel $p$-waves can be coupled to resonant bound states of
odd $\ell_\mathrm{c}=$ 1 or 3, although the latter would tend to be
quite weak and rarely observed.

\begin{table}[htdp]
  \caption{Classification of magnetic Feshbach resonances in collisions of ultracold atoms.  The type of the resonance is labeled by the partial wave $\ell_\mathrm{c}$ of the closed channel bound state rather than the entrance channel partial wave $\ell$.  Almost all cases known experimentally have $\ell=$ 0 or 1.  Note that identical bosons (fermions) in identical spin states can only interact with even (odd) partial waves.  All other cases permit both even and odd partial waves.}
    \begin{tabular}{@{\hspace{2em}}c@{\hspace{2em}}|@{\hspace{2em}}c@{\hspace{2em}}|@{\hspace{2em}}c@{\hspace{2em}}}
    \hline\hline
      Type & $\ell$ & $\ell_\mathrm{c}$  \\
      \hline
      $s$-wave resonance & 0, 2 $\ldots$ & 0 \\
      $p$-wave resonance& 1, 3 $\ldots$  & 1 \\
      $d$-wave resonance& 0, 2 $\ldots$ & 2 \\
      $f$-wave resonance& 1, 3 $\ldots$  & 3 \\
      $g$-wave resonance& 0, 2 $\ldots$ & 4 \\
      \hline\hline
    \end{tabular}
  \label{tab:resclass}
\end{table}

{ The long-range potential $V_{\mathrm{el},\alpha\beta}(R)$ is diagonal
for the interaction of two ground state alkali metal atoms for all combinations
$q_1q_2$ of Zeeman sublevels, and all channels have the same van der
Waals coefficient $C_6$.  Consequently each channel will have a spectrum of
vibrational and rotational levels for a van der Waals potential} as
described in Fig.~\ref{fig:C6boundstates}.  The $n=-1, -2,\ldots$
levels associated with closed spin channels $\alpha'$ can become
scattering resonances for entrance channel $\alpha$ if they exist
near energy $E_\alpha$.   The value of $n$ can be one of the values
comprising the set of approximate quantum numbers $q_\mathrm{c}$.
Some examples of the use of $n$ in resonance classification are
given by ~\cite{Marte2002} for $^{87}$Rb (see
Fig.~\ref{fig:Rb87_Marte}) or \cite{Kohler2006} for $^{85}$Rb. The
vibrational quantum number can be either $n$, counting down from the
top, or $v$, counting up from the bottom of the well.

The approximate spin quantum numbers in $q_\mathrm{c}$ are
determined by whatever set of quantum numbers blocks the Hamiltonian
matrix into nearly diagonal parts.  This will depend on the nature
of the coupling among the various angular momenta of the problem so
that no unique general scheme can be given.  For alkali-metal dimers
the spacing between vibrational levels, which is on the order of tens
of $E_{vdw}$ as seen from Fig.~\ref{fig:C6boundstates}, must be compared
to the spacing between the channel energies $E_\alpha$.  For example, in a
light molecule like $^6$Li$_2$ or $^{23}$Na$_2$ they are large compared to
the atomic hyperfine splitting $E_\mathrm{hf}=|E_{I+1/2}-E_{I-1/2 }|$ and
Zeeman interactions.  In this case, the vibrational levels are to a good
approximation classified according to the electronic spin coupling, $S=0$
or 1 of the respective $^1\Sigma^+_g$ and $^3\Sigma^+_u$ Born-Oppenheimer
potentials, with additional classification according to their nuclear
spin substructure.  \cite{vanAbeelen1999,Laue2002} give an example of
such a classification for $^{23}$Na$_2$, and \cite{Simonucci2005} give
an example for $^6$Li$_2$.

In contrast to light species, heavy species like Rb$_2$ or Cs$_2$
have vibrational spacings that are smaller than $E_\mathrm{hf}$, so
that near-threshold bound states of of the $^1\Sigma^+_g$ and
$^3\Sigma^+_u$  potentials are strongly mixed by the hyperfine
interaction.  The near-threshold molecular states do not correspond
to either $S=0$ or 1, but often can be characterized by the
approximate quantum number $f_\mathrm{c}$, where ${\bf f} = {\bf
f}_1 + {\bf f}_2$.  As with the $f_1$ or $f_2$ atomic quantum
numbers, $f_\mathrm{c}$ is not a good quantum number at large $B$
but can be used as a label according to the low field state with
which it adiabatically correlates.    \cite{Marte2002} give examples
of such resonance classification for $^{87}$Rb, and \cite{Chin2004}
and \cite{Kohler2006} do so for $^{133}$Cs and $^{85}$Rb
respectively. \cite{Hutson2008} describe improved computational methods
for calculating the coupling between bound state levels and characterize
a number of experimentally observed avoided crossings \cite{Mark2007a}
between Cs$_2$ levels having different approximate quantum numbers.

\begin{table}[htdp]
  \caption{Separated atom channel labels for the five $s$-wave $M=0$
    channels of $^{6}$Li.  The $(f_1f_2)$ quantum numbers are only exact at $B=0$.}
    \begin{tabular}{c@{\ \ \ \ }|@{\ \ \ \ }c@{\ \ \ \ }|@{\ \ \ \ }c}
    \hline\hline
      $\alpha$ & $(f_1\, f_2)$ & $m_{f_1},\, m_{f_2}$  \\
      \hline
      $ab$ &$(\frac{1}{2} \frac{1}{2})$ & $+\frac{1}{2},\,-\frac{1}{2}$  \\
      $ad$ &$(\frac{1}{2} \frac{3}{2})$ & $+\frac{1}{2},\,-\frac{1}{2}$  \\
      $be$ &$(\frac{1}{2} \frac{3}{2})$ & $-\frac{1}{2},\,+\frac{1}{2}$  \\
      $cf$ &$(\frac{3}{2} \frac{3}{2})$ & $-\frac{3}{2},\,+\frac{3}{2}$  \\
      $de$ &$(\frac{3}{2} \frac{3}{2})$ & $+\frac{1}{2},\,-\frac{1}{2} $ \\
      \hline\hline
    \end{tabular}
  \label{tab:sVblock}
\end{table}

\subsubsection{Some examples of resonance properties}\label{sssec:Li6res}

We will use the fermionic species $^6$Li to illustrate some basic
features of Feshbach resonances.  Figure \ref{fig:Li_Zeeman} shows
the atomic Zeeman levels.   The inset to
Fig.~\ref{fig:Li_2_potentials} shows the 5 channels and the
potentials $V_{\alpha\alpha}(R)$ at long range needed to describe
the $s$-wave collision of an $q_1=a$ atom with a $q_2=b$ atom.
These 5 channels summarized in Table~\ref{tab:sVblock} have the same
van der Waals $C_6$ coefficient and the same projection $M=0$.   Due
to the light mass, the last bound state of the van der Waals
potential must lie in a "bin" that is $39.5 E_\mathrm{vdw}/h=$ 24.3
GHz deep.

Figure \ref{fig:Li_2_potentials} shows that the last two $M=0$
coupled channels $s$-wave bound states for $B=0$ have the character
of the $n=-1$ or $v=38$ level of the $^1\Sigma_g^+$ potential.  They
have a binding energy of $\approx$1.38 GHz relative to the separated
atom energy $E_{ab}$, associated with the positive scattering length
of $a=45.17$ a$_0$ of the $S=0$ singlet potential
\cite{Bartenstein2005}.  The two levels have total nuclear spin
$I=0$ or 2 and projection $m_I=0$, where ${\bf I}={\bf I}_1+{\bf
I}_2$.  The next  bound states below threshold are three  $M=0$ spin
components of the $v=9$ level of the $^3\Sigma_u^+$ potential,  far
below threshold with binding energies $\approx$24 GHz near the
bottom of the $n=-1$ "bin".  These deeply bound levels are
associated with the large negative scattering length of $-2140$
a$_0$ for the $S=1$ potential \cite{Abraham1997,Bartenstein2005}.

\begin{figure}
\includegraphics[width=2.8in,clip,,angle=270]{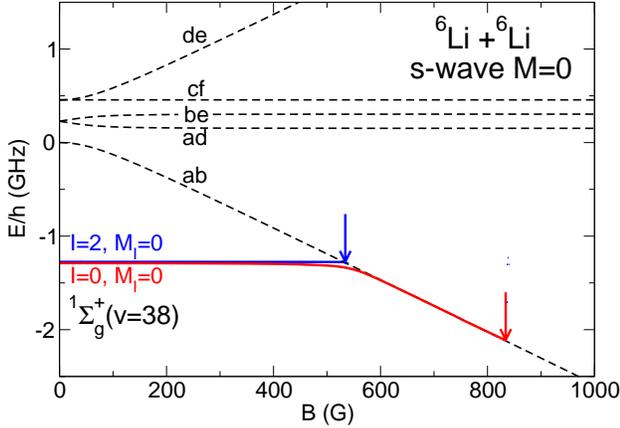}
\caption{Last coupled channels bound states of the $^6$Li$_2$ dimer
with $M=0$.  The arrows indicate the locations of the 543 G and 834
G Feshbach resonances, where the binding energy of a threshold bound
state equals $0$.  While the  low $B$ field $I=2$ $^1\Sigma^+_g(v=38)$ level
retains its spin character as it crosses threshold near 543 G, the
$I=0$ level mixes with the entrance channel and switches near
$\approx 550$ G to a bound level with $ab$ spin character, eventually
disappearing as a bound state when it crosses threshold at 834 G.}
\label{fig:Li_ab_bound}
\end{figure}

\begin{figure}
\includegraphics[width=\columnwidth,clip]{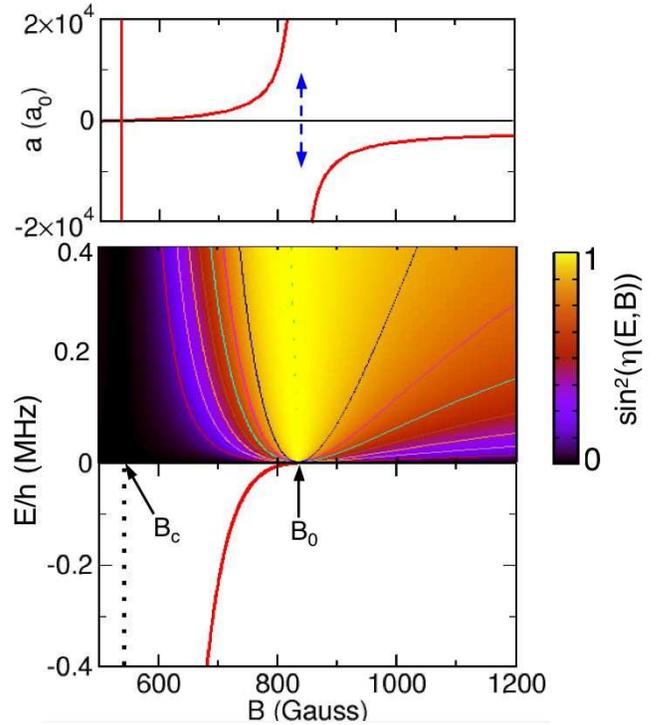}
\caption{Near-threshold bound and scattering state properties of the
$^6$Li $ab$ channel.  The upper panel shows the coupled channels
scattering length versus magnetic field $B$ using the model of
\cite{Bartenstein2005}.  The double-headed arrow indicates the point
of singularity $B_0$ for the broad resonance near 834 G, which is an
entrance channel dominated resonance with width $\Delta=300$ G and
$s_\mathrm{res}=59$ (see Sections~\ref{sssec:resscatt} and
\ref{sssec:resstrength}).  There also is a narrow resonance with a
singularity near 543 G.  The lower panel shows for $E<0$ the energy
of the bound state (solid line) that merges with the continuum at
$B_0$.  The zero of energy at each $B$ is the $ab$ channel energy
$E_{ab}(B)$.  An energy of $E/h=0.4$ MHz is equivalent to $E/k_B=19$
$\mu$K.  The universal bound state energy from Eq.~(\ref{II.A.4}) is
indistinguishable on the scale of this graph from the coupled
channels bound state energy.  The nearly vertical dotted line shows
the energy of the "bare" bound state $E_c(B)$ of
$^1\Sigma_g^+(v=38,I=0)$ character that crosses threshold at $B_c$
near 540 G.  The shaded contour plot for $E>0$ shows
$\sin^2{\eta(E,B)}$.  The broad light-colored region near the point
of resonance indicates the region where $\sin^2{\eta} \approx 1$ and
the cross section is near its maximum value limited by the unitarity
property of the $S$-matrix.  Since $s_\mathrm{res} \gg 1$ the width
$\Gamma(E)$ in Eq.~(\ref{II.B.6}) is larger than $E$ in the
near-threshold region so that there is no above-threshold
"resonance" feature in the collision cross section versus $E$. }
\label{fig:Li_ab_834}
\end{figure}

\begin{figure}
\includegraphics[width=\columnwidth,clip]{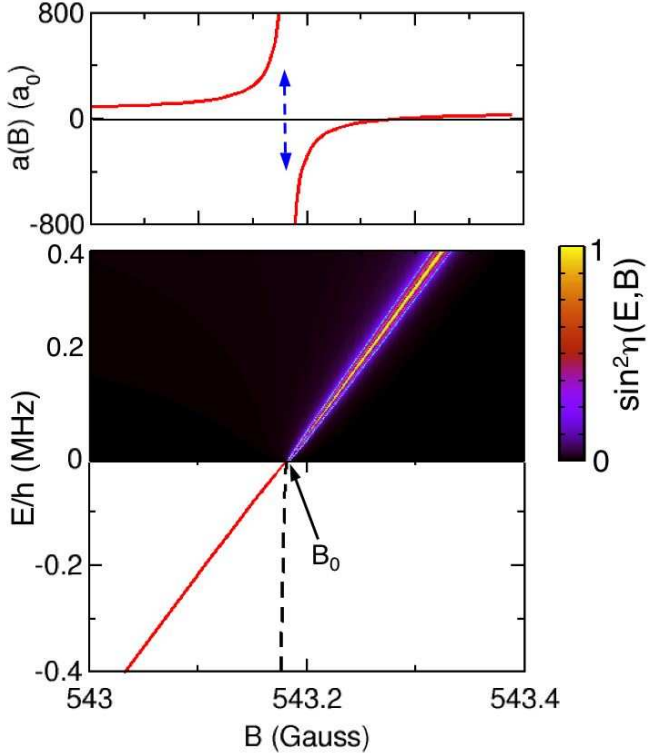}
\caption{Expanded view near 543 G of Fig.~\ref{fig:Li_ab_834}. The
upper panel shows the coupled channels scattering length versus
magnetic field strength $B$, where the double-headed arrow indicates
the calculated point of singularity $B_0$ for the narrow resonance
at $543.18$ G with a width of $\Delta=0.10$ G in excellent agreement
with the measured resonance at $543.26(10)$ G~\cite{Strecker2003}.
This is a closed channel dominated resonance with
$s_\mathrm{res}=0.001$.  The lower panel shows for $E<0$ the energy
of the coupled channels bound state that merges with the continuum
at $B_0$.  The dashed line shows the universal bound state energy
from Eq.~(\ref{II.A.4}).  Universality does not apply for detunings
over most of the width of the resonance but is only applicable in an
extremely narrow range very close to $B_0$.   The shaded contour
plot of $\sin^2{\eta(E,B)}$ for $E>0$ shows a very narrow and sharp
resonance emerging above threshold with a width $\Gamma(E)\ll E$,
with a linear variation of position with $B$, and with a very small
domain of unitarity of the $S$-matrix. } \label{fig:Li_ab_543}
\end{figure}

Figure \ref{fig:Li_ab_bound} shows how the channel energies and the
energies of the last two $M=0$ $s$-wave bound states of the
$^6$Li$_2$ molecule vary with magnetic field.   \cite{Simonucci2005}
give a detailed description of the molecular physics of these
multichannel bound states.  At high $B$ field, $E_{ab}$ varies
linearly with $B$ with a slope of nearly $dE_{ab}/dB = -2\mu_B$,
where $\mu_B$ is a Bohr magneton.  Since both bound states have
$S=0$ character near $B=0$, their magnetic moment vanishes, i.e,
$dE_c/dB = 0$ near $B=0$.    The $I=2$ state crosses $E_{ab}$ near
$B=543$ G, where it interacts weakly with the $ab$ entrance channel
and makes a very narrow Feshbach resonance, shown in
Fig.~\ref{fig:Li_ab_543}.  On the other hand, the energy of the
strongly interacting $I=0$ bound state changes dramatically above
about 540 G and becomes nearly parallel to the energy of the $ab$
entrance channel.  This state switches to $S=1$ character near the
540 G crossing region, and transforms into the $v=10$ level of the
$^3\Sigma_u^+$ potential at higher $B$.  This level becomes a very
weakly bound "universal" halo state of dominantly entrance channel
character above around 650 G, and does not disappear until it
reaches the $E_{ab}$ threshold near 834 G~\cite{Bartenstein2005},
where it makes a very broad Feshbach resonance, shown in
Fig.~\ref{fig:Li_ab_834}.

Figure ~\ref{fig:Li_ab_834} shows the near-threshold bound and
scattering state properties of the $^6$Li $ab$ channel between 500 G
and 1200~G, while Fig.~\ref{fig:Li_ab_543} shows an expanded view of
the narrow resonance near 543~G.  The energy range is typical of the
ultracold domain.  The two figures illustrate two extremes of
resonance behavior.  The 834~G resonance is strongly open channel
dominated with $s_\mathrm{res}=59$ (see Eq.~(\ref{II.B.5})) and is
well represented by a universal halo bound state of entrance channel
character over a large fraction of its width $\Delta$.  The 543.2~G
resonance is strongly closed channel dominated with
$s_\mathrm{res}=0.001$.  It  { exhibits open channel character
and universal behavior only over a negligible detuning range spanning
at most a few $\mu$G when $B$ is tuned near $B_0$.}

It is instructive to examine the wave functions for the coupled
channels bound states with the same binding energy near each
resonance.  For example, the binding energies in
Figs.~\ref{fig:Li_ab_834} and \ref{fig:Li_ab_543} are $\approx$ 200
kHz  near $700$ G and $543.1$ G respectively.   We calculate that
the projection $Z$ on the closed channel components, {
$Z=1-n_\alpha$ are  $0.002$ and $0.98$ for
these respective cases, where $n_\alpha=\int_0^\infty|\phi_{ab,ab}(R)|^2dR$
is the norm of the entrance channel component $\phi_{ab,ab}(R)$
of the bound state } from the coupled channels expansion in
Eq.~(\ref{eq:psi}).  The small projection $Z$ at 700 G is in good
agreement with the value measured by~\cite{Partridge2005}; see
Fig.~\ref{fig:molprobehulet}.  These
projections for levels with the same near-threshold binding energy
illustrate the very different character of entrance and closed
channel dominated resonances.

\begin{figure}
\includegraphics[width=2.8in,clip,,angle=270]{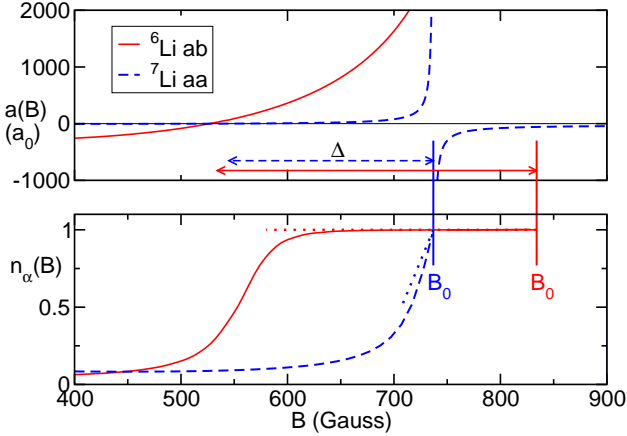}
\caption{{ Scattering length and entrance channel fraction for the $^6$Li 834 G open channel dominated resonance (solid lines) and the $^7$Li 737 G resonance (dashed lines), which is tending towards closed channel dominance.  The narrow $^6$Li resonance is not shown.  The upper panel shows the scattering length $a(B)$ versus magnetic field strength $B$, whereas the lower panel shows the norm $n_\alpha(B)$ of the entrance  channel spin component of the coupled channels wave function.  The vertical lines indicate the location $B_0$ of each resonance, and the horizontal double arrows indicate the width $\Delta$ of each.  The dotted lines on the lower panel indicates the slope of $n_\alpha(B)=1-Z(B)$ predicted near the resonance position $B_0$ by Eq. (\ref{2chan_mixing}) of Section~\ref{sssec:AnalMol}.   While both resonances have quite large $\Delta$, the $^6$Li 834 G one is clearly "open channel dominated" with $1-Z(B)$ remaining near unity when $|B-B_0|$ ranges over half of its width.  On the other hand, the 737 G $^7$Li resonance has $s_\mathrm{res} = 0.80$ and is tending towards being "closed channel dominated," since $1-Z(B)$ drops off rapidly from unity as $|B-B_0|$ increases from resonance, with $n_\alpha(B)>0.5$ only for $|(B-B_0)/\Delta|<0.11$.  Furthermore, this resonance has a universal bound state (not shown) only over a relatively small fraction of its width, with the calculated binding energy departing from Eq. (\ref{II.A.4}) by 10 percent when $|(B-B_0)/\Delta|=0.06$.}} \label{fig:Li7+Li6_A+Z}
\end{figure}

The width $\Delta$ itself does not determine whether a resonance is
entrance or closed channel dominated.  Rather, it is necessary to
apply the criterion in Eq.~(\ref{II.B.5}). A good example of this is
provided by the bosonic $^7$Li system, which has a very broad
resonance in the $aa$ channel  near 737 G with a width of 192
G~\cite{Khaykovich2002,Junker2008,Pollack2008}, where $a$ represents the state
which correlates with the $|f=1,m=1\rangle$ state at $B=0$. Because
the background scattering length is nearly two orders of magnitude
smaller for this $^7$Li case than for the 843 G $^6$Li resonance,
the $s_\mathrm{res}$ parameter for the $^7$Li $aa$ resonance is only
$0.80$ instead of 59.  Consequently, this broad $^7$Li resonance
{ is tending towards closed channel dominance according to
our classification scheme and only has a
small region of universality spanning a few G
when $B$ is tuned near $B_0$.  Figure~\ref{fig:Li7+Li6_A+Z}
shows the pronounced
differences between the $^6$Li $ab$ and $^7$Li $aa$ resonances, in spite
of the similar magnitudes of their widths; see also Section~\ref{sssec:AnalMol}.}

\begin{figure}
\includegraphics[width=\columnwidth,clip]{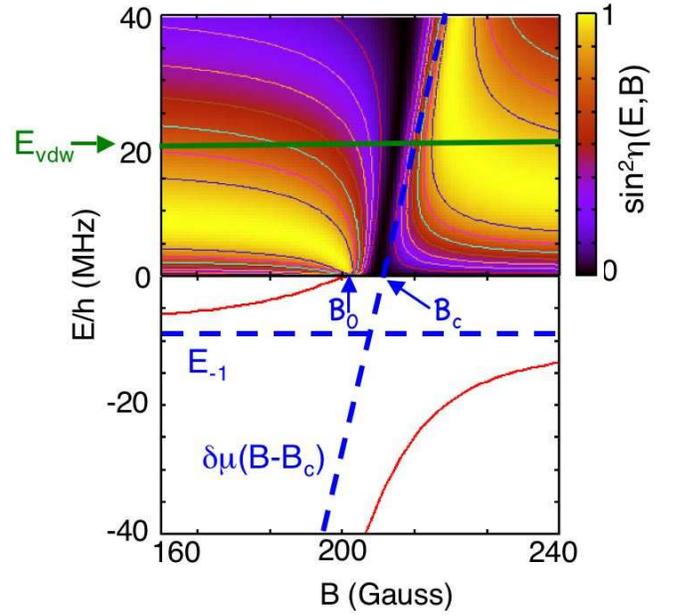}
\caption{ Bound states and scattering phase near the $^{40}$K $ab$
202 G resonance.  The energy is between $E/h=\pm 40$ MHz ($E/k_B=\pm
1.9$ mK) where the zero of energy is taken to be the separated atom
energy of an $a$ and a $b$ atom.   The horizontal solid line shows
$E_\mathrm{vdw}/h=21$ MHz.  The horizontal dashed line shows the
last  ``bare'' bound state energy $E_{-1}$ of the background
potential and the sloping dashed line shows the  ``bare'' resonance
level with energy $\delta \mu (B-B_c)$, where $ \delta \mu/h = 2.35$
MHz$/$G.  The resonance width $\Delta=7.8$ G~\cite{Greiner2003}.
The solid lines for $E <0$ indicate the coupled channels ``dressed''
energies of the $^{40}$K$_2$ molecule.  Away from resonance these
approach the ``bare'' energies.   The strong avoided crossing
between the two ``bare'' states leads to the shift in the point of
singularity $B_0$ from the ``bare'' crossing at $B_c$.  For positive
energies the interference between the background and resonant phases
is evident.  Since this is a open channel dominated resonance with
$s_\mathrm{res}=2$, no sharp resonant feature appears in
$\sin^2\eta(E,B)$ versus $E$ for $0<E<E_\mathrm{vdw}$.  A sharp
resonance feature only emerges when $E \gg E_\mathrm{vdw}$. }
\label{fig:K40_ab}
\end{figure}

In order to illustrate the difference between the ``bare'' and
``dressed'' resonance states introduced in
Section~\ref{sssec:resscatt}, Fig.~\ref{fig:K40_ab} shows the
coupled channel bound state energies and scattering phases in the
near-threshold region for the $^{40}$K $ab$ channel.  The $a$ and
$b$ states correlate at $B=0$ with the
$|f=\frac{9}{2},m=-\frac{9}{2}\rangle$ and
$|f=\frac{9}{2},m=-\frac{7}{2}\rangle$ atomic states of fermionic
$^{40}$K.  This resonance was observed by
~\cite{Loftus2002,Regal2003a} and has additionally been
characterized by ~\cite{Szymanska2005} and \cite{Nygaard2006}.  The
actual eigenstates of the  ``dressed'' system (solid lines) result
from the avoided crossing of the ramping closed channel  ``bare''
state energy $E_c=\delta \mu (B-B_c)$ and the last  ``bare'' bound
state at $E_{-1}$ of the background  potential (dashed lines).  The
shift in the location of the singularity in $a(B)$ at $B_0$ from the
threshold crossing of the  ``bare'' state at $B_c$ is given from
Eq.~(\ref{II.B.7}):
\begin{equation}
  B_0-B_c=\Delta\frac{r_\mathrm{bg}(1-r_\mathrm{bg})}{1+\left ( 1-r_\mathrm{bg} \right )^2}  \,.
  \label{eq:shift}
\end{equation}
This same formula predicts the large difference between $B_0$ and
$B_c$ evident in Fig.~\ref{fig:Li_ab_834} for the broad $^6$Li $ab$
resonance.

\begin{figure}
\includegraphics[width=\columnwidth,clip]{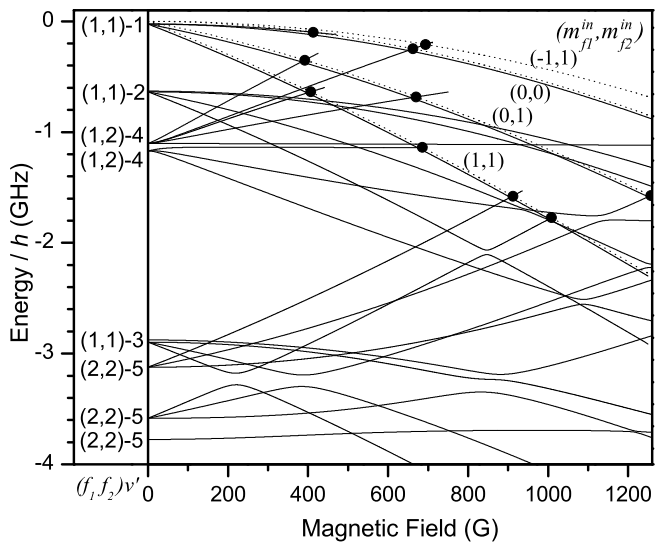}
\caption{Coupled channels $^{87}$Rb $s$-wave bound states. The
dotted lines show the channel energies for four different entrance
channels with $f_1=f_2=1$, labeled by the projection quantum numbers
$(m_{1},m_{2})$.  The bound-state energies (solid lines) for these
channels are shown as a function of magnetic field $B$ with quantum
numbers $(f_1 f_2)n$ assigned at $B=0$, where $n$ is the vibrational
quantum number in the $f_1 f_2$ channel counting down from the
separated atom limit in that channel.  A Feshbach resonance occurs
when a bound state with quantum number $M=m_{1}+m_{2}$ crosses a
dissociation threshold having the same $M$ (solid dots).
From~\cite{Marte2002}. } \label{fig:Rb87_Marte}
\end{figure}

Finally, we give an example of resonances for a heavy species
$^{87}$Rb where classification of near-threshold bound states using
the electronic spin $S=$ 0 and 1 quantum numbers is not possible.
The bin size $\approx 0.240$ GHz for the last bound state is much
less than the $^{87}$Rb ground state hyperfine splitting
$E_{hf}/h=6.835$ GHz so that the last few bound states of the
$^{87}$Rb$_2$ molecule are mixed by the hyperfine interaction.
Fig.~\ref{fig:Rb87_Marte} shows the coupled channels $s$-wave bound
states calculated by ~\cite{Marte2002} for channels with $M=m_1+m_2$
$=$ 2, 1, and 0.  Unlike the $^6$Li case in
Fig.~\ref{fig:Li_ab_bound}, there are a number of bound states
within a few GHz of threshold.  The levels are labeled at $B=0$ by
the spin quantum numbers $(f_1 f_2)$ of the separated atoms  and the
vibrational quantum number $n$ counting down from the separated atom
dissociation limit.   The figure shows the last 3 vibrational levels
of the lowest $(f_1 f_2)=(11)$ separated atom limit.  The $B=0$
energy of the $(12)$ separated atom limit is $E_{hf}=6.835$ GHz and
only the $n=-4$ vibrational level appears in this range.  Similarly
only the $n=-5$ level appears for the $(22)$ separated atom limit.
The closed channel dominated resonance ($s_\mathrm{res}=0.17$) near
1007 G in the $(11)$ $M=2$ channel has been used to make molecules
in atomic gases~\cite{Durr2004a} and lattices~\cite{Thalhammer2006}.

\subsection{Simplified models of resonance scattering}\label{ssec:simpmod}

While coupled channels models are very valuable for understanding
the near threshold molecular physics of scattering resonances and
for highly quantitative predictive calculations for a range of $B$
field and multiple spin channels, they can be quite complicated to
set up and use.  Consequently, it is highly desirable to have
simplified models that are accessible to experimental and
theoretical researchers.  Fortunately, a variety of high quality
models are available, each valid over a limited domain of energy.

The key to practical approximations for the near threshold bound and
scattering states for ultracold neutral atom interactions is the
separation of the length and energy scales associated with the
separated atoms on the one hand and the molecular interactions on
the other.  The molecular interactions are characterized by various
energy scales associated with the van der Waals
potential, the potential at $R_\mathrm{ex}$, or the minimum of the
potential. This scale should be compared with the hyperfine, Zeeman,
and kinetic energies of the ultracold atoms.  For ranges of
internuclear separation $R$ where the molecular energy scale is much
larger than the atomic one, the phases and amplitudes of the coupled
channels wave function components $\phi_{\alpha \beta}$ are nearly
independent of energy and partial wave over energy range on the
order of the atomic scale. In effect, the short range wave function
provides an energy-independent  boundary condition for connecting to
the near threshold asymptotic bound or scattering states, which are
strongly energy dependent.    While this separation of scales can be
made explicit in methods based on long range coupled channels
calculations~\cite{Tsai1997,vanAbeelen1999} or multichannel quantum
defect
methods~\cite{Julienne1989,Burke1998,Vogels1998,Raoult2004,Julienne2006},
it remains implicit as the basis for many other approximation
schemes.

{ A variety of simplified treatments show that it is
sufficient to use the basic framework in
Section~\ref{sssec:resscatt} to parameterize threshold resonances in
ultracold atoms.  Thus, resonances are characterized by a reduced
mass $\mu$, a background scattering length $a_\mathrm{bg}$, a
tunable position $E_0$ selected by an external field, and an energy-
and tuning-insensitive width $\Gamma_0$.  Resonances that decay,
whether by emission of light or by relaxation to lower energy open
channels, can readily be treated by introducing the decay width
$\gamma$~\cite{Fedichev1996a,Bohn1999,Kohler2005,Hutson2006}.
% We are primarily interested in nondecaying resonances in this Review.
The review by ~\cite{Kohler2006} describes how two-channel models
can be especially effective when the resonance parameters are already known.
}

\subsubsection{Contact potential model}\label{sssec:contact}

The simplest approximation for resonance scattering is to use the
Fermi pseudopotential~\cite{Huang1957}
\begin{equation}
 V({\bf R}) = \frac{2\pi \hbar^2}{\mu} a(B) \delta({\bf R}) \frac{\partial}{\partial R} R \cdot \,,
 \label{eq:Fermips}
 \end{equation}
with a strength proportional to the scattering length $a(B)$.  This
zero-range delta-function pseudopotential is an excellent
approximation for the full molecular interaction when $k|a(B)| \ll
1$ and $k\bar{a} \ll 1$ and becomes exact in the limit $E \to 0$. It
can be used for positive or negative $a$, and its phase shift is
$\tan\eta(E,B)=-ka(B)$.  For $a >0$ it has a bound state given by
the universal energy of Eq.~(\ref{II.A.4}).

If the resonance parameters $a_\mathrm{bg}$, $\Delta$, and $B_0$ are
known, the effect of tuning near a resonance can then be fully
incorporated using $a(B)$ from Eq.~(\ref{II.A.21}).  For an entrance
channel dominated resonance with $s_\mathrm{res} \gg 1$, so that the
universal binding energy in Eq.~(\ref{II.A.4}) applies (see
Eq.~(\ref{II.B.3})), the scattering length is the only parameter
needed to treat near-threshold bound states and
scattering~\cite{Kohler2006}. However, more robust approximations
are needed, since universality will only apply for detunings that at
most span a range on the order of the width $\Delta$ and can be much
less, depending on $s_\mathrm{res}$.

\subsubsection{Other approximations}\label{sssec:OtherApprox}

Although the underlying molecular physics often involves a number of
coupled channels, many resonances are isolated in energy and
magnetic field.  Then the properties of the "bare" resonance level
are determined by energy scales large compared to the the small
kinetic energies of the ultracold domain, and the level can be
accurately approximated as coming from a single bound state channel,
as in the Fano treatment summarized in Section~\ref{sssec:resscatt}.
A number of groups have developed a variety of simplified methods
for characterizing the properties of ultracold scattering
resonances, but we can not review this work exhaustively or in
detail.  For example, \cite{Moerdijk1995}, \cite{Timmermans1999}
and \cite{Kokkelmanns2002} introduce the standard Feshbach formalism
of separating the system into  bound and scattering subspaces, $Q$
and $P$, to characterize magnetically tunable resonances for ground
state alkali metal atoms.   \cite{Goral2004} use a Green's function
formalism and introduce a separable potential that is chosen to
accurately represent the two-body scattering and bound states of the
background channel.    \cite{Marcelis2004} are especially interested
in representing the case of a large negative $a_\mathrm{bg}$, which
is relevant to the $^{85}$Rb system.  \cite{Mies2000a} use the
resonances of two $^{23}$Na atoms to show how to reduce a coupled
5-channel problem to an effective 2-channel problem using a
Lennard-Jones pseudopotential with the correct van der Waals
coefficient.  \cite{Nygaard2006} illustrates this method for the
$^{40}$K system.

One model that shows great promise for practical and accurate
fitting of resonance data is the asymptotic bound state (ABM) model
based on the work of ~\cite{Moerdijk1995}.  Rather than solving for
the bound states of a set of coupled equations in order to locate
resonance positions, it uses an expansion in the last bound states
of the Born-Oppenheimer potentials. The model is far less
computationally demanding than full coupled channels calculations.
\cite{Stan2004} used a simplified version of this model to
characterize measured resonances due to the triplet molecular state
in the $^6$Li$+^{23}$Na system.  Recently \cite{Wille2007} used this
model to quantitatively characterize a number of resonances of
$^6$Li$+^{40}$K that involved strong mixing of the singlet and
triplet molecular states.

\subsubsection{van der Waals resonance model}\label{sssec:anvdwmod}

By introducing the van der Waals $C_6$ coefficient as an additional
model parameter, the properties of bound and scattering states can
be extended away from their very near-threshold domain  into the
domain where $k\bar{a} \gg 1$ and the binding energy is much larger
than $E_\mathrm{vdw}$ or $\bar{E}$.  The reason is that there is a
large range of $R$, namely $R_\mathrm{ex}<R<R_\mathrm{vdw}$, where
the potential is accurately represented as $-C_6/R^6$ and is much
larger in magnitude than $E_\mathrm{vdw}$.  The properties of the
van der Waals potential have been discussed in
Section~\ref{sssec:vdw}.

Feshbach resonances are characterized by a width $\Gamma(E)$ and
shift $\delta E(E)$.  These are given in the $E \to 0$ limit by
Eqs.~(\ref{II.B.6}) and (\ref{II.B.7}), which depend on the
dimensionless resonance strength $s_\mathrm{res}$ and
$r_\mathrm{bg}$.   The $E \to 0$ result can be generalized to finite
energy  by introducing two standard functions
$C_\mathrm{bg}(E)^{-2}$ and $\tan \lambda_\mathrm{bg}(E)$ of
multichannel quantum defect theory
(MQDT)~\cite{Julienne1989,Mies2000,Raoult2004}
\begin{eqnarray}
 \frac{1}{2}  \Gamma(E)&=& \frac{\bar{\Gamma}}{2} \, C_\mathrm{bg}(E)^{-2}     \label{II.C.1} \\
 \delta E(E) &=&  \frac{\bar{\Gamma}}{2} \, \tan\lambda_\mathrm{bg}(E)  \,.
 \label{II.C.2}
\end{eqnarray}
where for the van der Waals background potential~\cite{Julienne2006}
\begin{equation}
  \frac{\bar{\Gamma}}{2} = (\bar{E}s_\mathrm{res} ) \frac{1}{1+\left ( 1-r_\mathrm{bg} \right )^2}
  = \Gamma_0 \frac{r_\mathrm{bg}}{1+\left ( 1-r_\mathrm{bg} \right )^2}
  \label{II.C.3}
\end{equation}
is proportional to $s_\mathrm{res}$ and is independent of energy.
The MQDT functions have the following limiting form as $E\to 0$:
$C_\mathrm{bg}(E)^{-2}=k\bar{a}(1+(1-r_\mathrm{bg})^2)$ and
$\tan\lambda_\mathrm{bg}(E)=1-r_\mathrm{bg}$.  When $E \gg \bar{E}$,
$C_\mathrm{bg}(E)^{-2} \to 1$ and $\tan\lambda_\mathrm{bg}(E) \to
0$.  Consequently, $\Gamma(E)=\bar{\Gamma}$ and $\delta E(E)$
vanishes when $E$ becomes large compared to $\bar{E}$.  If
$|r_\mathrm{bg}| \gg 1$, the $C_\mathrm{bg}(E)^{-2}$ function has a
maximum and $|\tan\lambda_\mathrm{bg}(E)|$ has decreased to half its
$E=0$ value at $E \approx \hbar^2/(2\mu \left
(a_\mathrm{bg}-\bar{a})^2 \right )$.

The functions $C_\mathrm{bg}(E)$ and $\tan\lambda_\mathrm{bg}(E)$,
as well as $\eta_\mathrm{bg}(E)$, depend on only three parameters,
$C_6$, $\mu$, and $a_\mathrm{bg}$~\cite{Julienne2006}.  The
near-threshold phase $\eta(E)$ in Eq.~(\ref{II.A.15}) can be
evaluated over a wide range of energy on the order of $\bar{E}$ and
larger from a knowledge of these three parameters plus
$s_\mathrm{res}$, the magnetic moment difference $\delta \mu$, and
the resonant position $B_0$.  The $\sin^2\eta(E,B)$ function
evaluated using Eqs.~(\ref{II.C.1}) and (\ref{II.C.2}) are virtually
indistinguishable from the coupled channels results shown in
Figs.~\ref{fig:Li_ab_834}-\ref{fig:K40_ab}.

\subsubsection{Analytic 2-channel square well model}\label{sssec:SquareWell}

A very simple square well model, because it is analytically
solvable,  can capture much of the physics of near-threshold bound
and scattering states.  \cite{Bethe1935} used such a model to
successfully explain the threshold scattering of cold neutrons from
atomic nuclei, where the neutron de Broglie wavelength was very
large compared to the size of the nucleus. \cite{Kokkelmanns2002}
and \cite{Duine2004} have introduced 2-channel square well models to
represent Feshbach resonances in ultracold atom scattering.

\begin{figure}
\includegraphics[width=2.8in]{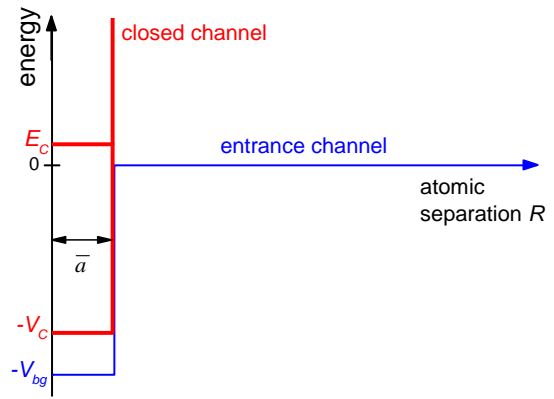}
\caption{Two-channel square well model of a magnetic Feshbach
resonance. The potential  for the ``bare'' entrance background
channel (blue line) has a well depth $-V_\mathrm{bg}$. The spatial
width is chosen to be $\bar{a}$ so as to simulate the length scale
of a van der Waals potential. The potential for the ``bare'' closed
channel (red line) has a well depth $-V_\mathrm{c}$ relative to the
separated atoms and is infinite for $R>\bar{a}$.  The value of
$V_\mathrm{bg}$ is chosen so that the background scattering length
is $a_\mathrm{bg}$. The value of $V_\mathrm{c}$ is chosen so that
there is a bound state with energy $E_c$ near $E=0$.   We assume
that this energy varies with magnetic field as $E_c=\delta \mu
(B-B_c)$, where $\delta \mu$ is a relative magnetic moment and
$E_c=0$ at $B=B_c$.  } \label{fig:2chan_fig1}
\end{figure}

Figure~\ref{fig:2chan_fig1} shows the ``bare'' background and closed
channel potentials for a square well model where for convenience of
analysis we take the width of the wells to be the van der Waals
length $\bar{a}$. The background entrance channel and the closed
channel are designated by $|bg\rangle$  and $|c\rangle$
respectively. Using a two-state coupled channels expansion as in
Eq.~(\ref{eq:psi}), $|\psi(R,E)\rangle = |c \rangle
\phi_\mathrm{c}(R,E)/R + |bg \rangle \phi_\mathrm{bg}(R,E)/R $, the
potential matrix in Eq.~(\ref{eq:V}) is
\begin{eqnarray}
\mathbf{V}&=&\left(
\begin{array}{cc}
-V_\mathrm{c} & W \\
W & -V_\mathrm{bg}  \end{array} \right)\mbox{ for $R<\bar{a}$} \label{box_V} \\
&=&\left(
\begin{array}{cc}
\infty & 0 \\
0 & 0 \end{array} \right)\,\,\,\,\,\,\,\,\,\mbox{ for $R>\bar{a},$}
\nonumber
\end{eqnarray}
The off-diagonal matrix element $W$ describes the weak coupling between
the two channels.

In order to simulate a magnetically tuned Feshbach resonance, the
model parameters need to be chosen so as to give the correct
parameters for that resonance.  The well depth $V_\mathrm{bg}$ is
chosen so that the background channel scattering length is
$a_\mathrm{bg}$.  The well depth $V_c$ is chosen so that the well
has a ``bare'' bound state at $E_c$. The tuning of the bound state
as $E_c=\delta \mu (B-B_c)$ can be simulated by varying $V_c$
linearly with the external magnetic field $B$. Finally,  weak
coupling requires $|W |\ll |V_\mathrm{bg}-V_\mathrm{c}|$. The
coupling parameter $W$ can then be chosen to give the right
resonance width $\Gamma(E)=2ka_\mathrm{bg} \delta \mu \Delta$ at low
energies (see Eq.~(\ref{II.A.16})), using the known resonance width
$\Delta$. Analytically calculating the matrix element defining
$\Gamma(E)$ in Eq.~(\ref{II.A.13}) relates $W$ to $\Delta$ as
follows,
\begin{equation}
\frac{2V_\mathrm{c}W^2}{(V_\mathrm{bg}-V_\mathrm{c})^2}
=\frac{r_\mathrm{bg}}{(1-r_\mathrm{bg})^2} \delta \mu \Delta \,.
\label{box_Gamma0}
\end{equation}
With the chosen parameters, the square well model yields analytic
form for the scattering phase shift as in Eq.~(\ref{II.A.15}) and
the scattering length as in Eq.~(\ref{II.A.21}).

The square well model also permits an analytic evaluation of the
weakly bound state below the continuum. Assuming an eigenstate
$|\psi_b\rangle$ exists at energy $-E_b=-\hbar^2 k_b^2/(2\mu)<0$ and
$|a_\mathrm{bg}|\gg \bar{a}$, we get
\begin{equation}
k_b
=\frac{1}{a_{bg}-\bar{a}}+\frac{\Gamma_\mathrm{sq}/2}{\bar{a}(E_b+E_c)}
\,, \label{2chan_em}
\end{equation}
where $\Gamma_\mathrm{sq}/2 =\delta\mu\Delta
r_\mathrm{bg}(1-r_\mathrm{bg})^{-2}$. \cite{Marcelis2004} derived a
similar result for a contact potential.   Note that when the
coupling term $W \rightarrow 0$ so that $\Gamma_\mathrm{sq} \to 0$,
the solutions $E_b=-E_c$ and $E_b=\hbar^2/[2\mu(a_{bg}-\bar{a})^2]$
correspond to the bare states of the square well in the closed and
open channel (for $a_{bg}>\bar{a}$), as expected.  Since the
resonant singularity in the scattering length occurs when $E_b \to
0$, taking this limit of Eq.~(\ref{2chan_em}) allows us to calculate
the resonance energy shift $\delta E=\delta\mu (B_0 - B_c)$ as
\begin{equation}
\delta E = \frac{\Gamma_{sq}}2(1-r_\mathrm{bg})\,.
\label{2chan_shift}
\end{equation}

Both Eq.~(\ref{2chan_shift}) and Eq.~(\ref{2chan_em}) can also be
derived from the van der Waals model with $\Gamma_{sq}$ replaced by
$\bar{\Gamma}=\Gamma_{sq}[1+(r_\mathrm{bg}-1)^{-2}]$. Note that
$\Gamma_{sq}$ and $\bar{\Gamma}$ are nearly the same for
$|r_{bg}|\gg1$. The modified version of Eq.~(\ref{2chan_shift}) is
equivalent to Eq.~(\ref{II.B.7}) and Eq.~(\ref{eq:shift}), derived
from the van der Waals model.

\cite{Lange2008} extends the above model to precisely determine the
scattering length and the resonance parameters in the magnetic field
regime where multiple Feshbach resonances overlap.

\subsubsection{Properties of Feshbach molecules}\label{sssec:AnalMol}

A variety of properties of Feshbach molecules can be calculated by
solving Eq.~(\ref{2chan_em}) for the binding energy $E_b$.  For
example, the closed channel fraction $Z$ of the eigenstate can be
found by differentiating $E_b$ with respect to $E_c$; see
Eq.~(\ref{eq:Zeq}). In the limit $B \to B_0$ where $E_b$ vanishes
and $a\rightarrow +\infty$, we have
\begin{equation}
Z= \frac1{\zeta} \left |\frac{B-B_0}{\Delta} \right |\,.
 \label{2chan_mixing}
\end{equation}
where the dimensionless proportionality constant
\begin{equation}
  \zeta=\frac{1}{2} s_\mathrm{res}
 | r_\mathrm{bg}|=\frac{r_{bg}^2}2\frac{|\delta\mu\Delta |}{\bar{E}}
\label{zeta}
\end{equation}
determines the rate at which the Feshbach molecular state deviates
from the entrance channel dominated regime or, equivalently, the
halo molecule regime, when $B$ is tuned away from $B_0$.
Eq.~(\ref{2chan_mixing}) shows that having a small closed channel
fraction $Z \ll 1$ requires the magnetic field to be close to
resonance, $|B-B_0|\ll \zeta | \Delta |$.   { Figure~\ref{fig:Li7+Li6_A+Z}
of Section~\ref{sssec:Li6res}
compares $1-Z$ from Eq.~(\ref{2chan_mixing}) for the respective open
and closed channel dominated $^6$Li 834 G and $^7$Li 737 G resonances.}

\begin{figure}
\includegraphics[width=3.3 in]{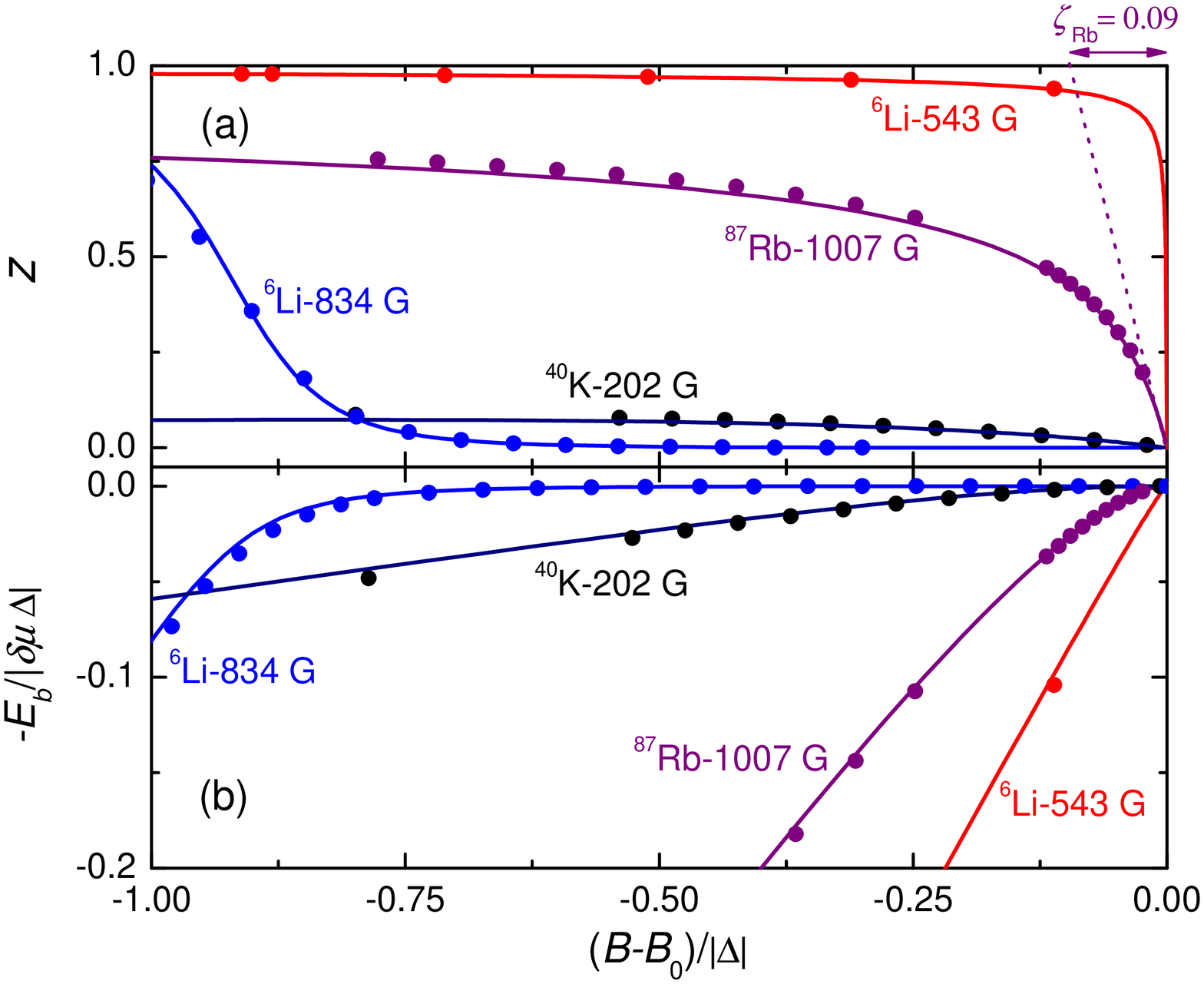}
\caption{(Color) Closed channel fraction $Z$ (panel a) and energy
$-E_b$ (panel b) of weakly-bound molecular states as a function of
magnetic field for selected Feshbach resonances. The open circles
come from coupled channels calculations as described in
Section~\ref{sssec:coupchan}. The solid curves are calculated based
on Eq.~(\ref{2chan_em}), Eq.~(\ref{eq:Zeq}), and the parameters in
Table~\ref{tab:resonances}. The slope of $Z$ at $B=B_{0}$ defines
the $\zeta$ parameter; an example for the $^{87}$Rb 1007~G resonance
is shown as the dashed line with $\zeta_{Rb}=0.09$. This $^{87}$Rb
resonance and the $^6$Li 543 G one are closed channel dominated and
have $s_{red}, \zeta < 1$.  The $^6$ Li 834 G and the $^{40}$K 202 G
resonances are open channel dominated resonances with $s_{res},
\zeta >1$.} \label{fig:E&Z}
\end{figure}

For open channel dominated resonances, where $s_\mathrm{res} \gg 1$,
it is usually true that $|r_\mathrm{bg}|\geq1$ and $\zeta \gg 1$,
and $Z$ remains small over a large fraction of the resonance width
$\Delta$. The bound state wave function takes on primarily entrance
channel character over this range.  See the examples of the $^{6}$Li
834 G or $^{40}$K 202 G resonances in Fig.~\ref{fig:E&Z}. Closed
channel dominated resonances have $s_\mathrm{res} \ll 1$ and small
$\zeta \ll 1$. Consequently, $Z$ remains small only over a small
range of the resonance; see the example of the $^{87}$Rb 1007 G or
$^6$Li 543 G resonances in Fig.~\ref{fig:E&Z}. A small
$\zeta_{Rb}=0.09$ for the Rb 1007 G resonance implies that the
molecular state is entrance channel dominated only within $\sim 9\%$
of the resonance width.
Table~\ref{tab:resonances} in the Appendix lists $\zeta$ for several
other resonances.

Expanding Eq.~(\ref{2chan_em}) at small binding energies, the
molecular binding energy has the following form in the threshold
limit:
\begin{eqnarray}
E_b=\frac{\hbar^2}{2\mu(a-\bar{a}+R^*)^2}
\label{2chan_threshold1},
\end{eqnarray}
 where
$R^*=\bar{a}/s_\mathrm{res}$. This expression applies in the limit
that $a \gg \bar{a}$ and $a \gg 4R^*$.  The binding energy $E_b$
shows two corrections to the universal $1/a^2$ threshold law in
Eq.~(\ref{II.A.4}). One is the finite range correction $\bar{a}$
shown in Eq.~(\ref{II.B.3}); the other one, $R^*$, introduced by
\cite{Petrov2004}, is unique to Feshbach resonances. The correction
$R^*$ is negligibly small for open channel dominated resonances with
$s_\mathrm{res} \gg1$.  Closed closed channel dominated resonances
can have large $R^* \gg \bar{a}$.  Such resonances only have a
regime with a bound state of predominant entrance channel character
near $B_0$ where $a \gg 4R^*$. This condition is consistent with the
one given in the previous paragraphs, namely,
$|B-B_0|\ll\zeta|\Delta|$.

\section{Finding and characterizing Feshbach resonances}\label{sec:species}

Magnetically tunable Feshbach resonances have been experimentally
observed in essentially all alkali-metal species, in some mixtures
of different alkali-metal atoms, as well as for Cr
atoms. { Experimental data on the resonance positions have in many cases enabled the construction of accurate models to describe the near-threshold behavior, including scattering properties and molecular states.}
We will first review the various experimental approaches to
identify and characterize Feshbach resonances in
Sec.~\ref{ssec:species_A} and then discuss observations of
Feshbach resonances for alkali-metal atoms in
Sec.~\ref{ssec:species_B} and various other systems in
Sec.~\ref{ssec:species_C}.

\subsection{Experimental methods}\label{ssec:species_A}

Experimental approaches to detect magnetic Feshbach resonances can be classified into several types.
After some general considerations in Sec.~\ref{sssec:genconsid},
we will discuss detection by inelastic collisional trap loss in
Sec.~\ref{sssec:lossspect}, by elastic collision properties in
Sec.~\ref{sssec:elcoll}, and loss in the presence of optical radiation in
Sec.~\ref{sssec:spectmol}.  Finally, Section~\ref{ssec:rfspec} discusses precision radio-frequency spectroscopy of Feshbach molecules.

\subsubsection{General considerations}\label{sssec:genconsid}

\noindent \emph{a. What is the magnetic field range to be explored?}
The typical spacing between two Feshbach resonances can be estimated
from the ratio of the energy splitting between closed channel
molecular levels and the relative magnetic moment $\delta \mu$
between the entrance channel and the closed channel.  The
vibrational energy splitting between near-threshold bound states is
determined by the long-range van der Waals potential to be on the
order of 100$E_{\rm vdw}$; see Sec.~\ref{sssec:vdw}.  For
alkali-metal atoms $\delta \mu$ is on the order of two Bohr magneton
$2 \mu_B=2.8$\,MHz/G and 6 times larger for $^{52}$Cr.  For atoms
with a small hyperfine splitting compared to 100$E_{\rm vdw}$,
Feshbach resonances are induced by the last bound states.  This
leads to a typical $s$-wave Feshbach resonance separation of  $\sim
10,000$~G for $^6$Li and $\sim100$~G for $^{52}$Cr. For atoms with
hyperfine splittings much larger than 100$E_{\rm vdw}$, resonances
can be induced by much deeper bound states { in the
closed channel} (see Fig.~\ref{fig:Rb87_Marte} for $^{87}$Rb), and
the expected spacings can be estimated accordingly.

The density of Feshbach resonances increases when higher
partial-wave scattering and multiple closed hyperfine channels,
defined in Sec.~\ref{sssec:coupchan}, are included. The relevant
number of channels is determined by the angular momentum dependence
of the molecular potentials and identical particle statistics. For
alkali-metal atoms there are on the order of 10 closed channels for
the lowest partial wave $\ell=0$. The number of channels increases
rapidly for higher partial waves. The ultralow temperature { usually} limits scattering to $s$- and sometimes $p$-wave entrance channels,
but coupling to molecular states with up to $\ell_\mathrm{c}=4$ has
been observed; see Tables~\ref{tab:resclass} and
\ref{tab:resonances}.

For species without unpaired electrons, e.g., Sr and Yb, one expects
no or very limited magnetic tunability because there is no electron
contribution to the magnetic moment and the nuclear contribution is
very small. In these systems, Feshbach resonances can possibly be
optically induced; see Sec.~\ref{ssec:optfesh}.

\noindent \emph{b. What is the required magnetic field resolution?}
The width of the resonance generally determines the magnetic field
resolution required for detection. Many $s$-wave Feshbach resonances
have widths larger than 1~G; see Table IV in the Appendix. High
partial wave Feshbach resonances are typically much narrower because
of the weaker Feshbach coupling strength. Usually, a resolution in
the milligauss range is required to detect $d$- or $g$-wave Feshbach
resonances.

\noindent \emph{c. How to trap atoms for collision studies?} Optical
dipole traps, reviewed by \cite{Grimm2000}, are the main tool to
confine cold atoms for collision studies related to Feshbach
resonances. Optical potentials { trap} atoms in any
sub-level of the electronic ground state { and permit}
investigation of collisions in any corresponding spin channel. For
many experimental applications, the lowest atomic state is of
particular interest, which is a high-field seeking state and can
therefore not be trapped magnetically. Optical dipole traps allow
for the application of arbitrary homogenous magnetic fields without
affecting the trapping potential. In contrast, magnetic traps can
only confine atoms in low-field seeking states, and the application
of a magnetic bias field for Feshbach tuning can strongly influence
the trap parameters. This limits the application of Feshbach tuning
in magnetic traps to very few situations.

\noindent \emph{d. How low a temperature is needed to observe the
resonances?} In most experiments, a temperature of a few $\mu$K is
sufficiently low to observe a clear resonant structure in inelastic
collisional loss. Collision studies can be performed with thermal
samples, BECs { or degenerate Fermi gases}. Elastic collision
measurements are more complex. Enhancement of elastic collision
rates near Feshbach resonances is more prominent at lower
temperatures $<1\mu$K. On the other hand, suppression of elastic
collision rates due to a zero crossing of the scattering length can
be easily seen well above $1\mu$K; see Sec.~\ref{sssec:elcoll}.
%{ Finally, it should be noted that locations of narrow resonances %can be temperature dependent. }

\subsubsection{Inelastic loss spectroscopy}\label{sssec:lossspect}

Resonant losses are the most frequently observed signatures of
Feshbach resonances in cold-atom experiments. These losses can be
induced by two-body or three-body processes. Loss occurs because of
the release of internal energy into the motion when colliding atoms
end up in a lower internal state or when a molecule is formed. The
gain in kinetic energy is on the order of the Zeeman energy, the
hyperfine energy, or the molecular vibrational energy, depending on
the inelastic channel, and is generally so large that all atoms
involved in the collisions are lost. Near a Feshbach resonance,
inelastic loss is strongly enhanced because the Feshbach bound
states have strong couplings to inelastic outgoing channels.

Two-body and three-body collision loss can be quantified based on
the evolution of the atom number $N(t)$, which for a single species
satisfies
\begin{equation}
\dot{N}(t)=-\frac{N(t)}{\tau}-\int \left [ L_2n^2({\bf r},t)+L_3n^3({\bf
r},t) \right ] d^3r , \label{in_n1}
\end{equation}
where $\tau$ is the one-body lifetime, typically determined by
background gas collisions, $n({\bf r},t)$ is the position- and
time-dependent atomic density distribution and $L_2$ ($L_3$) is the
thermally averaged two-body (three-body) loss coefficient.

\begin{figure}
\includegraphics[width=3in]{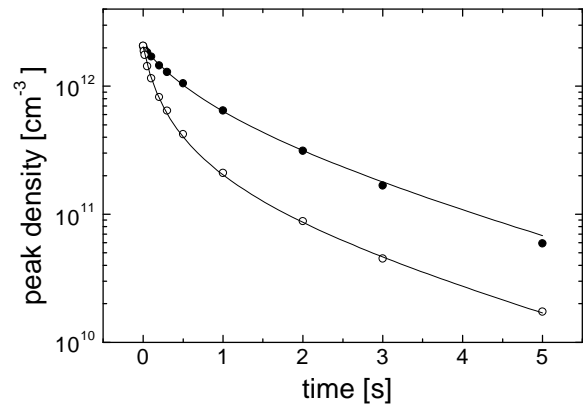}
\caption{Time evolution of the peak atomic density in a cloud of
$^{133}$Cs atoms  in the $|f=4, m=-4\rangle$ state at $T=5.3\,\mu$K.
The solid circles show the off-resonant evolution at  $B=140\,$G
whereas the open circles show the on-resonant evolution at
$B=205$\,G (open circles), where a Feshbach resonance is located.
The fit is based on Eq.~(\ref{in_n2}) with $L_3=0$, and the peak
density corresponds to $2\sqrt{2}\bar{n}$. From \cite{Chin2004}.}
\label{fig:in_decay}
\end{figure}

The loss equation can be further simplified under the assumption
that thermalization is much faster than inelastic loss. For example, for
a thermal cloud with temperature $T$ in a 3D harmonic trap, one
finds
\begin{eqnarray}
\dot{\bar{n}}(t)&=&-\frac{\bar{n}(t)}{\tau}-L_2\bar{n}(t)^2-(4/3)^{3/2}L_3\bar{n}(t)^3,
\label{in_n2}
\end{eqnarray}
where $\bar{n}=N\bar{\omega}^3(4\pi k_B T/m)^{-3/2}$ is the mean
density, $m$ the atomic mass, and $\bar{\omega}$ the geometric mean
of the three trap vibrational frequencies. Examples of
density-dependent loss curves are shown in Fig.~\ref{fig:in_decay}.

\begin{figure}
\includegraphics[width=2.5in]{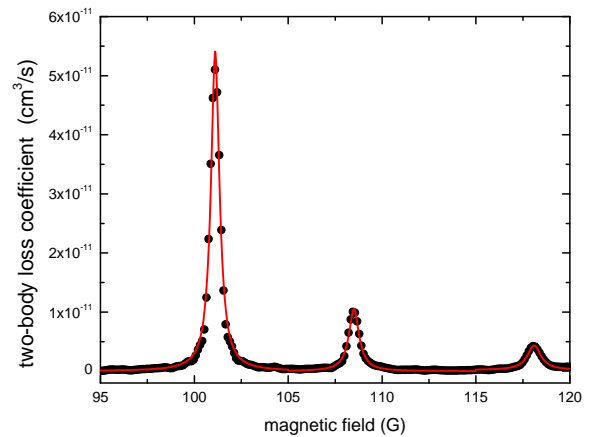}
\caption{Two-body inelastic loss coefficient of cesium atoms in the
$|f=3, m=-3\rangle$ state as a function of magnetic field. The loss
coefficient is extracted from the atomic density evolution, as shown
in Fig~\ref{fig:in_decay}. Three resonances are identified here. The
solid line is a Lorentzian fit. From \cite{Chin2004}.}
\label{fig:in_inelastic}
\end{figure}

The trap loss coefficient $L_2$ is related to the inelastic loss
coefficient in Sec.~\ref{sssec:collrate} by $L_2=K_{\rm loss}(T)$, where we
have assumed both atoms are lost in one collision event. Near a
Feshbach resonance, $L_2$ is enhanced and has a Lorentzian profile
at low temperatures; see Sec.~\ref{sssec:resscatt} for more details
on inelastic scattering resonances. Two-body collision loss has been
observed in many cold atom system; see Fig.~\ref{fig:in_inelastic}
for an example.

Three-body loss, as described by the loss coefficient $L_3$ in
Eq.~(\ref{in_n1}), is also strongly enhanced near Feshbach
resonances. In many experiments cold atoms are polarized in the
lowest ground state and two-body inelastic collisions (in the $aa$
channel) do not occur so that three-body recombination loss is the
dominant trap loss process. Three-body recombination occurs when
three atoms interact and form a diatomic molecule and a free atom.
In this process, the molecular binding energy is released into the
kinetic energies of the outgoing molecule and the third atom, which
except for very small molecular binding energies leads to immediate
trap loss.

In the first experimental report on atomic Feshbach resonances, \cite{Inouye1998}
observed very fast trap loss of a sodium BEC near a Feshbach
resonance. In this experiment,  three-body recombination is
the leading trap loss process. Recombination losses induced by
Feshbach resonances have been observed and studied in numerous later
experiments, for example, those by \cite{Roberts2000} on $^{85}$Rb, by \cite{Smirne2007,Marte2002} on $^{87}$Rb and by \cite{Weber2003} on $^{133}$Cs.

For bosonic atoms with large scattering length $a \gg \bar{a}$ and
low temperatures, $L_3$ scales generally as $a^4$ {
\cite{Fedichev1996a, Nielsen1999, Esry1999, Braaten2006}, but with additional quantum features (resonance and interference effects) as discussed
in Sec.~\ref{ssec:efimov} on Efimov physics.} For fermionic atoms the
situation is more complicated because of Pauli suppression effects;
see Sec.~\ref{ssec:fermi}, but generally a loss feature accompanies
a Feshbach resonance.

\subsubsection{Elastic collisions}\label{sssec:elcoll}

Elastic collisions refer to scattering processes in which the
colliding atoms only change their motional state, but not the
internal state. The cross section $\sigma_{\rm el}$ for elastic
$s$-wave collisions follows from Eq.~(\ref{II.A.6}). Neglecting the
effective range correction ($r_0=0$ in Eq.~(\ref{II.A.3})), one
obtains the simple expression
\begin{eqnarray}
\sigma_{\rm el}(E)=g \frac{4\pi a^2}{1+k^2a^2}, \label{in_n}
\end{eqnarray}
where  $a$ depends on magnetic field $B$ and $g$ is the symmetry
factor introduced in Sec.~\ref{sssec:collrate}. Near a Feshbach
resonance, the scattering length becomes very large. The elastic
$s$-wave cross section is $g$~$4\pi a^2$ at very low energy as $k
\to 0$ and approaches its upper bound $g\,4\pi/k^2$ in the unitarity
limit at finite $k$ where $ka \gtrsim 1$.  The latter can be reached
in the $\mu$K regime if $a$ becomes very large.

A strong enhancement of the thermally averaged elastic collision rate $n \langle \sigma_{\rm el} v\rangle$  can indicate the occurrence of a Feshbach resonance.
One experimental approach is to measure the thermalization rate,
which is proportional to the elastic collision rate as
$\kappa n \langle \sigma_{\rm el} v\rangle$. Here $\kappa$ is
numerically calculated as $2.7$ in the low temperature $k \to 0$ limit
\cite{Monroe1993}, $10.5$ in the unitarity limit \cite{Arndt1997}.
Similarly, \cite{DeMarco1999} found $\kappa=$4.1 for $p$-wave collisions in Fermi gases.
Finding Feshbach resonances based on analyzing thermalization rates
was reported by \cite{Vuletic1999} on $^{133}$Cs atoms and by
\cite{Loftus2002} on $^{40}$K.

\begin{figure}
\includegraphics[width=3in]{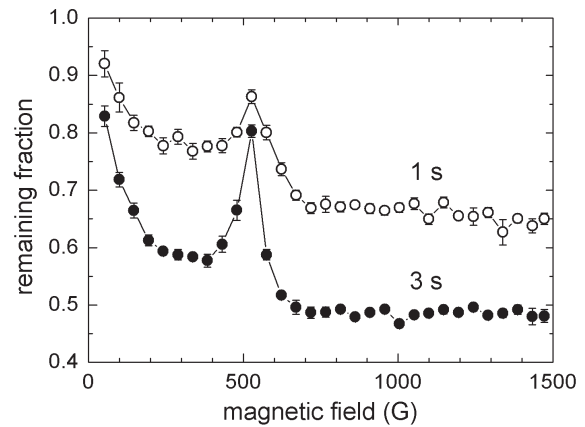}
\caption{Fraction of trapped atoms remaining after evaporation of $^6$Li atoms in an optical dipole trap after 1 s (open circles) and 3 s (closed circles) hold times. The
peak near $530$~G indicates a minimum of evaporative loss. This
minimum results from the zero crossing of the scattering length
induced by the broad 834-G Feshbach resonance. From \cite{Jochim2002}.}
\label{fig:in_eva}
\end{figure}

Near a resonance the thermalization rate can be limited under
hydrodynamic conditions, which are reached when the cross section is
so large that the collision rate in the trap exceeds the trap
frequency \cite{Vuletic1999}.  The maximum collision rate is also
bounded by the unitarity limit when $a$ is large. In both cases
resonance structure that is due to elastic scattering can become
less evident near $B_0$.

Another efficient method to identify Feshbach resonances based on
elastic collisions is to locate the zero crossing of the scattering
length, that is, the magnetic field for which the scattering length
vanishes near resonance where $B=B_0 + \Delta$; see
Eq.~(\ref{II.A.21}).  This can be monitored by measuring atom loss
resulting from  elastic collisions during the process of
evaporation. Thermalization and evaporation loss are suppressed at
the zero crossing. Schemes to locate the zero crossing have been
applied to $^{85}$Rb by \cite{Roberts1998}, $^{133}$Cs by
\cite{Chin2000}, $^{40}$K by \cite{Loftus2002}, $^6$Li by
\cite{Jochim2002, Ohara2002} and to a { $^{40}$K-$^{87}$Rb
mixture by \cite{Zaccanti2006}}. An example is shown in
Fig.~\ref{fig:in_eva}. While the zero crossing is evident, no
resonance feature is seen near $B_0=834$ G.

Finally, resonant changes of elastic scattering can also be revealed
through the detection of collision shifts in atomic clock
experiments \cite{Marion2004} and by measurements of the mean-field
interaction in Bose-Einstein condensates
\cite{Regal2003,Inouye1998,Cornish2000}; see discussion in
Sec.~\ref{sssec:meanfield}.

\subsubsection{Radiative { Feshbach} spectroscopy}\label{sssec:spectmol}

{ Radiative Feshbach spectroscopy makes use of red- or
blue-detuned light to detect the variation of the collisional wave
function near a Feshbach resonance.}  The amplitudes of both the
open and closed channel components of the wave function are strongly
modified for distances on the order of or less than $\bar{a}$ when
$B$ is tuned near resonance.  This is evident for the closed channel
component since the outer turning point will be on the order of
$\bar{a}$; see Sec.~\ref{sssec:vdw}. The optical transition induces
loss of atoms by excitation of atom pairs at such separations.

\cite{Courteille1998} adopted the idea of radiative spectroscopy to
identify a Feshbach resonance in $^{85}$Rb. In this experiment, a
photoassociation laser beam \cite{Jones2006} is held at a fixed
frequency to the red of the strong $S \to P$ atomic transition and
serves as a sensitive probe to measure the resonance position $B_0$.
The light excites the colliding pair of atoms to an excited
molecular level in a state with an attractive potential. The excited
level decays by spontaneous emission, giving rise to atom loss.  The
experiment monitors the atom loss as $B$ is varied near $B_0$, thus
locating the resonance.

\begin{figure}
\includegraphics[width=3in]{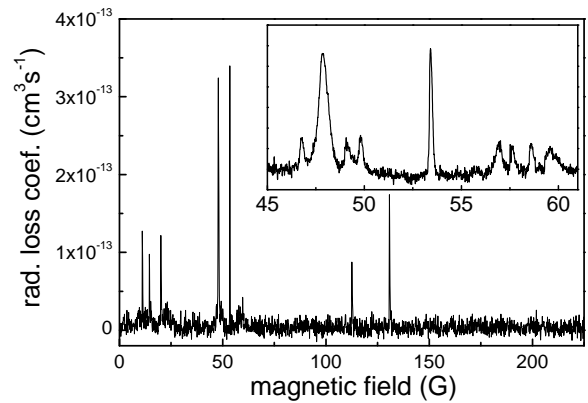}
\caption{Radiatively induced atom loss versus magnetic field due to
a far blue-detuned laser beam  applied to a sample of $^{133}$Cs
atoms confined in an optical dipole trap at a  temperature of
3~$\mu$K.   The laser wavelength is tuned 5~nm above the free atomic
transition to excite the colliding atoms to a repulsive molecular
state. Once excited, the atoms quickly dissociate and are lost from
the trap. Multiple narrow $d$- and $g$-wave Feshbach resonances were
observed. The inset shows an expanded view with more resonances
resolved. From \cite{Chin2003}.} \label{fig:in_radiative}
\end{figure}

In contrast, \cite{Chin2003} applied a laser with far blue detuning
to detect Feshbach resonances in cesium samples. The blue-detuned
light excites a molecular state with a repulsive potential so that
the atoms are repelled, accelerated from one another, and lost from
the trap. This method requires less detailed knowledge of molecular
structure than the previous method.    In this experiment, multiple
narrow $d$- and $g$-wave resonances were identified in two different
collision channels of cesium atoms; see Fig.~\ref{fig:in_radiative}.

\begin{figure}
\includegraphics[width=3in]{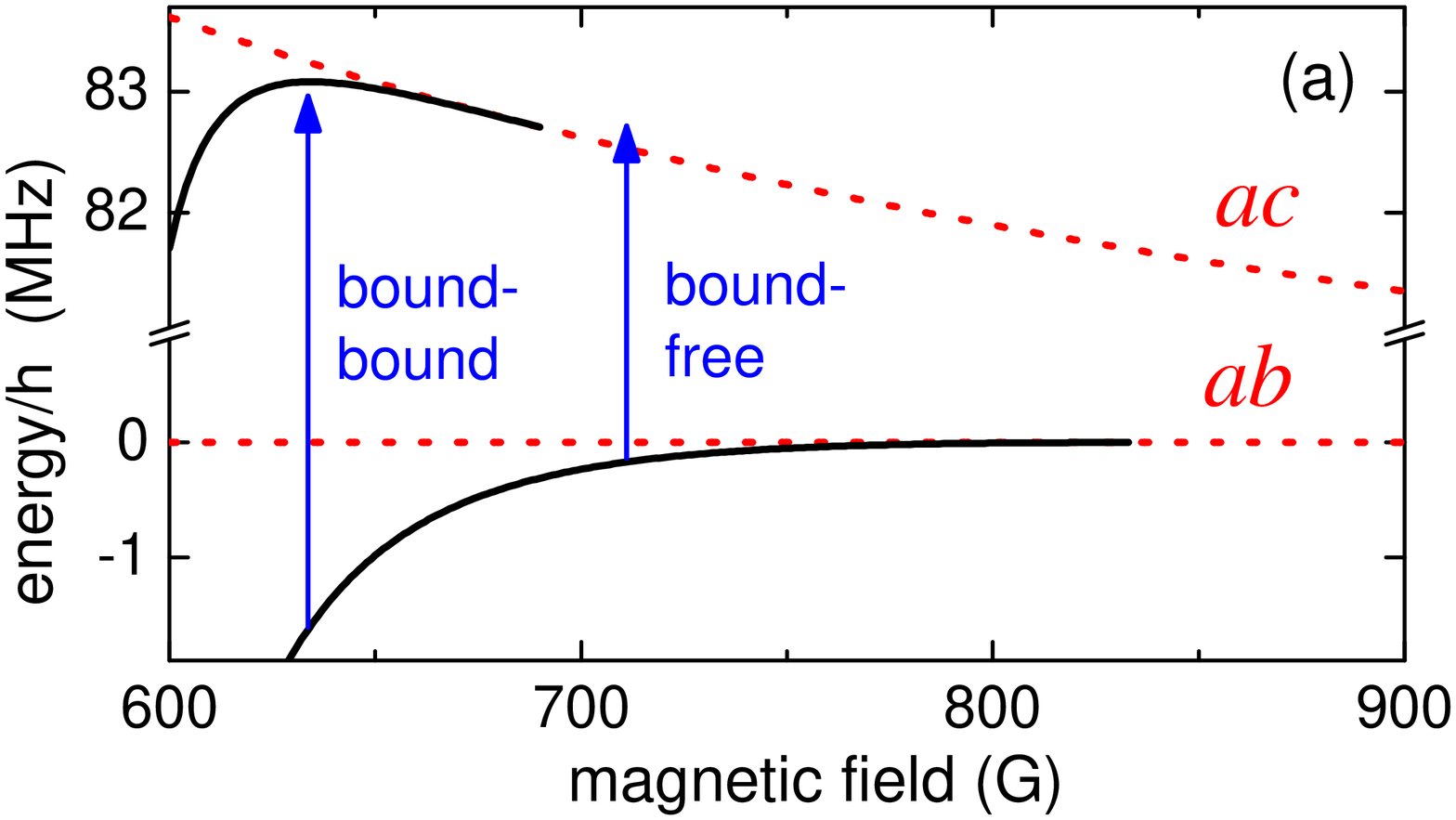}
\includegraphics[width=3in]{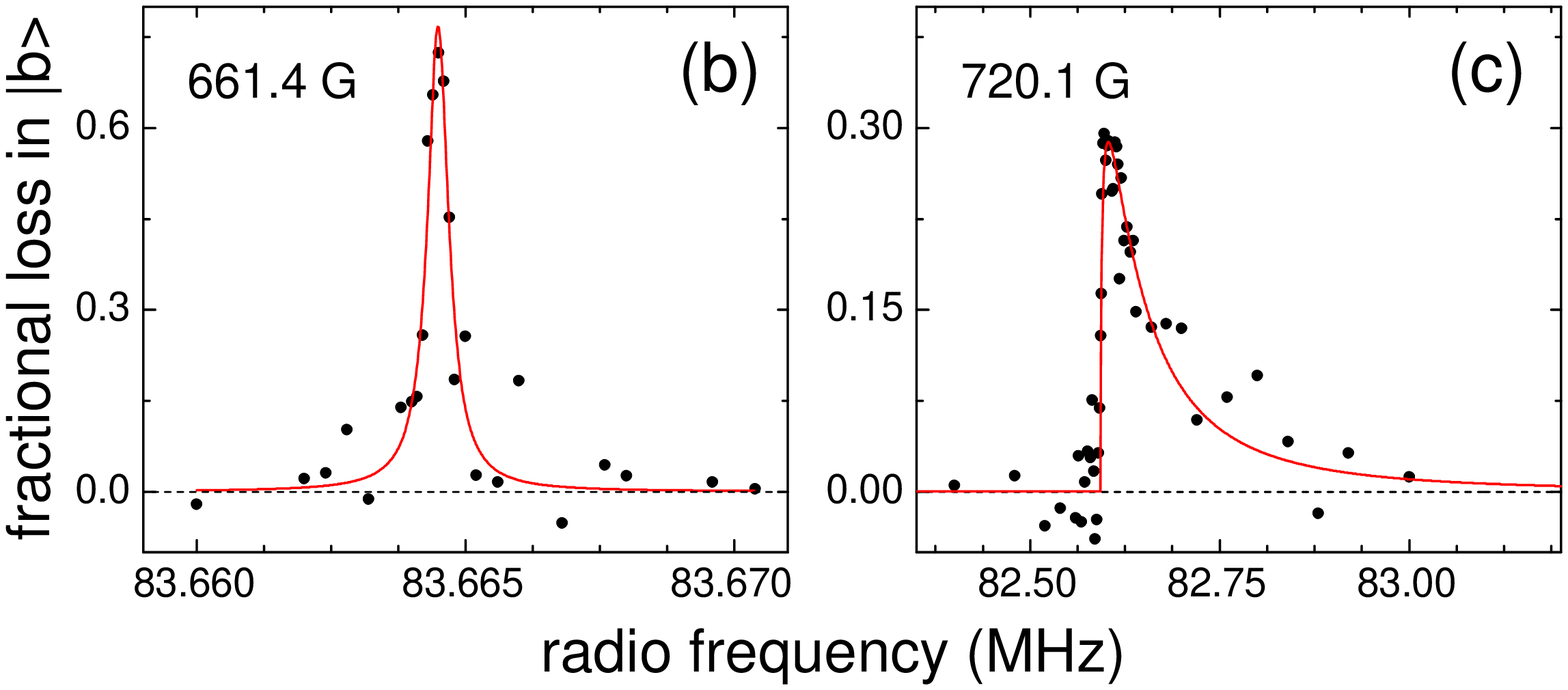}
\caption{Investigation of Feshbach resonances in the lowest two
scattering channels in $^6$Li with radio-frequency (rf) spectroscopy
on ultracold molecules. In this experiment, $10^5$ molecules,
initially prepared in the $ab$ channel and confined in a dipole
trap, are radiatively excited to the $ac$ channel. The rf excitation
lineshape shows a markedly different behavior below and above 690~G.
Panel (a) shows the molecular bound state (solid) and channel
energies (dotted) for the two channels. Below 690~G, a weakly bound
state exists in the $ac$ channel, and a bound-bound transition is
possible (blue arrow). Above 690~G, no weakly bound state exists in
the $ac$ channel, and only a bound-free transition that dissociates
the molecules is possible (blue arrow). Panel (b) shows the observed
narrow and symmetric bound-bound excitation lineshape at 661.4 G.
Panel (c) shows the observed broad and asymmetric bound-free
lineshape at 720.1 G. The disappearance of a bound state in $ac$
indicates that a Feshbach resonance occurs in this channel. Adapted
from \cite{Bartenstein2005}.} \label{fig:in_rf}
\end{figure}

\subsubsection{Binding energy { measurements}}\label{ssec:rfspec}

The detection methods discussed in the previous subsections provide
good ways to determine the existence of resonances and their
positions. Measurements of the magnetic-field dependent binding
energies of near-threshold Feshbach molecules can yield information
to precisely determine the scattering properties near a specific
resonance{; see Fig.~\ref{fig:intro_e&a} and
Sec.~\ref{sssec:AnalMol}}.

 \cite{Regal2003a, Bartenstein2005} employed
radio-frequency (rf) spectroscopy on Feshbach molecules to measure
very small molecular binding energies. An example of the rf
spectroscopy is {shown} in Fig.~\ref{fig:in_rf}. In this
experiment, weakly bound molecules are first prepared near the
Feshbach resonance; see Sec.~\ref{ssec:mol_formation}. An rf field
is then applied to drive either a ``bound-free'' transition, which
dissociates the molecules, or a ``bound-bound'' transition, which
converts them into a different molecular state. Based on the
lineshape functions calculated by \cite{Chin2005a}, binding energy
of Feshbach molecules can be measured to 1~kHz. This kind of precise
data can be combined with theoretical modeling to determine the
position and the width of the resonance.

Other efforts to spectroscopically probe weakly bound states include oscillating magnetic field spectroscopy employed by \cite{Papp2006,
Thompson2005}, and rf and microwave techniques by \cite{Mark2007a,
Zirbel2008}. Using the theoretical models described in
Sec.~\ref{sssec:SquareWell}, \cite{Lange2008} show how Feshbach
resonance parameters can be extracted from molecular
binding energy measurements.

\subsection{Homonuclear alkali-metal systems}\label{ssec:species_B}

Feshbach resonances have been found and characterized in essentially
all single-species alkali-metal systems. The scattering properties
show vast differences between the various species and also between
different isotopes of the same species. Each system is unique and
has particular properties. Here, we give a brief account for each
species or isotope, ranging from first observations to the best
current knowledge, and we discuss the
characteristic properties. A table of important resonances can be
found in the Appendix (Table \ref{tab:resonances}).

\subsubsection{Lithium}\label{sssec:Li}

Lithium has two stable isotopes, one fermionic ($^6$Li) and one
bosonic ($^7$Li).

$^6$Li -- In 1998 \cite{Houbiers1998} predicted Feshbach resonances
in cold collisions of fermionic $^6$Li. Experimental evidence for a
prominent resonance was established by monitoring inelastic decay
\cite{Dieckmann2002} and elastic collision properties
\cite{Jochim2002, Ohara2002}. These experiments showed large
variations in loss and thermalization rates as a function of the
magnetic field strength for equal mixtures of the lowest two
hyperfine ground states $a$ and $b$ prepared in an optical dipole
trap. Because of the fermionic nature of $^6$Li even partial-wave
scattering and, in particular, $s$-wave scattering only occurs
between atoms in unlike hyperfine states. Consequently, if the gas
is sufficiently cold such that the $\ell>0$ centrifugal barriers are
higher than the temperature of the gas, { $s$-wave collisions in the $ab$ channel represent the essential thermalization mechanism}. \cite{Jochim2002, Ohara2002} reported a strongly suppressed thermalization rate near 530~G, indicating the zero crossing of the scattering length in the
$ab$ channel. Most recently, \cite{Du2008} located the zero crossing
to $527.5 \pm 0.2$\,G. \cite{Dieckmann2002} observed enhanced
inelastic loss near 680\,G, about 150\,G below the actual resonance
location ($B_0 = 834$\,G). Note that this particular resonance is
extraordinarily broad ($\Delta \approx -300$\,G). Further
experiments by \cite{Bourdel2003,Gupta2003} provided confirmation of
this Feshbach resonance.

In order to further pinpoint the resonance position,
\cite{Bartenstein2005} conducted radio-frequency spectroscopy on
Feshbach molecules in the $ab$ channel as described in
Sec.~\ref{sssec:spectmol}. The resonance could be located with an
uncertainty of 1.5\,G, more than two orders of magnitude smaller
than its width. Broad resonances in two other entrance channels
($ac$ and $bc$) were reported as well.

\cite{Strecker2003} identified a narrow $s$-wave resonance at
543.25~G in the $ab$ channel, and \cite{Zhang2004, Schunck2005}
observed three $p$-wave resonances near 200\,G, one in each of the
channels $aa$, $ab$, and $bb$. All of these observed resonances are
induced by $s$- and $p$-wave rotational levels of the $v=38$ bound
state of the singlet X$^1\Sigma_g^+$ potential, as shown in
Fig.~\ref{fig:Li_ab_bound} for $s$-wave resonances.

The two $s$-wave resonances in the $ab$ channel illustrate the
concept of resonance strength, as described in
Sec.~\ref{sssec:resstrength}. The broad resonance at 834~G is
strongly entrance-channel dominated, while the narrow resonance at
543~G is an extreme case of a closed-channel dominated resonance.
The former has an extraordinarily large magnetic field range of
universal behavior, while that for the latter is vanishingly small (detunings of a few $\mu$G or less); see Sec.~\ref{sssec:AnalMol} and Fig.~\ref{fig:E&Z}. This has important consequences for molecules formed near these resonances.

The very broad resonance at 834~G in the $ab$ channel is used to
create molecular Bose-Einstein condensates and fermionic
superfluids. This will be extensively discussed in
Sec.~\ref{ssec:fermi}.

$^7$Li -- \cite{Moerdijk1994} predicted Feshbach resonances in
collisions of bosonic $^7$Li atoms in $|f=1, m=1\rangle$ ($aa$ channel).
\cite{Strecker2002} identified a zero crossing of the scattering
length induced by a broad Feshbach resonance. This resonance was used to study the formation of bright solitons by \cite{Khaykovich2002, Strecker2002} at small and negative scattering lengths near the zero crossing; see also Sec.~\ref{sssec:solitons}. { In these early experiments, the resonance position was estimated to 720\,G. Later measurements by \cite{Junker2008} and \cite{Pollack2008} accurately pinpointed the resonance position and the zero crossing to 736.8(2)\,G and 543.6(1)\,G, respectively.}

{
\cite{Gross2008} observed two resonances in the state $|f=1, m=0\rangle$ ($bb$ channel). A narrower one was found at 831(4)\,G with a width of 7\,G, while a broader one (width 34\,G) was located at $\sim$\,884\,G . In between these two resonances, a zero crossing was found at 836(4)\,G.}

\subsubsection{Sodium}\label{sssec:Na}

%Sodium has only one stable isotope ($^{23}$Na).

\cite{Inouye1998} pioneered experimental research in locating
Feshbach resonances. In an optically trapped BEC of $^{23}$Na (the
only stable isotope) with all atoms in the lowest hyperfine state
$|f=1, m=1\rangle$, they identified resonances at 853~G and 907~G. A
third resonance in a different channel at 1195~G was later reported
by \cite{Stenger1999}. All three resonances are narrow and $s$ wave
in nature. The 1998 experiment showed both strongly enhanced trap
loss and the dispersive tuning of the scattering length near the
907-G resonance; see Fig.~\ref{fig:firstobsNa}.

The experimental determination of the resonances has enabled
detailed models of the interaction potentials between ultracold Na
atom \cite{vanAbeelen1999}. These models were further refined by
\cite{Samuelis2000} based on conventional molecular spectroscopy in
combination with photoassociation data.

The Feshbach resonance at 907~G was used to create ultracold Na$_2$
molecules \cite{Xu2003} and to demonstrate coherent molecule optics
\cite{Abo2005}; see also Sec.~\ref{sssec:mol_timedep}.

\subsubsection{Potassium}\label{sssec:K}

Potassium has three stable isotopes, two of them are bosonic
($^{39}$K and $^{41}$K) and one is fermionic ($^{40}$K).

$^{39}$K -- In 1996 \cite{Boesten1996a} predicted Feshbach resonance
locations in collisions between $^{39}$K atoms. Their results were
based on spectroscopic data of binding energies of ro-vibrational
states of the X$^1\Sigma_g^+$ and $^3\Sigma_u^+$ potentials.  The
data, however, { were} not sufficiently complete to give
quantitative resonance locations, but it did show that the
likelihood of resonances was large. Another early prediction of
resonance locations was made by \cite{Bohn1999a}.

Feshbach resonances in $^{39}$K were observed { at 402~G}
and analyzed by \cite{Derrico2007}, and applied to create a tunable
BEC of this species \cite{Roati2007}. The zero crossing of the
scattering length near the broad 402-G resonance has found
intriguing applications for atom interferometry with non-interacting
condensates \cite{Fattori2007}; see also Sec.~\ref{sssec:nonint}.
{ This resonance has an intermediate character between
that of an entrance channel and a closed channel dominated resonance.}

$^{40}$K -- Early predictions on Feshbach resonances in $^{40}$K
were made by \cite{Bohn1999a}. The first experimental observation
was reported by \cite{Loftus2002}, who demonstrated resonant control
of elastic collisions via the 202-G resonance in a mixture of the
lowest two spin states ($ab$ channel). One year later, the same
group reported on a $p$-wave resonance at 199\,G \cite{Regal2003};
they measured the resonantly enhanced elastic collision rate of
atoms in the second-lowest hyperfine state ($bb$ channel) at a
temperature of 3~$\mu$K. At an even lower temperature and by
monitoring the collision-induced heating rate, \cite{Ticknor2004}
found that this resonance is actually a doublet. This doublet
structure in the $p$-wave resonance is due to a small energy
splitting between the $|\ell=1, m_\ell=0\rangle$ and $|\ell=1,
m_\ell=\pm1\rangle$ molecular states. The anisotropic nature of the
$p$-wave resonances has found interesting applications in low-dimensional traps \cite{Gunter2005}.

In spin mixtures with $s$-wave interactions, \cite{Regal2003b,
Regal2003} identified 10~G wide $s$-wave Feshbach resonances in the
$ab$ and $ac$ channels at $201.6(6)$~G and $224.21(5)$~G,
respectively, by monitoring the thermalization rate and mean-field
shifts. These resonances provide a convenient tool to study strongly
interacting Fermi gases and fermionic condensates; see
Secs.\ref{ssec:fermi}.

$^{41}$K -- { A Feshbach resonance was recently observed by \cite{Kishimoto2008} at 51.4~G in the $cc$ channel. The observation confirmed a  theoretical prediction by \cite{Derrico2007}, which was based on experimental data available for the other potassium isotopes.}

\subsubsection{Rubidium}\label{sssec:Rb}

$^{85}$Rb -- \cite{Courteille1998} reported a Feshbach resonance in
a magnetically trapped thermal sample by observing enhanced
photoassociative loss; see Sec.~\ref{sssec:spectmol}. This result
confirmed the prediction by \cite{Vogels1997}. \cite{vanAbeelen1998}
suggested using photoassociation as a probe to identify Feshbach
resonances.
%the $ee$ collision of $^{85}$Rb by the photoassociation method.
%Atoms loss due to a photoassociation laser beam serves as a probe
%for the atomic scattering wavefunction.  Near a Feshbach resonance
%the wavefunction changes dramatically, which then leads to a
%strongly enhanced photoassociation rate.
By monitoring inelastic loss, \cite{Roberts1998} determined the
position of this 10-G wide resonance to be at 155~G.
\cite{Claussen2003} then used a BEC to perform a high-precision
spectroscopic measurement of the molecular binding energy and
determined the resonance parameters $B_0$ and $\Delta$ within
20\,mG.

Attainment of BEC in $^{85}$Rb crucially depended on the existence
of the 155-G resonance; see Sec.~\ref{sssec:attainment}. Only in a
10-G window near the resonance the scattering length is positive and
\cite{Cornish2000} were able to Bose condense $^{85}$Rb. As the
first available BEC with widely tunable interactions the system has
received considerable attention; see Sec.~\ref{sssec:solitons}.
Coherent atom-molecule coupling \cite{Donley2002} and the formation
of ultracold $^{85}$Rb$_2$ molecules \cite{Thompson2005} were
reported based on this Feshbach resonance.

$^{87}$Rb -- \cite{Marte2002} conducted a systematic search for
Feshbach resonances in bosonic $^{87}$Rb. For atoms polarized in
various combinations of magnetic sublevels in the $f=1$ hyperfine
manifold, more than 40 resonances were observed between 300~G and
1200~G by monitoring atom loss in an optical dipole trap. These
resonances are induced by $s$- and $d$-wave bound states and are all
very narrow. For the $s$-wave states, the underlying molecular
structure is shown in Fig.~\ref{fig:Rb87_Marte}. The widest and most
often used resonance is located at 1007~G and has a width of 0.2~G.
In different experiments, \cite{Erhard2004,Widera2004} observed a
low-field resonance near $9$~G in the $ae$ channel.

Several resonances in $^{87}$Rb have been used to form ultracold
Feshbach molecules. They are formed by ramping the magnetic field
through the resonance \cite{Durr2004a}; see
Sec.~\ref{sssec:mol_timedep}. The great potential of combining
Feshbach resonances with optical lattices has been demonstrated in a
series of experiments with $^{87}$Rb
\cite{Thalhammer2006,Winkler2006,Volz2006,Syassen2007};
see Sec.~\ref{ssec:ola}.

\subsubsection{Cesium}\label{sssec:Cs}

Cesium, for which the isotope $^{133}$Cs is the only stable one, was proposed as the first alkali-metal species in which
Feshbach resonances could be observed \cite{Tiesinga1993}. With the limited
state of knowledge of the interaction potentials no
quantitative predictions could be made. The
first observation of Feshbach resonances in cesium collisions was
published seven years later \cite{Vuletic1999}.

\cite{Vuletic1999, Chin2000, Chin2004} reported on more than 60
Feshbach resonances of various types using several detection
schemes. These include $s$-, $p$-, $d$-, $f$- and $g$-wave
resonances in 10 different scattering channels. \cite{Vuletic1999}
used cross-axis thermalization rates to identify resonances in the
lowest $aa$ channel and used trap loss measurements in the $gg$
channel.  \cite{Chin2000} reported many more resonances by preparing
the atoms in other internal states and by monitoring the evaporation
rates to more efficiently measure the elastic cross section. The
resonances provided \cite{Leo2000} with the essential information to
precisely determine the interaction potentials of ultracold cesium.

Narrow resonances were observed by \cite{Chin2003, Chin2004} using
radiative Feshbach spectroscopy; see Sec.\ \ref{sssec:spectmol}. In
these experiments $^{133}$Cs atoms were illuminated by a far
blue-detuned laser beam, whose wavelength was optimized to only
remove atoms near a resonance.  These resonances are induced by
$g$-wave bound  states and are only strong enough to be observed due
to the large second-order spin-orbit coupling of cesium atoms. The
observed resonances are at low magnetic fields, which can be understood
as a consequence of the large background scattering length in the $aa$
collision channel; see Sec.~\ref{sssec:vdw}. In addition, $l$-wave
Feshbach molecules have been produced by \cite{Mark2007a, Knoop2007};
{ their coupling to the $s$-wave continuum is too weak to lead to observable resonances in collision experiments.} Finally, \cite{Lee2007} predicted a broad $s$-wave resonance at
$\sim$800~G in the $aa$ channel. This is a magnetic field regime that has not been experimentally explored yet.

\begin{figure}
\epsfxsize=3.3 in \epsfbox{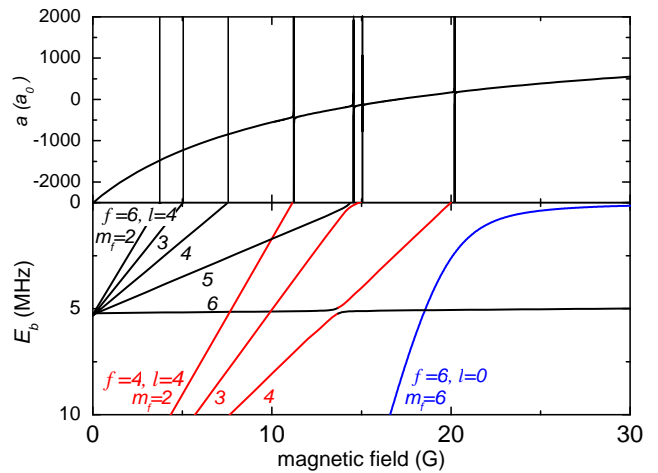}
\caption{Scattering length and bound state energies for cesium
atoms in the lowest internal state as a function of magnetic field.
From \cite{Chin2004}.} \label{fig:cs_aa_slength}
\end{figure}

Figure~\ref{fig:cs_aa_slength} shows the scattering length in the
$aa$ channel. It has a zero crossing at 17.1\,G \cite{Chin2004,
Gustavsson2007} and multiple narrow resonances below 50~G. The
gradual change in $a$ from $-2500$ $a_0$ to 500 $a_0$ across 30 G in
the Figure is actually the tail of a broad resonance with $B_0=-12$
G; see Appendix and Table~\ref{tab:resonances}. The negative $B_0$
follows from fitting to Eq.~(\ref{II.A.21}). Both the narrow and
broad resonances have provided favorable conditions for many
exciting experiments. This included the attainment of BEC with Cs
atoms, a Bose condensate \cite{Weber2003}, the formation of Cs$_2$
\cite{Herbig2003, Chin2003}, the observation of resonances between
ultracold molecules \cite{Chin2005}, and studies on Efimov physics;
see Sec.~\ref{ssec:efimov}.

\subsection{Heteronuclear and other systems}\label{ssec:species_C}

Most of the experimental and theoretical attention so far has been focused
on locating and using magnetic Feshbach resonances in single-species
alkali-metal atom gases. Over the last five years, however, considerable progress has also been made in locating Feshbach resonance in other atomic species and in mixtures of alkali-metal atoms. These systems are investigated for various reasons. Of particular interest is the
promise of more exotic quantum many-body behavior \cite{Bloch2007,Menotti2007,Micheli2006}.

Heteronuclear systems provide the path to prepare mixtures of bosonic and fermionic quantum degenerate gases. Intriguing applications include the creation of fermionic molecules in an atomic Bose-Fermi mixture \cite{Ospelkaus2006b, Zirbel2007} and novel quantum phases of fermions with unequal masses \cite{Petrov2007}.
Feshbach resonances provide means to tune the interactions between different species in order to explore quantum phases in various regimes.

Atoms with magnetic moments interact via the long-range magnetic
dipole-dipole interaction $V_{ss}$ in addition to the van der Waals
and more { short-range} interactions.  For alkali-metal
atoms the effect of this dipole-dipole interaction on collective
behavior is small.  In atomic species with much larger magnetic
moments, however, the dipole-dipole interaction can have a
significant impact on the many-body behavior of the gas
\cite{Goral2000,Santos2003,Santos2006}. In an atomic species like
chromium (Sec.~\ref{sssec:Cr}), magnetic Feshbach resonances can be
used to tune the relative strength of the short-ranged interactions
and the long-range dipole-dipole interaction
\cite{Yi2002,Lahaye2007}.

Another way to create exotic many-body systems is by pairing
different atomic species into Feshbach molecules, which can then be
converted into deeply bound molecules with a large electric dipole
moment. Such a moment gives rise to large dipolar interactions,
orders of magnitude larger than possible with magnetic dipole
moments. These molecules, which are either bosonic or fermionic,
have many applications in dipolar molecular quantum gases
\cite{Micheli2006,Buchler2007} and quantum computation
\cite{Demille2002}.

\subsubsection{Chromium}\label{sssec:Cr}

\begin{figure*}[t]
\includegraphics[scale=1]{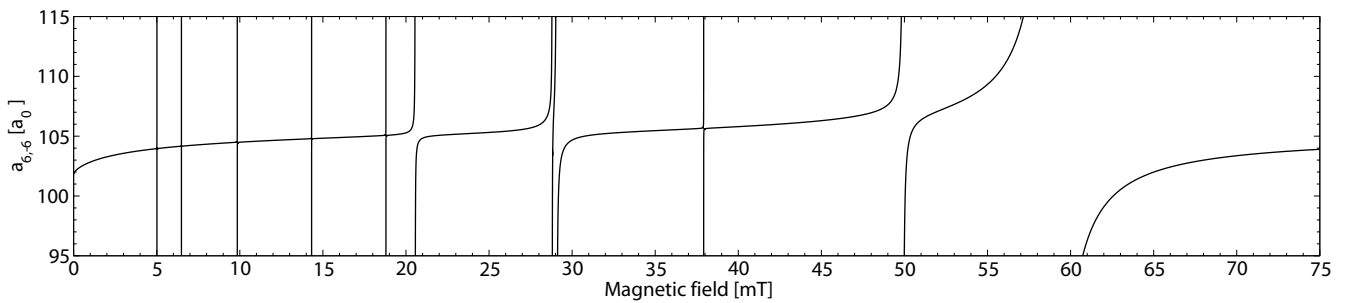}
\caption{Scattering length of $^{52}$Cr atoms versus magnetic
field. The feature near 29~mT is a pair of nearly-degenerate
Feshbach resonances. From \cite{Werner2005}.}
 \label{fig:Cr}
\end{figure*}

In 2005 \cite{Griesmaier2005} announced BEC
of $^{52}$Cr atoms. This species has a magnetic moment that is six times
larger than that for alkali-metal atoms. Subsequently,
\cite{Stuhler2005} showed that the expansion of a cigar-shaped
chromium condensate depended on the relative orientation of the
magnetic moment and the elongated condensate, which for the first
time showed the effects of the dipole-dipole interaction in a
quantum-degenerate gas.

\cite{Werner2005} measured fourteen magnetic Feshbach resonances in
$^{52}$Cr in the energetically lowest magnetic sublevel.  Feshbach
resonances were detected by measuring, after a fixed hold time, the
number of remaining atoms as a function of magnetic field.  Atom
loss could only have occurred by enhanced three-body recombination
near the resonance. Figure~\ref{fig:Cr} shows the $s$-wave
scattering length as a function of magnetic field derived from the
observed locations of the Feshbach resonances and a multi-channel
scattering model of the collision.  Most of the resonances are
$g$-wave resonances with the exception of one of the two
nearly-degenerate resonances at 29~mT and those at 50~mT and 59~mT.
These three resonances have $d$-wave character. Moreover,
\cite{Werner2005} also assigned two resonances as originating from $d$-wave collisions and coupling to an $s$-wave closed channel, { i.e.\ $\ell = 2$ and $\ell_c=0$ according to the notation discussed in Sec.~\ref{sssec:feshcoupl}. Note that these resonances do not show up in the $s$-wave scattering length as displayed in Fig.~\ref{fig:Cr}. One of these unusual resonances was investigated in some detail in \cite{Beaufils2008}.}

\subsubsection{Mixed species}\label{sssec:mixed}

K+Rb -- The first mixed system to receive a detailed effort to
locate Feshbach resonances was a mixture fermionic $^{40}$K and
bosonic $^{87}$Rb atoms \cite{Simoni2003}.  Based on experimentally
determined inelastic and elastic rate coefficients at zero magnetic
field they predicted the location of fifteen Feshbach resonances
with an uncertainty ranging from 10 G to 100 G.

\begin{figure}
\includegraphics[width=3.0in]{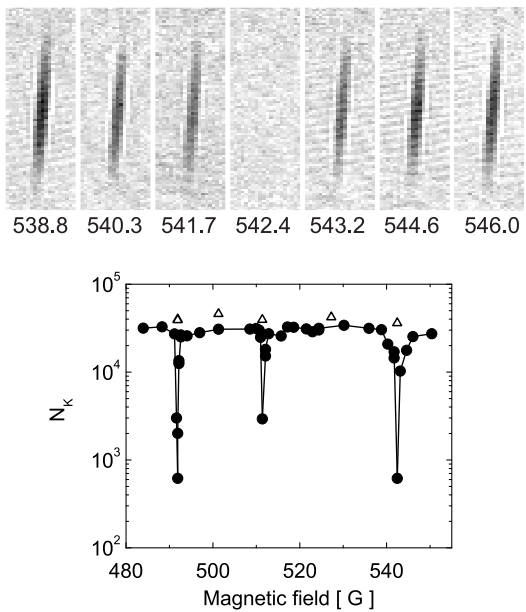}
\caption{Observation of Feshbach resonances in $^{40}$K+$^{87}$Rb.
The top panel shows in-trap absorption images of $^{40}$K atoms
after a fixed hold time of $\approx 1$ s at various magnetic fields.
The label on each image gives the magnetic field in Gauss. No
$^{40}$K atoms could be seen at 542.2 G. The bottom panel shows the
number of remaining $^{40}$K atoms after the fixed hold time as a
function of magnetic field. Narrow features are observed at 492 G,
512 G, and 543 G. Adapted from \cite{Inouye2004}.} \label{fig:KRb}
\end{figure}

\cite{Inouye2004} presented the first direct determination of the
magnetic field location of three Feshbach resonances.  The two
atomic species, each in their energetically lowest hyperfine state,
were optically trapped.  As shown in
Fig.~\ref{fig:KRb} the Feshbach resonances were detected by
measuring, after a fixed hold time, the number of remaining atoms as
a function of magnetic field.  The atom loss is due to three-body
recombination.

\cite{Ferlaino2006} confirmed the positions of these three
resonances and found nine additional resonances.  Theoretical
modeling uniquely assigned each resonance and
determined the scattering lengths of the X$^1\Sigma^+$ and
a$^3\Sigma^+$ Born-Oppenheimer potentials to about 2\%.  The
difference between the experimental Feshbach locations and those of
the best-fit theory was less than 1 G for all resonances. The
experimental uncertainty in the resonance locations was 0.2 G.

\cite{Ferlaino2006} also predicted the location of resonances of
several isotopic combinations. For the bosonic $^{39}$K and
$^{87}$Rb collision with both atoms in their lowest hyperfine state
\cite{Roati2007} confirmed the location of one such resonance.  It
was found at 317.9~G well within the uncertainties quoted by
\cite{Ferlaino2006}. \cite{Klempt2007} observed a number of new
Feshbach resonances, constructed an accurate potential model, and
predicted resonances in other isotopic KRb combinations. { \cite{Simoni2008} presented a refined near-threshold model for scattering and bound-state calculations for all isotopic combinations of K and Rb.}

Feshbach resonances in mixtures of $^{40}$K and $^{87}$Rb have been applied to the creation of Feshbach molecules in optical traps \cite{Zirbel2007,Zirbel2008}
and in single sites of an optical lattice \cite{Ospelkaus2006b}; see Sec.~\ref{sssec:ola_mol}. Recently these fermionic molecules have been transferred to more deeply bound levels \cite{Ospelkaus2008}; see Sec.~\ref{sssec:transfer}.
Furthermore, interspecies interaction tuning has been exploited to study the collective behavior of a $^{40}$K+$^{87}$Rb Bose-Fermi mixture \cite{Ospelkaus2006a} and to realize a tunable double species BEC in a $^{41}$K+$^{87}$Rb Bose-Bose mixture \cite{Thalhammer2008}.

Li+Na -- \cite{Stan2004} have observed three magnetic Feshbach
resonances in the interaction between a degenerate fermionic $^6$Li
gas and a Bose-Einstein condensate of Na.  The resonances were
observed by detecting atom loss when sweeping the magnetic field at
constant rate through the resonances.  Atom loss could only have
occurred from three-body recombination or from molecule formation
during the sweep.  The observed resonances at 746.0 G, 759.6 G, and
795.6 G are $s$-wave resonances. { In a recent theoretical study based on this experimental input data, \cite{Gacesa2008} derive precise values of the triplet and singlet scattering lengths for both the $^6$Li-Na and the $^7$Li-Na combination. Moreover they predict a variety of additional Feshbach resonances within an experimentally attainable field range.}

$^6$Li+$^{40}$K -- \cite{Wille2007} described the observation of
thirteen Feshbach resonances in fermionic $^6$Li and $^{40}$K in
various hyperfine states. Their theoretical analysis, which relies on the model developed by \cite{Stan2004} and discussed in Sec.~\ref{sssec:OtherApprox}, indicated that the resonances were either $s$- or $p$-wave resonances. This isotopic combination is a prime candidate for the study of strongly interacting Fermi-Fermi mixtures.

Na+Rb -- Predictions of Feshbach resonance locations based on
analysis of high-resolution Fourier spectroscopy of the molecular
X$^1\Sigma_g^+$ and a$^3\Sigma_u^+$ states in a 600~K beam of NaRb
molecules are described in \cite{Bhattacharya2004} and
\cite{Pashov2005}.  For example \cite{Pashov2005} predict for the
ultra-cold collision between $^{23}$Na and $^{85}$Rb, both in the
energetically lowest hyperfine state, $s$-wave Feshbach resonances
at 170~G and 430~G with an uncertainty of about 50~G. This uncertainty
is sufficiently small that the predictions will be helpful for
planning experiments which can accurately locate the
resonances.

Initial experiments on other combinations of mixed atomic species have been performed. Feshbach resonances have been reported in the $^6$Li+$^{87}$Rb system by \cite{Deh2008}, in the $^7$Li+$^{87}$Rb system by \cite{Marzok2008}, and in the $^{87}$Rb+$^{133}$Cs system by \cite{Pilch2008}.

\subsubsection{Isotopic mixtures}\label{sssec:isomix}

A special case of a mixed system is that where different isotopes of
the same element are combined.  In particular, isotopic mixtures of Rb,
K, and Li have been studied. Isotopic mixtures of K or Li are of
particular interest as both fermionic and bosonic isotopes exist.

Feshbach resonances in isotopic mixtures of rubidium have recently
been observed.  \cite{Papp2006} found two $s$-wave Feshbach
resonances in the collision of $^{85}$Rb and $^{87}$Rb when both
isotopes are in their lowest hyperfine state.  Their magnetic field
locations of 265.44(0.15)~G and  372.4(1.3)~G are consistent with
the predictions of \cite{Burke1998}.

\cite{vanKempen2004} have predicted the location of $^6$Li+$^7$Li
Feshbach resonances. When both isotopes are in the lowest hyperfine
state resonances occur between 200~G and 250~G as well as between
550~G and 560~G. Four of these resonances have been observed by \cite{Zhang2005}.

In predictions for isotopic mixtures ``mass scaling'' is often used.
As the interatomic potentials are to good approximation independent
of the isotopic composition of the dimer, the only thing that
changes is the (reduced) mass of the dimer. Corrections are due to
the breakdown of the Born-Oppenheimer approximation.
\cite{vanKempen2002,Seto2000} showed that for Rb the experimental
data on $^{85}$Rb$^{85}$Rb, $^{85}$Rb$^{87}$Rb and
$^{87}$Rb$^{87}$Rb are consistent with mass scaling.  A
new analysis that includes the 2006 observations of \cite{Papp2006}
is needed.

For the lithium system \cite{vanKempen2004} showed that mass scaling
is insufficient to explain the observed data for the homonuclear
$^6$Li+$^6$Li and  $^7$Li+$^7$Li systems.  Consequently,
\cite{vanKempen2004} quote a 1 G uncertainty for the location of
$^6$Li+$^7$Li Feshbach resonances from the breakdown of mass
scaling.

\section{Control of atomic quantum gases}
\label{sec:qugas}

Tuning two-body interactions via Feshbach resonances is the
experimental key to control collective phenomena in degenerate
quantum gases. This has found numerous applications, both with atomic Bose-Einstein condensates \cite{Cornell2002,Ketterle2002} and with degenerate Fermi gases \cite{Varenna2006}.

The different decay properties of Bose and Fermi gases near Feshbach
resonances play a crucial role for the experiments. For Bose gases,
resonant two-body scattering in general leads to rapid decay via
three-body collisions, as we have discussed in context with loss
spectroscopy on Feshbach resonances in Sec.\ \ref{sssec:lossspect}.
Three-body decay limits the practical applicability of Feshbach
tuning to Bose gases, restricting the experiments to the dilute gas
regime where the scattering length is small compared to typical
interparticle separations. In contrast, Fermi gases can be
remarkably stable near { $s$-wave} Feshbach resonances
\cite{Petrov2004a}.

For atomic Bose-Einstein condensates in the dilute gas regime
(Sec.~\ref{ssec:bec}), the collective behavior can be described in a
mean-field approach. In strongly interacting Fermi gases (Sec.~\ref{ssec:fermi}), the role of the
scattering length is much more complex. Here Feshbach tuning can be
used to control the nature of fermionic pairing in different
superfluid regimes.

\subsection{Bose-Einstein condensates}
\label{ssec:bec}

In experiments on atomic BEC, the role of Feshbach tuning can be
divided into two parts. Firstly, the control of collision
properties can be essential for the attainment of BEC. Secondly, the possibility to control the mean-field
interaction opens up a variety of interesting applications.

\subsubsection{Attainment of BEC}
\label{sssec:attainment}

Some atomic species offer favorable collision properties for the attainment
of BEC without any necessity of interaction tuning; $^{87}$Rb and
$^{23}$Na are the most prominent examples
\cite{Cornell2002,Ketterle2002}. In other cases, however, Feshbach
tuning is essential either to produce large condensates ($^{7}$Li),
or to achieve BEC at all ($^{85}$Rb, $^{133}$Cs, $^{39}$K).

Let us first consider the general question, what is a ``good''
scattering length $a$ for making a BEC. First, $a$ should be
positive, because condensates undergo collapse at negative
scattering length when the number of condensed atoms exceeds a
relatively small critical value \cite{Bradley1997}.
Moreover, $a$ should not be too small, because a sufficiently large elastic
collision rate is required for evaporative cooling towards BEC; the
cross section for elastic collisions between identical bosons is
$8\pi a^2$. Finally, the scattering length should not be too large to
avoid rapid decay by three-body collisions
\cite{Roberts2000,Weber2003a} as three-body decay
scales as $a^4$ \cite{Fedichev1996a,Braaten2006}. In practice these
conditions result in typical values for a ``good'' value of $a$
between a few ten and a few hundred times the Bohr radius $a_0$. In
detail, the optimum value for $a$ depends on the confinement
properties of the trap and behavior of inelastic
decay.

For $^{7}$Li, early magnetic trapping experiments showed BEC in the
internal state $f=2, m=2$, where the scattering length $a=-27\,a_0$
is small and negative \cite{Bradley1995,Bradley1997}; here the
number of condensate atoms was limited through collapse to only a
few hundred. Later experiments \cite{Strecker2002,Khaykovich2002} on
optically trapped $^{7}$Li in the internal state $f=1, m=1$
exploited the 737-G Feshbach resonance to tune the scattering length from its very small
background value ($a_{\rm bg} \approx +5\,a_0$) to sufficiently
large positive values, typically in a range between $+40\,a_0$ and
$+200\,a_0$. Evaporative cooling then resulted in condensates with
up to $3\times10^5$ atoms. { More recently, \cite{Gross2008} exploited Feshbach tuning in the state $f=1, m=0$ for the all-optical production of a BEC. They obtained favorable conditions for efficient evaporative cooling at 866\,G, where the scattering length is about $+300\,a_0$.}

Bose-Einstein condensation of $^{85}$Rb was achieved by evaporative cooling in a magnetic trap \cite{Cornish2000} exploiting the broad resonance at 155\,G
(Sec.~\ref{sssec:Rb}) in the state $f=2, m=-2$ to tune the scattering length to positive values. The large negative background scattering
length $a_{\rm bg} = -443\,a_0$ would limit the number of condensate atoms to less than one hundred. Two stages of cooling were performed. The first stage used a magnetic field of 250\,G, where the scattering length $a$ is close to its background value; the second was close to the resonance at 162.3\,G, where $a = +170\,a_0$. This procedure optimized the ratio of elastic to inelastic collision rates \cite{Roberts2000} for the temperatures occurring during these stages.

Feshbach tuning played a crucial role for the attainment of BEC in
$^{133}$Cs \cite{Weber2003,Kraemer2004}. The condensate was produced
in an optical trap in the state $f=3, m=3$. In this lowest internal
state, two-body decay is energetically forbidden, and the scattering
length $a(B)$ shows a large variation at low magnetic fields
(Fig.~\ref{fig:cs_aa_slength}), which is due to a broad Feshbach
resonance at $-12$\,G. In a first evaporative cooling stage a
shallow large-volume optical trap was employed, and a large
scattering length of $a=+1200\,a_0$ at $B=73$\,G provided a
sufficiently large elastic collisions rate at rather low atomic
densities. The second cooling stage employed a much tighter trap.
Here, at much higher densities, an optimum magnetic field of $21$\,G
was found, where $a=+210\,a_0$. Highly efficient evaporation led to
the attainment of BEC. Later experiments revealed a minimum of
three-body decay in this magnetic field region \cite{Kraemer2006}; see also Sec.~\ref{sssec:efimov_Cs}.

For the attainment of BEC in $^{39}$K, \cite{Roati2007} employed
a mixture with $^{87}$Rb atoms with both species being in their internal ground states. First, an interspecies Feshbach resonance near 318\,G was used to optimize sympathetic cooling; the interspecies scattering length was tuned to $+150\,a_0$ by choosing a magnetic field of 316\,G. Then, after the removal of the Rb atoms, final evaporative cooling towards BEC was performed near the 402-G resonance of $^{39}$K (Sec.~\ref{sssec:K}) with the scattering length tuned to a positive value of $+180\,a_0$.

\subsubsection{Condensate mean field}
\label{sssec:meanfield}

Trapped atomic BECs in the dilute-gas regime are commonly described
\cite{Dalfovo1999} by the Gross-Pitaevskii equation for the
condensate wave function $\Phi$,
\begin{equation}
i \hbar { \partial \over \partial t} \Phi =
 \left( - { \hbar^2 \nabla^2 \over 2m } + V_{\rm ext}\,
+V_{\rm mf} \right) \Phi \, , \label{eq:GPequation}
\end{equation}
where $V_{\rm ext}$ is the external trapping potential. Interactions are taken into account by the mean-field
potential
\begin{equation}
V_{\rm mf} = { 4 \pi \hbar^2 a \over m}\, n\, , \label{eq:vint}
\end{equation}
where the atomic number density $n$ is related to $\Phi$ by
$n=|\Phi|^2$. This mean field enters the Gross-Pitaevskii equation as a nonlinearity and leads to many interesting phenomena.

In the Thomas-Fermi regime of large condensates with $a>0$ one can
neglect the kinetic energy term and obtain the equilibrium density
distribution of a BEC
\begin{equation}
n = {m \over 4 \pi \hbar^2 a} \left( \mu - V_{\rm ext}\right) \, ,
\end{equation}
which applies for $\mu > V_{\rm ext}$; otherwise $n=0$. For a given
particle number $N$, the chemical potential $\mu$ follows from the
normalization condition $N = \int n d^3r$.

\begin{figure}
\includegraphics[width=3.2in]{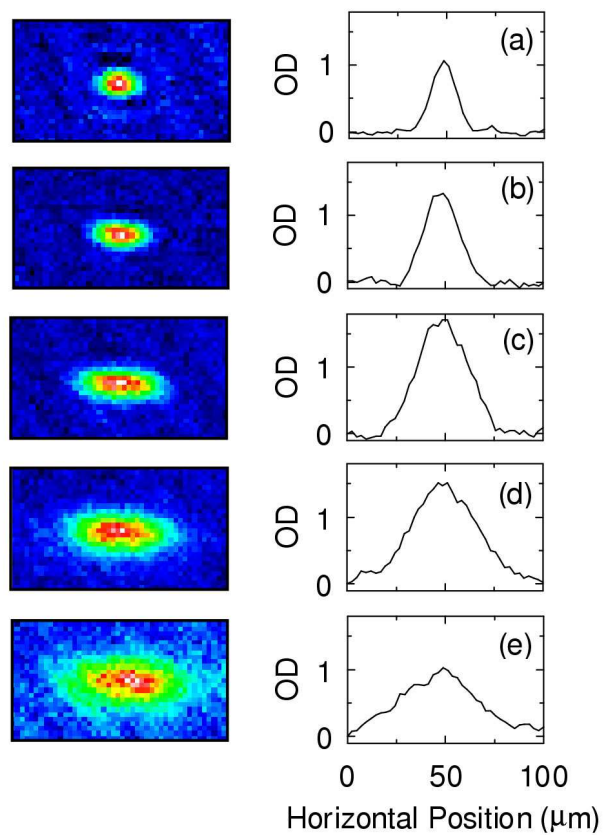}
\caption{In-situ images and density profiles showing the variation
of the equilibrium size of a magnetically trapped $^{85}$Rb BEC
close to a Feshbach resonance. With decreasing magnetic field (a-e), the scattering length increases from a very small positive value at $165.2$\,G (zero crossing at 165.8\,G) to a very large value of $a \approx +6000\,a_0$ at $156.4$\,G (resonance position at 155.0\,G). From \cite{Cornish2000}.} \label{fig:rb85size}
\end{figure}

In many cases of practical relevance, stable condensates with
positive $a$ are confined in harmonic traps and are in
the Thomas-Fermi regime. The condensate is then characterized by the
Thomas-Fermi radius $r_{\rm TF}$, given as the radius at which the external trapping potential equals the chemical potential, and the peak number density
$n_0$. These two quantities follow the scaling laws
\begin{equation}
%\mu \propto a^{2/5}, \,\,\,
r_{\rm TF} \propto a^{1/5}, \,\,\, n_0 \propto a^{-3/5}.
\end{equation}
Figure \ref{fig:rb85size} illustrates how the size of a trapped
$^{85}$Rb condensate increases when the Feshbach resonance at 155\,G
is approached. The {\it in-situ} measurements were used to
experimentally determine $a(B)$ \cite{Cornish2000}. The results are
in good agreement with later measurements of the molecular binding
energy, which allowed for a more precise determination of the
scattering properties near the Feshbach resonance
\cite{Claussen2003}.

The mean-field approach is valid for scattering lengths which are
small compared to the typical interparticle distance. The prospect
to observe beyond-mean-field effects in BECs
\cite{Dalfovo1999} has been an important motivation for experiments
near Feshbach resonances at large $a$. In atomic Bose gases, however, the
fast inelastic decay makes it very difficult to observe such phenomena.  \cite{Papp2008} finally demonstrated beyond-mean-field behavior by Bragg spectroscopy on a $^{85}$Rb BEC. In molecular BECs created in atomic
Fermi gases (Sec.~\ref{sssec:molbec}), the collisional stability facilitates the observation of beyond-mean-field behavior by simpler means. For example, \cite{Altmeyer2007} observed beyond-mean-field behavior by studying collective oscillations of a $^6$Li$_2$ molecular BEC.

\subsubsection{Controlled collapse and bright solitons}
\label{sssec:solitons}

For negative scattering lengths the condensate mean field is
attractive. The resulting nonlinearity can then lead to a condensate
collapse and to the formation of bright matter-wave solitons. To study phenomena of this kind by Feshbach tuning, the general experimental
strategy is to first produce a stable BEC at positive $a$. Then, the
attractive interaction is introduced by changing $a$ to negative
values. Experiments of this class have been performed with $^{85}$Rb
\cite{Roberts2001,Donley2001,Cornish2006}, $^7$Li
\cite{Khaykovich2002,Strecker2002}, and $^{133}$Cs
\cite{Weber2003,Rychtarik2004}.

Exploiting the 155-G Feshbach resonance in $^{85}$Rb,
\cite{Roberts2001} investigated the stability of a BEC with
attractive interactions. They first produced the condensate in a
magnetic trap at a moderate positive scattering length. They then
slowly changed the atom-atom interaction from repulsion to
attraction by ramping the magnetic field into the region of negative
scattering length. With increasing attractive interaction they
observed an abrupt transition in which atoms were ejected from the
condensate. These measurements of the onset of condensate collapse
provided a quantitative test of the stability criterion for a BEC
with attractive interactions.

\begin{figure}
\includegraphics[width=2.8in]{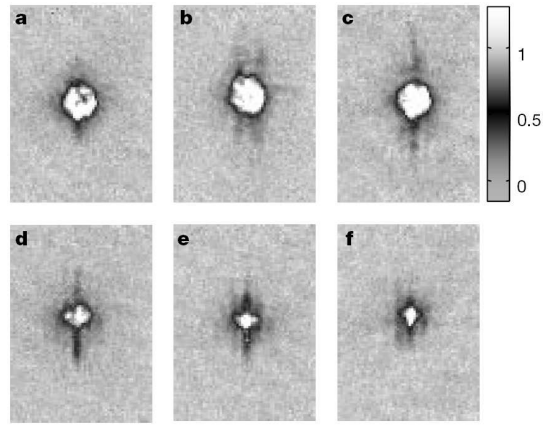}
\caption{Striking phenomena have been observed in the controlled
collapse of a $^{85}$Rb BEC. The images show the formation of
`jets', where streams of atoms with highly anisotropic velocities
are ejected by the collapsing condensate. Reprinted by permission from Macmillan Publishers Ltd: Nature \cite{Donley2001}, copyright 2001.}
\label{fig:bosenova}
\end{figure}

The controlled collapse in $^{85}$Rb following a sudden change of
$a$ led to the spectacular observation of a ``Bosenova,'' a
condensate implosion with fascinating and unexpected properties
\cite{Donley2001}. An anisotropic burst of atoms was observed that
exploded from the condensate during the early stage of collapse
(Fig.~\ref{fig:bosenova}), leaving behind a highly excited
long-lived remnant condensate. Strikingly, the number of atoms in
the remnant BEC was significantly larger than the critical number
for a collapse. The surprising fact that the remnant BEC did not
undergo further collapse was later explained by its fragmentation
into bright solitons \cite{Cornish2006}.

Condensate collapse experiments were also used to detect the
presence of a small BEC of Cs atoms in an optical surface trap
\cite{Rychtarik2004}. While a thermal gas did not show loss when the
scattering length was suddenly switched to negative values, the
condensate showed up in the sudden onset of collapse-induced loss.

\begin{figure}
\includegraphics[width=2.8in]{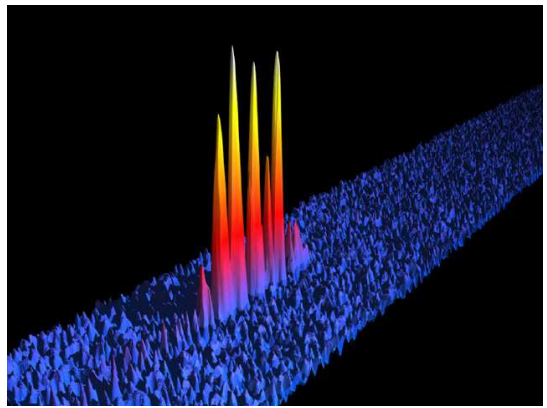} \caption{A train of
matter-wave solitons created from an optically trapped BEC of $^7$Li
atoms \cite{Strecker2002}. The individual solitons contain up to about 5000 atoms. Figure courtesy of Randall Hulet.} \label{fig:solitontrain}
\end{figure}

Bright solitons were observed in experiments on $^7$Li atoms near
the broad 736-G Feshbach resonance. \cite{Khaykovich2002} produced a BEC by evaporative cooling at $a\approx40\,a_0$ in an optical trap. They then released the BEC into a one-dimensional optical waveguide and studied the propagation of the resulting matter-wave packet for an ideal gas ($a=0$) and a gas with a small attractive mean-field interaction ($a=-4\,a_0$). In the latter case, they observed the dispersion-free propagation that is characteristic for
a soliton. In a similar experiment \cite{Strecker2002} created a
train of solitons, see Fig.~\ref{fig:solitontrain}, from an
optically trapped $^7$Li BEC by abruptly switching the scattering
length from $200\,a_0$ to $-3\,a_0$. They also observed the propagation
of the solitons in the trap and their mutual repulsion.
These spectacular experiments on bright solitons
highlight the analogy between bright matter-wave solitons and
optical solitons in fibers and thus the intimate connection between
atom optics with BECs and light optics.

\subsubsection{Non-interacting condensates}
\label{sssec:nonint}

The {\em zero crossing} of the scattering length near
a Feshbach resonance can be used to realize non-interacting ideal-gas condensates.
BECs of $^7$Li, $^{39}$K, $^{85}$Rb, and $^{133}$Cs are good candidates to reduce $|a|$ to very small values on the order of $a_0$ or smaller.

\begin{figure}
\includegraphics[width=2.5in]{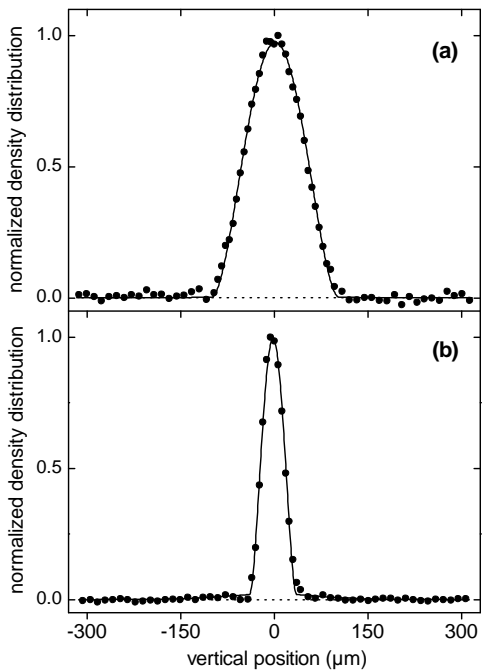} \caption{Density
profiles of a Cs BEC after 100\,ms of expansion at (a) $a=210\,a_0$
and (b) $a=0$. The expansion energy of the non-interacting
condensate is as low as $k_B \times 50$\,pK. From
\cite{Kraemer2004}.} \label{fig:nonintBEC}
\end{figure}

To explore non-interacting condensates with $^{133}$Cs
\cite{Weber2003,Kraemer2004} exploited the zero crossing near 17\,G
and studied expansion after release from the trap. The scattering
length was abruptly changed to zero when the
condensate was released from the trap. An extremely slow expansion
was observed with a release energy as low as $k_B \times
50$\,pK; see Fig.~\ref{fig:nonintBEC}. The surprising observation that the release energy is a factor of five below the kinetic energy associated with the motional ground state of the trap is explained by the fact that the initial size of the
expanding matter-wave packet is determined by the repulsive
condensate self-interaction before release, which is larger than the
bare ground state of the trap. The momentum spread is thus
significantly smaller. In contrast, a slow change of $a$ to zero
before release would have ideally resulted in a wavepacket with position and
momentum spread corresponding to the bare ground state.

Besides the small equilibrium size of the condensate, a vanishing
scattering length has profound consequences for the collective
behavior of a BEC. The sound velocity ($\propto na^{1/2}$) is
vanishingly small so that all excitations will become particle-like,
and not phonon-like. Moreover, the healing length ($\propto
na^{-1/2}$) becomes very large, which may be applied in experiments
on rotating condensates \cite{Madison2000} to increase the core size
of vortices.

Non-interacting condensates are promising for the observation that are masked by interacting effects, as e.g.\ phenomena on a lower energy scale.
In atom interferometry the mean-field interaction of a condensate is a substantial systematic error source, as \cite{Gupta2002} observed in the context of photon recoil
measurements. \cite{Roati2004} have studied Bloch oscillations in an
optical lattice under the influence of gravity. They showed that,
for interacting bosons, the oscillations lost contrast much faster
than for identical fermions without $s$-wave interaction. A
non-interacting BEC combines the advantages of an ultralow momentum
spread with very long observation times. Two recent experiments have
reported on long-lived Bloch oscillations with BECs of $^{133}$Cs
\cite{Gustavsson2007} and $^{39}$K \cite{Fattori2007}, which is an
important advance towards high-precision atom interferometry. This
could, for example, open up new possibilities for precision
measurements of gravitational effects \cite{Carusotto2005}.

Another intriguing application of the zero crossing of a Feshbach
resonance was demonstrated by \cite{Lahaye2007} with a BEC of
$^{52}$Cr atoms. This species exhibits a very large magnetic
dipole-dipole interaction because of its magnetic moment of
6\,$\mu_B$. When the isotropic contact interaction is reduced by
tuning the scattering length close to zero, the magnetic dipole interaction dominates. In $^{52}$Cr this was achieved near the 589-G Feshbach resonance
(Fig.~\ref{fig:Cr}). The resulting dipolar quantum gas represents a
model system for a ``quantum ferrofluid'', the anisotropic
properties of which have been attracting considerable interest \cite{Menotti2007}. In further work on $^{52}$Cr BEC near the zero of the scattering length, \cite{Lahaye2008} investigated the controlled collapse of the system and demonstrated its complex dynamics and \cite{Koch2008} studied the stability of the dipolar condensate depending on the trap geometry. { The effect of the magnetic dipole interaction has also been observed in non-interacting condensates made of alkali atoms with magnetic moments of the order of $1 \mu_B$, for $^{39}$K by \cite{Fattori2008} and for $^7$Li by \cite{Pollack2008}. The recent observation of Anderson localization of matter waves in a disordered optical potential by \cite{Roati2008} represents a further exiting application of a non-interacting condensate.}

\subsection{Degenerate Fermi gases}\label{ssec:fermi}

In experiments on ultracold Fermi gases \cite{Varenna2006}, Feshbach
resonances { serve as a} key to explore many-body physics in the strongly interacting regime \cite{Bloch2007}. This regime is realized when the scattering length exceeds the interparticle spacing and connects the field of ultracold atoms to fundamental questions in various fields of physics, like high T$_c$-superconductors, nuclear matter, neutron stars, and the quark-gluon plasma. The first Feshbach resonance in a Fermi gas was observed by  \cite{Loftus2002}. \cite{Ohara2002a} produced the first strongly
interacting Fermi gas. Since then the research field has undergone rapid developments with many exciting achievements \cite{Varenna2006}.

{ The decay properties of ultracold Fermi gases are strongly influenced by Pauli's exclusion principle \cite{Esry2001, Suno2003, Petrov2003, Petrov2004a}. Three-body recombination processes in one- and two-component Fermi gases necessarily involve identical particles. This generally leads to a suppression of loss as compared to Bose gases or systems with three nonidentical particles. The majority of recent experiments on Fermi gases \cite{Varenna2006} has focussed on two-component spin mixtures of $^6$Li or $^{40}$K  with resonant $s$-wave interactions, realized near broad, entrance-channel dominated resonances. In such systems, it is possible to realize a resonant $s$-wave interaction ($a\rightarrow \pm \infty$) practically without any decay. Nevertheless, these resonances are accompanied by subtle loss features \cite{Dieckmann2002}, which do not appear at the resonance center, but at side where the scattering length is positive. In contrast to the remarkable stability near this special $s$-wave scenario, three-body collisions near a $p$-wave Feshbach resonance usually lead to significant loss \cite{Regal2003, Zhang2004, Schunck2005}.}

Here, as prominent examples for the application of Feshbach tuning, we review the attainment of BEC of molecules (Sec.~\ref{sssec:molbec}) and studies of the BEC-BCS crossover and the observation of fermion superfluidity (Sec.~\ref{sssec:crossover}).

\subsubsection{BEC of molecules}\label{sssec:molbec}

Bose-Einstein condensation of molecules were created in atomic Fermi
gases of $^6$Li \cite{Jochim2003a,Zwierlein2003, Bourdel2004} and
$^{40}$K \cite{Greiner2003}. The molecules are very weakly bound
dimers at the side of an entrance-channel dominated $s$-wave
Feshbach resonance where the scattering length is positive and very
large. These dimers are formed in a halo state
(Sec.~\ref{sssec:halo}), which is stable against inelastic decay in
atom-dimer and dimer-dimer collisions \cite{Petrov2004a}. This
stability originates from basically the same Pauli suppression
effect that also affects three-body decay in an atomic Fermi gas.
{ Such weakly bound molecules can be detected by converting
them back to atoms or by direct absorption imaging;
see Sec.~\ref{sssec:mol_diss}.}

In a spin mixture of $^6$Li in the lowest two internal states, the route to molecular BEC is particularly simple \cite{Jochim2003a}. Evaporative cooling towards BEC can be performed in an optical dipole trap at a constant magnetic field of about 764\,G near the broad 834-G Feshbach resonance; here $a=+4500a_0$ and $E_b = k_B \times 1.5\,\mu$K. In the initial stage
of evaporative cooling the gas is purely atomic and $a$ is the
relevant scattering length for elastic collisions
between the atoms in different spin states. With decreasing
temperature the atom-molecule equilibrium (Sec.~\ref{sssec:mol_atmoleq}) favors the formation of molecules and, in the final evaporation stage, a purely molecular
sample is cooled down to BEC.  The large atom-dimer and dimer-dimer scattering lengths of $1.2\,a$ and $0.6\,a$ along with strongly suppressed loss (Sec.~\ref{sssec:mol_coll}) facilitate an efficient evaporation process. In this way, molecular BECs are achieved with a
condensate fraction exceeding 90\%.

\begin{figure}
\includegraphics[width=3.2in]{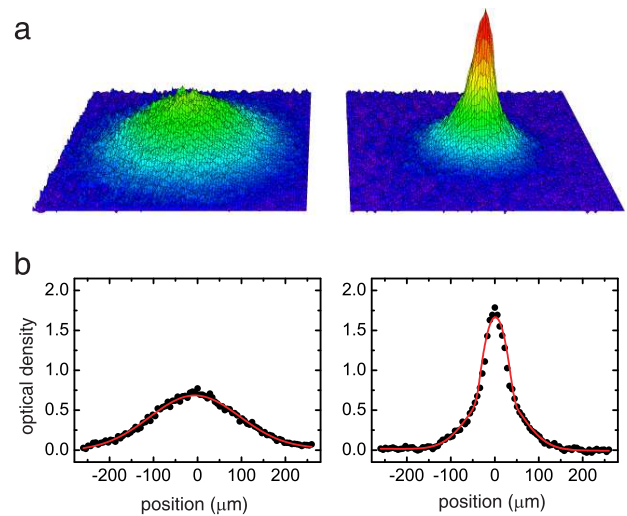}
\caption{Emergence of a molecular BEC in an ultracold Fermi gas of
$^{40}$K atoms, observed in time-of-flight absorption images. The
density distribution on the left-hand side (upper graph, 2D surface
plot; lower graph, 1D cross section) was taken for a weakly interacting Fermi gas which was cooled down to 19\% of the Fermi temperature. After ramping
across the Feshbach resonance no BEC was observed as the sample was
too hot. The density distribution on the right-hand side was
observed for a colder sample at 6\% of the Fermi temperature. Here
the ramp across the Feshbach resonance resulted in a bimodal
distribution, revealing the presence of a molecular BEC with a
condensate fraction of 12\%. Reprinted by permission from Macmillan Publishers Ltd: Nature \cite{Greiner2003}, copyright 2003.}
\label{fig:mbec}
\end{figure}

The experiments in $^{40}$K followed a different approach to achieve
molecular BEC \cite{Greiner2003}. For $^{40}$K the weakly bound
dimers are less stable because of less favorable short-range
three-body interaction properties. Therefore the sample is first
cooled above the 202-G Feshbach resonance, where $a$ is large and negative,
to achieve a deeply degenerate atomic Fermi gas.
A sweep across the Feshbach resonance
then converts the sample into a partially condensed cloud of
molecules. Figure~\ref{fig:mbec} demonstrates the emergence of the
molecular BEC in $^{40}$K.

The molecular BEC can be described in the mean-field approach
outlined in Sec.\ \ref{sssec:meanfield} by simply replacing the
atomic with the molecular mass ($m \rightarrow 2m$), and the atomic
with the molecular scattering length ($a \rightarrow 0.6a$). The
mean field of the molecular condensate was experimentally studied in
\cite{Bartenstein2004,Bourdel2004}. However, because of the large
scattering length, molecular BECs show considerable
beyond-mean-field effects \cite{Altmeyer2007}.

\subsubsection{BEC-BCS crossover and fermion superfluidity}
\label{sssec:crossover}

At a Feshbach resonance in a two-component Fermi
gas, different regimes of fermion pairing and superfluidity can be
experimentally realized. Pairing on the side with $a>0$ can be
understood in terms of molecule formation, and superfluidity results
from molecular Bose-Einstein condensation. On the other side of the
resonance ($a<0$), pairing is a many-body effect and the ground
state of the system at zero temperature is a fermionic superfluid.
In the limit of weak interactions, this regime can be understood in
the framework of the well-established Bardeen-Cooper-Schrieffer
(BCS) theory, developed in the 1950s to describe superconductivity.
Both limits, BEC and BCS, are smoothly connected by a crossover
through a regime where the gas is strongly interacting. This BEC-BCS
crossover has attracted considerable attention in many-body quantum physics for more than two decades, { and has recently reviewed in \cite{Chen2005, Varenna2006, Giorgini2007}}. A theoretical
description of this challenging problem is very difficult and
various approaches have been developed. With tunable Fermi gases,
a unique testing ground has become available to quantitatively
investigate the crossover problem.

The interaction regime can be characterized by a dimensionless
parameter $1/(k_F a)$, where $k_F$ is the Fermi wave number of a
non-interacting gas, related to the Fermi energy by $E_F = \hbar^2
k_F^2/(2m)$. For $1/(k_Fa) \gg 1$, the molecular BEC regime is realized.
For $1/(k_Fa) \ll -1$, the system is in the BCS regime. In between
($1/|k_Fa| \lesssim 1$), the Fermi gas is strongly interacting. In
the experiments, $E_F/k_B = 1\,\mu$K gives a typical value
for the Fermi energy and $1/k_F \approx 4000\,a_0$ sets the typical
length scale. The realization of a strongly interacting gas thus
requires $|a| \gtrsim 4000\,a_0$, which for the particularly broad
$^6$Li resonance (Fig.~\ref{fig:Li_ab_834}) is obtained over a more
than 100-G wide magnetic field range.

A particularly interesting situation is the exact resonance case, where $1/(k_Fa)=0$. Here, $a$ is no longer a relevant quantity to describe the problem and scattering is only limited by unitarity (Sec.~\ref{sssec:collrate}). Consequently, $k_F^{-1}$ and $E_F$ define the relevant scales for length and
energy, and the Fermi gas acquires universal properties
\cite{Varenna2006,Giorgini2007}. For example, the size and shape of
a harmonically trapped ``universal Fermi gas'' can be obtained just as a rescaled version of a
non-interacting Fermi gas.

Experimentally, various properties of strongly interacting Fermi
gases have been explored in the BEC-BCS crossover
\cite{Varenna2006}. All these experiments have been performed on
two-component spin mixtures of $^6$Li near the 834-G resonance or of
$^{40}$K near the 202-G resonance. In both cases, the resonances
have entrance-channel dominated character, where the two-body
interaction can be modeled in terms of a single scattering channel
and universality applies (Sec.~\ref{sssec:resstrength}).
This condition is particularly well fulfilled for the $^6$Li
resonance, where the resonance is exceptionally strong.

\cite{Regal2004} introduced fast magnetic-field sweeps to observe the condensed fraction of pairs in the crossover. Starting with an ultracold $^{40}$K Fermi gas in the
strongly interacting regime, they performed {\em fast} Feshbach
ramps into the BEC regime. The ramps were fast compared to the time
scale of establishing a thermal atom-molecule equilibrium by collisions; see Sec.~\ref{sssec:mol_atmoleq}. However, the Feshbach ramps
were slow enough to adiabatically convert fermion pairs formed in
the strongly interacting regime into molecules. After the ramp, the
observed molecular condensate reflected the fermion condensate
before the ramp. The fast-ramp method was applied by
\cite{Zwierlein2004} to observe fermion condensates in $^6$Li.

For $^6$Li \cite{Bartenstein2004} showed that {\it slow} Feshbach
ramps allow conversion of the gas in a reversible way from the molecular
BEC to the BCS regime.
Here the gas adiabatically follows and stays in thermal equilibrium. They also observed in-situ profiles of the trapped, strongly
interacting gas and measured its changing
size for variable interaction strength.

Collective modes in the BEC-BCS crossover were studied in $^6$Li
gases. \cite{Kinast2004} reported on ultraslow damping in a universal
Fermi gas with resonant interactions, providing evidence for
superfluidity. \cite{Bartenstein2004a} measured how the frequencies
of collective modes in the crossover changed with variable
interaction parameter $1/(k_Fa)$. They also observed a breakdown of
hydrodynamic behavior on the BCS side of the resonance, which marks
a transition from the superfluid to the normal phase. Precision
measurements of collective modes also revealed beyond-mean-field
effects in the molecular BEC regime \cite{Altmeyer2007}.

\begin{figure}
\includegraphics[scale=0.35, bb=50 179 712 675]{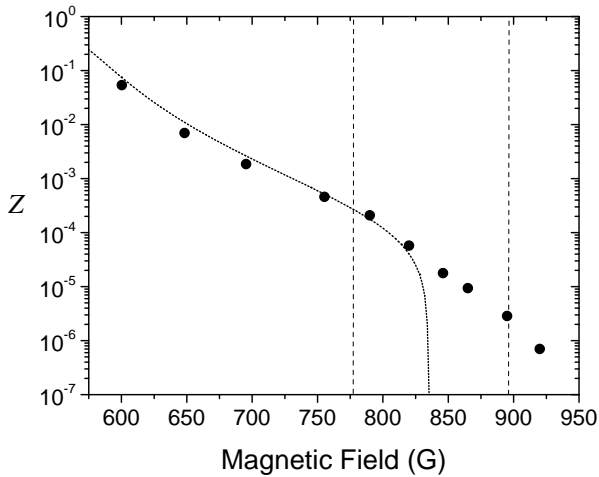}
\caption{Measurements of the closed-channel fraction $Z$ for pairs in the BEC-BCS crossover. The experiment uses the 834-G resonance in $^6$Li and a photoassociation probe that only couples to closed-channel singlet component. The vertical dashed lines indicate the boundaries of the strongly interacting regime, $k_F|a| > 1$. The dotted line shows the result of a coupled channels calculation for molecules; see Sec.~\ref{sssec:AnalMol} and Fig.~\ref{fig:E&Z}. Comparison with the experimental data shows that two-body physics describes the situation well up to close to the 834-G resonance. For higher fields, $Z$ shows strong many-body effects. Above the resonance, where two-body pairs cannot exist, the many-body system shows the closed-channel admixture of the many-body pairs.
From \cite{Partridge2005}.} \label{fig:molprobehulet}
\end{figure}

\cite{Chin2004a} performed spectroscopy on fermion pairs by
using a radio-frequency technique. They measured the binding energy of the pairs in the crossover. They showed how an effective pairing gap continuously evolved from the molecular regime, where it simply reflects the dimer
binding energy, into a many-body regime of pairing; see
also more recent work by \cite{Schunck2008}.

\cite{Partridge2005} used an optical molecular-probe technique based on photoassociation to measure the closed-channel contribution
of pairs in the crossover. Their results, displayed in Fig.~\ref{fig:molprobehulet}, show that this fraction is very small in the strongly interacting regime. These observations strongly support single-channel descriptions for the crossover along with the concept of universality. The results also demonstrate that fermionic pairing reaches from the strongly interacting regime well into the weakly interacting BCS regime.

Superfluidity of a $^6$Li Fermi gas in the BEC-BCS crossover was
observed by \cite{Zwierlein2005}. They produced a rotating Fermi gas
and observed the formation of vortex arrays. Here Feshbach tuning was applied not only
to explore different regimes in the crossover, but also to increase
the vortex cores in the expanding Fermi gas after release from the
trap; the latter was essential to observe the vortices by optical
imaging.

Currently, there is considerable interest in exploring novel regimes of pairing and superfluidity. \cite{Zwierlein2006} and \cite{Partridge2006} performed
experiments with unbalanced mixtures of two spin states, i.e.\ polarized Fermi gases. This led to deeper insight into phenomena like phase separation \cite{Partridge2006,Shin2008}. Ultracold Fermi-Fermi mixtures of different species, like $^6$Li and $^{40}$K, have recently become available \cite{Taglieber2008,Wille2007,Voigt2008}, adding the mass ratio and independent control of the trapping potentials for both components to our tool box to explore the broad physics of strongly interacting fermions.

\section{Ultracold Feshbach molecules}
\label{sec:molecules}

Cold molecules are at the center of a rapidly developing research
field \cite{Polar2004,Hutson2006a}, offering many new opportunities
for cold chemistry, precision measurements,
many-body physics, and
quantum information. The coldest
attainable molecules, at temperatures in the nanokelvin range, are
created by Feshbach association in ultracold atomic gases. Here
Feshbach resonances serve as the experimental key to couple pairs of
colliding atoms into molecules.

In 2002, \cite{Donley2002} observed coherent oscillations between
atom pairs and Feshbach molecules in a BEC of $^{85}$Rb atoms. The
oscillation frequency reflected the small binding energy of the
dimer in a weakly bound state and provided indirect evidence for the
creation of molecules. In 2003, several groups reported on more
direct observations of Feshbach molecules in Fermi gases of $^{40}$K
\cite{Regal2003a} and $^6$Li
\cite{Strecker2003,Cubizolles2003,Jochim2003} and BECs of Cs
\cite{Herbig2003}, $^{87}$Rb \cite{Durr2004a}, and Na \cite{Xu2003}.
This rapid development culminated in the attainment of molecular BEC
at the end of 2003 (Sec.~\ref{sssec:molbec}), and has paved the
way for numerous applications.

A comprehensive review of Feshbach molecules and their theoretical
background has been given by \cite{Kohler2006}. We do not consider
optical methods of making molecules, since
this was recently reviewed by \cite{Jones2006}. Here we discuss various formation methods based on magnetic Feshbach resonances (Sec.~\ref{ssec:mol_formation}) and the main properties of Feshbach molecules (Sec.~\ref{ssec:mol_prop}).

\subsection{Formation}
\label{ssec:mol_formation}

Various schemes to create ultracold molecules near Feshbach
resonances have been developed in the last few years, most of them
relying on the application of time-varying magnetic fields.
Section~\ref{sssec:mol_timedep} describes the use of magnetic field
ramps while Sec.~\ref{sssec:mol_oscillatory} discusses the application of
oscillatory fields. These schemes have been applied to a variety of
bosonic and fermionic atom gases.  Section~\ref{sssec:mol_atmoleq}
describes the formation of collisionally stable molecules of
fermionic atoms through three-body recombination and thermalization.

Here we restrict our discussion to ultracold gases
confined in macroscopic traps. The microscopic trapping sites of an
optical lattice, where atom pairs can be tightly confined,
constitute a special environment for molecule formation. This will
be reviewed separately in Sec.~\ref{ssec:ola}.

\begin{figure}
\includegraphics[width=2.2in]{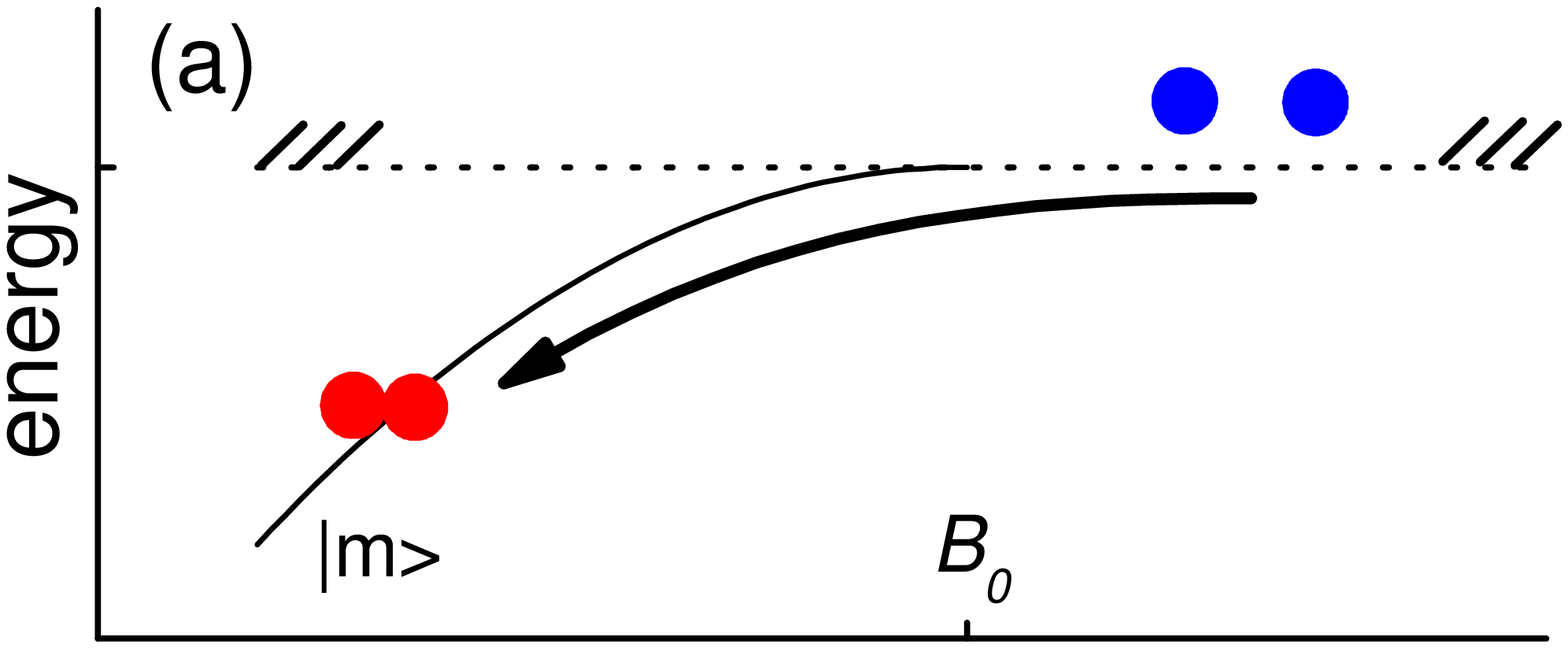}
\includegraphics[width=2.22in]{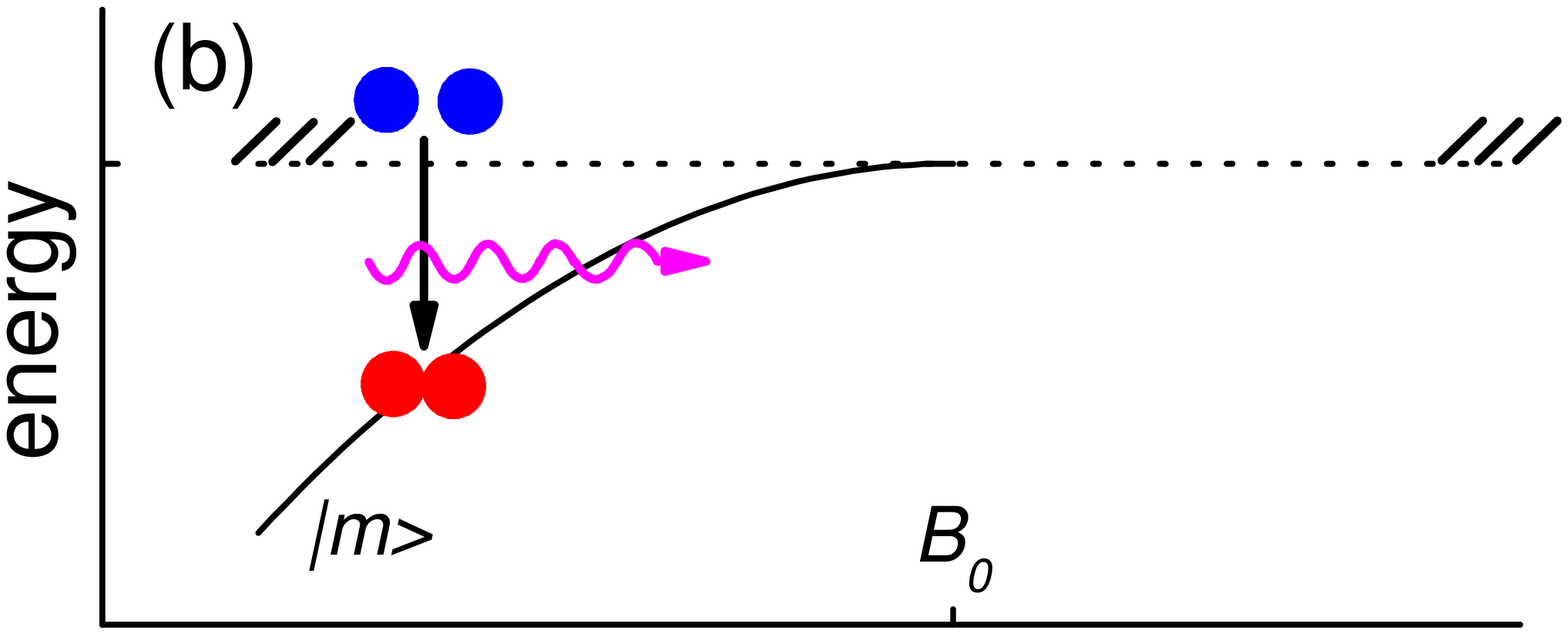}
\includegraphics[width=2.2in]{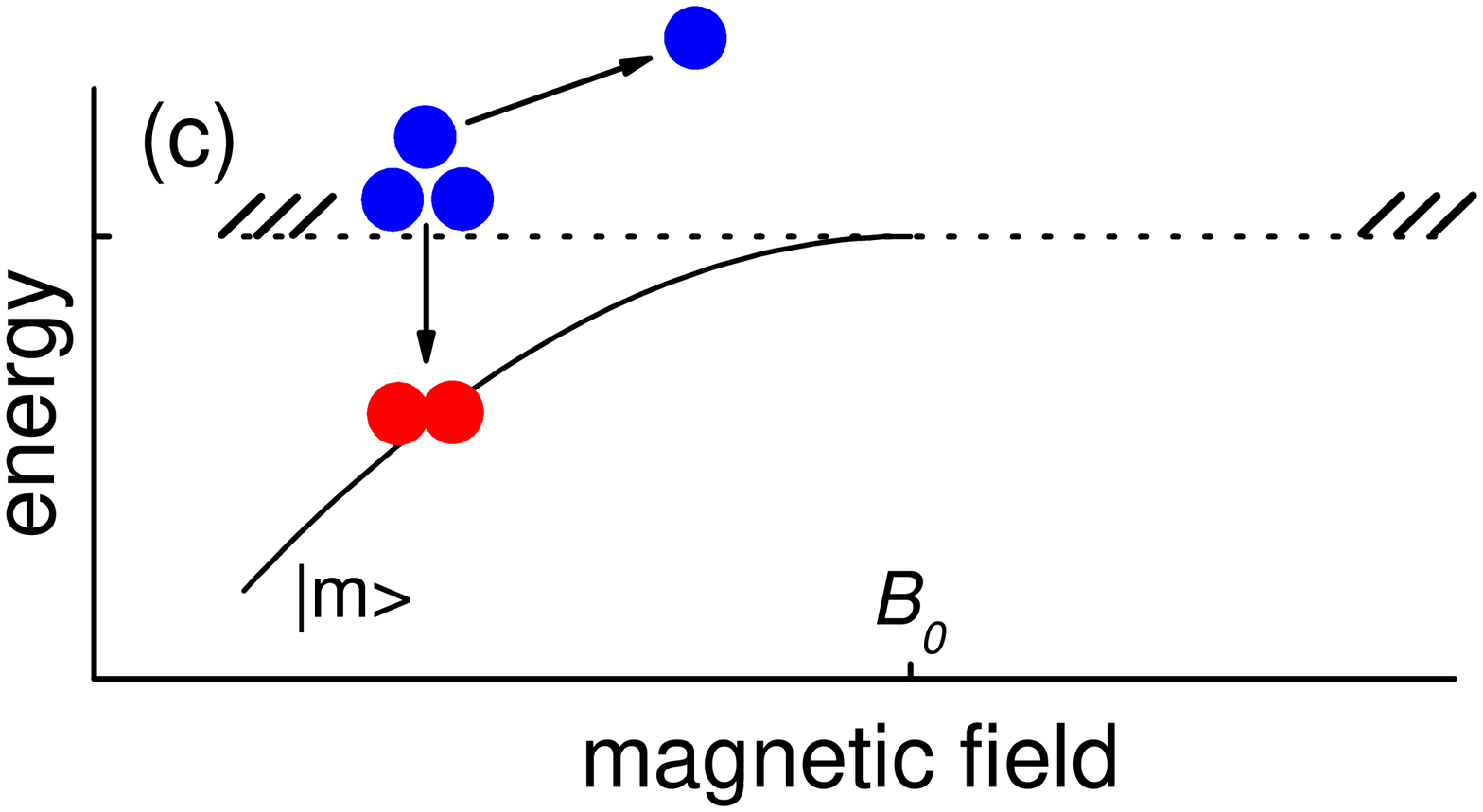}
\caption{Illustration of experimental schemes to create ultracold
molecules. The solid line marks the weakly bound molecular state
$|m\rangle$, which dissociates into the continuum (indicated by the
dotted line) at resonance $B=B_0$. In (a), the magnetic field is ramped across the
resonance, which adiabatically converts two interacting atoms into
one molecule; in (b), an oscillatory magnetic field drives the
transition from the scattering state to the molecular state; in (c),
three-body recombination results in molecule formation.} \label{mol_over}
\end{figure}

\subsubsection{Feshbach ramps}
\label{sssec:mol_timedep}

Ramping an external magnetic field across a Feshbach resonance is the most commonly adopted scheme to form Feshbach molecules. This scheme, usually referred to as a Feshbach ramp, was proposed by \cite{vanAbeelen1999b,Timmermans1999,Mies2000a}. In a simplified
picture, illustrated in Fig.~\ref{mol_over}(a), the resonant coupling between the scattering state and the molecular state opens up a way to adiabatically convert interacting atom pairs into molecules. The atomic gas is prepared at a field $B$ away from resonance where the two atoms do not have a weakly bound
state. In Fig.~\ref{mol_over}(a), this corresponds to $B>B_0$. The
field is then ramped to a final $B<B_0$ to make a Feshbach molecule.

\begin{figure}
\includegraphics[width=2.2in]{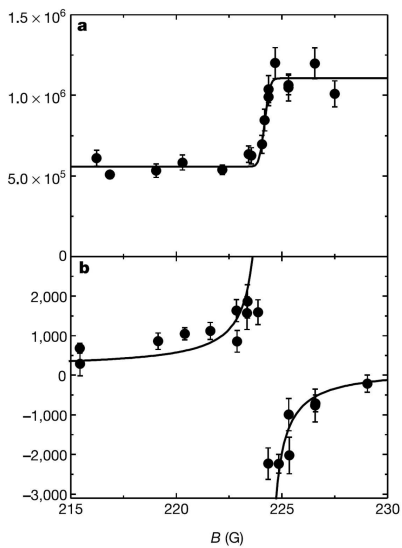}
\caption{Molecule formation in a $^{40}$K gas near the 224-G
Feshbach resonance. (a) The number of atoms measured after a ramp
across the resonance as a function of the final magnetic field shows
the disappearance of atoms having formed molecules. For comparison
(b) shows the magnetic-field dependence of the $s$-wave scattering
length in units of Bohr radius, as measured by radio-frequency
spectroscopy. From \cite{Regal2003a}.} \label{fig:jin_firstmol}
\end{figure}

\cite{Regal2003a} created ultracold molecules in a degenerate Fermi
gas of $^{40}$K, exploiting the resonance at 224\,G in the $ac$
channel. In the 10-G ramp across the resonance with a ramp speed of
25\,G/ms, about $50\%$ of the atoms were converted to molecules. The
experimental signatures, shown in Fig.~\ref{fig:jin_firstmol}(a),
were a disappearance of atoms when the field was ramped below 224~G
and a recovery of the atoms when the field was ramped back. Since the Feshbach bound state exists below 224~G,
their observation strongly suggests formation and dissociation of
Feshbach molecules below and above 224~G, respectively.  For the
202-G resonance in the $ab$ channel of $^{40}$K, \cite{Hodby2005}
reported conversion efficiencies of up to $80\%$. This resonance was
also used for the formation of a molecular BEC; see
Sec.~\ref{sssec:molbec}.

Feshbach ramps were also applied to degenerate Fermi gases of $^6$Li
in the lowest two spin states ($ab$ channel), using both the narrow
resonance at 543\,G \cite{Strecker2003} and the broad resonance at
834\,G \cite{Cubizolles2003}. These two resonances in $^6$Li are
extensively discussed in Sec.~\ref{sssec:Li6res}. Both experiments
revealed a remarkable collisional stability of the molecules, see
Sec.~\ref{sssec:mol_coll}.

In bosonic gases, an efficient atom-molecule conversion by a Feshbach
ramp has to overcome inelastic collision loss during the ramp
process; see Sec.~\ref{sssec:mol_coll}. In general, ramps applied to
bosonic atom samples do not only create weakly bound Feshbach molecules,
but they also lead to atom loss that cannot be recovered by a reverse
field ramp. These atoms are lost presumably to deeply bound
molecular states as a result of atom-molecule inelastic collisions.
To optimize the Feshbach molecule fraction, which corresponds to the recoverable fraction, it is crucial to optimize the ramp speed. Furthermore, a fast separation of the molecules from the remaining atoms is essential. The latter can, for example, be achieved by Stern-Gerlach or optical methods.

\cite{Herbig2003} created a pure sample of about 3000 $g$-wave
molecules from a BEC of $^{133}$Cs atoms by a Feshbach ramp across the
narrow resonance at 20\,G. The presence of an inhomogeneous magnetic
field during the ramp facilitated an immediate Stern-Gerlach
separation of the molecules from the atomic cloud. The Feshbach ramp
removed about 60\% of the atoms from the BEC, but only $20\%$ were found in the weakly bound molecular state and thus could be recovered by a reverse field ramp. Expansion measurements on the
Cs$_2$ cloud showed temperatures below 5\,nK and suggested
phase-space densities close to or exceeding unity. In later
experiments on $^{133}$Cs, \cite{Chin2005,Mark2007,Knoop2007} produced
Feshbach molecules in a variety of different internal states and
partial waves including $s$- to $l$-wave ($\ell=8$) states.

\begin{figure}
\begin{turn}{90}
\includegraphics[width=1.1in]{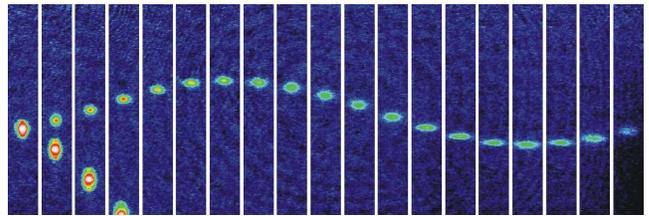}
\end{turn}
\caption{Motion of $^{87}$Rb$_2$ molecules in a magnetic field
gradient. Right after the Feshbach ramp, the molecules are separated
from the atoms because of the different magnetic moments. While the
atom cloud leaves the observation region (see first four images) the
molecules undergo an oscillatory motion, which is due to a changing magnetic moment caused by an avoiding level crossing in the molecular states.
The images correspond to steps of 1\,ms, and the field of view of each image is $0.24 \times 1.7$\,mm. From \cite{Durr2004a}.} \label{fig:rb2osci}
\end{figure}

\cite{Durr2004a} applied a similar scheme to a $^{87}$Rb BEC and
formed 7000 $s$-wave molecules near the 1007-G resonance with a 7\%
conversion efficiency. A magnetic gradient was applied right after
the Feshbach ramp to separate atoms and molecules. The experiments
showed remarkable oscillations of the molecular cloud in the
gradient field (Fig.~\ref{fig:rb2osci}), which appear as a result of changing
magnetic moment near an avoided crossing between two
molecular states. The creation of $^{87}$Rb Feshbach molecules has
led to fascinating applications in optical lattices; see
Sec.~\ref{sssec:ola_mol}.

\cite{Xu2003} demonstrated molecule formation with a 4\% conversion
efficiency in a large BEC of $^{23}$Na atoms by a Feshbach ramp
across the 907\,G resonance .   A large number
of more than $10^5$ Feshbach molecules was observed.
Expansion measurements on the Na$_2$ cloud showed that
it was in the quantum-degenerate regime. By diffraction off an optical standing wave, \cite{Abo2005} demonstrated the matter-wave coherence of Na$_2$ molecules.

\begin{figure}
\includegraphics{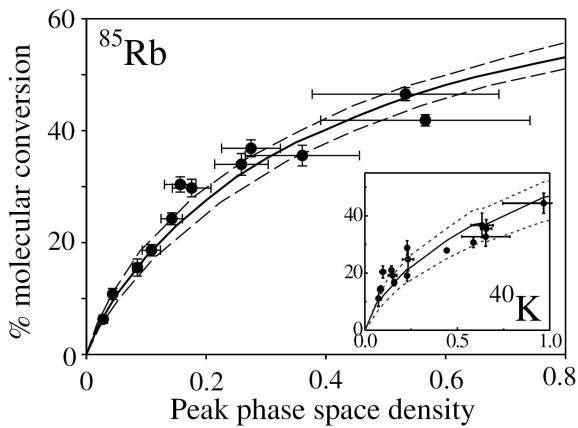} \,\,\,
\caption{Illustration of the dependence of the atom-molecule
conversion efficiency on the atomic phase-space density. The main
figure shows the results obtained in a near-degenerate bosonic gas
of $^{85}$Rb atoms, while the inset shows corresponding results on a
fermionic gas of $^{40}$K atoms. The lines refer to a theoretical
model that is based on the phase-space overlap of atom pairs in the
trapped gases. From \cite{Hodby2005}.} \label{fig:Rb2K2}
\end{figure}

\cite{Hodby2005} presented a model of the atom-molecule conversion
efficiency that is valid for both near-degenerate fermionic and
bosonic atoms. Figure~\ref{fig:Rb2K2} shows a comparison of this
model with data from {experiments}  on $^{40}$K and
$^{85}$Rb. The atom-molecule conversion efficiency follows a
Landau-Zener like behavior
\begin{equation}
\frac{P}{P_{\rm max}} = 1-\exp\left(-\alpha \, n \, \frac{\hbar}{m}
\left|\frac{\Delta \, a_{\rm bg}}{\dot{B}}\right|\right),
\label{eq:mol_LZ}
\end{equation}
where $P_{\rm max}$ is the maximum conversion efficiency, solely
determined by the phase-space density of the atomic cloud. This
expression reveals a simple dependence on the resonance parameters $\Delta$ and $a_{\rm bg}$, the atomic number density $n$, and
the ramp speed $\dot{B}$; the dimensionless prefactor $\alpha$ is
discussed in \cite{Kohler2006}.
%{ Various theoretical aspects of the two- and many-body physics of atom-molecule conversion are discussed in \cite{Kokkelmans2002, Borca2003, Andreev2004, Altman2005, Vonstecher2007}.}

Improved conversion techniques have been reported for atomic BECs.
For the case of the narrow 20-G $g$-wave resonance of $^{133}$Cs,
\cite{Mark2005} successfully converted $30\%$ of the atoms into
Feshbach molecules.  Their scheme relied on fast switching of the
magnetic field from 21\,G right on the resonance, followed by a hold
time of $\sim$10\,ms and a further switch to 18\,G. This scheme was found to be
superior to using an optimized linear ramp.

Feshbach ramps have also been applied to create $p$-wave molecules
in ultracold Fermi gases. Formation of $p$-wave molecules in a
$^6$Li gas was reported by \cite{Zhang2004} based on the 185-G
resonance in the $ab$ channel. { \cite{Fuchs2008} measured the binding energies of such $p$-wave Feshbach molecules in three channels ($aa$, $ab$, $bb$) using an oscillating magnetic field; see Sec.~\ref{ssec:rfspec}. \cite{Inada2008} studied the collisional properties of these molecules in all three channels.}
\cite{Gaebler2007} created $p$-wave molecules in a $^{40}$K gas by fast switching of the magnetic field to the $p$-wave Feshbach resonances at 198.4 and 198.8~G { (both in the $bb$ channel)} and studied the lifetimes of the molecules.

Similarly, Feshbach ramps have also been applied to ultracold atomic
mixtures to create heteronuclear molecules, { such as $^{40}$K$^{87}$Rb \cite{Ni2008} and $^{6}$Li$^{40}$K \cite{Voigt2008}}. An isotopic rubidium mixture was used to associate $^{85}$Rb$^{87}$Rb molecules by \cite{Papp2006}. A variety of other heteronuclear molecule systems,
are currently under investigation in different laboratories.
Feshbach ramps have become a standard approach to create ultracold
molecules and serve as a starting point to investigate the dynamics and the
interaction properties of Feshbach molecules.

\subsubsection{Oscillatory fields}
\label{sssec:mol_oscillatory}

Another powerful method to produce ultracold Feshbach molecules is
based on a modulation of the magnetic field strength
\cite{Thompson2005a,Hanna2007}.  The oscillating field induces a
stimulated transition of two colliding atoms into a bound molecular
state; see Fig.~\ref{mol_over}(b).  Heating and atom loss are
reduced since association occurs at a bias field $B$ away from the
resonance position $B_0$.  For a $^{85}$Rb BEC near the 155-G
resonance, \cite{Thompson2005a} report { on high conversion efficiencies
for molecules with binding energies on the order
of 10 kHz; the molecule formation was inferred from the observation of a resonant loss signal. \cite{Lange2008} used the same method to explore weakly bound molecular states of Cs atoms in the $aa$ channel} in an energy range of up to 300 kHz.

In a $^{40}$K spin mixture, $p$-wave molecules were produced with an
oscillating magnetic field near the 198-G resonance doublet
in the $bb$ channel \cite{Gaebler2007}. { The method was also applied to produce heteronuclear $s$-wave molecules in mixtures of the two Rb isotopes \cite{Papp2006} and of $^{41}$K and $^{87}$Rb atoms \cite{Weber2008}.}

In contrast to the magnetic-field modulation method, radio-frequency transitions in the range of tens of MHz that involve a change of spin channel can also be used to associate two atoms to make a Feshbach molecule.  This is the inverse of the dissociation process described in Sec.~\ref{ssec:rfspec}.    In this way \cite{Zirbel2008, Klempt2008} achieved association of $^{87}$Rb and $^{40}$K in a dipole trap. Similar association experiments in an optical lattice are described in Sec.~\ref{sssec:ola_mol}.

\subsubsection{Atom-molecule thermalization}
\label{sssec:mol_atmoleq}

A particular situation for molecule formation arises in a spin
mixture of $^6$Li near the 834-G Feshbach resonance. At the
low-field side of the resonance there is a broad field range, where
the $s$-wave scattering is large and positive. Here a weakly bound
state exists with a pronounced halo character. The molecular state
shows an extraordinary stability against inelastic decay, which
opened the way to efficiently create molecular BECs by
straightforward evaporative cooling at a constant magnetic field
near 764\,G \cite{Jochim2003a,Zwierlein2003}.

The formation of molecules in this region can be understood in terms
of a chemical atom-molecule equilibrium
\cite{Chin2004b,Kokkelmans2004}, where exoergic three-body
recombination events compete with endoergic two-body dissociation
processes. From a balance of these processes one can intuitively
understand that molecule formation is favored at low temperatures
and high number densities, i.e.\ at high phase-space densities.
Indeed, for a non-degenerate gas, the atom-molecule equilibrium
follows a simple relation \cite{Chin2004b}
\begin{equation}
\phi_{\rm mol} = \phi_{\rm at}^2 \exp\left(\frac{E_b}{k_{\rm
B}T}\right), \label{eq_equilibrium}
\end{equation}
where $\phi_{\rm mol}$ and $\phi_{\rm at}$ denote the molecular and
atomic phase-space densities, respectively. The Boltzmann factor, determined by the ratio of the molecular binding energy $E_b$ and the thermal energy $k_B T$, enhances the fraction of molecules and can
partially compensate for a low atomic phase-space density.

\begin{figure}
\includegraphics[width=3in]{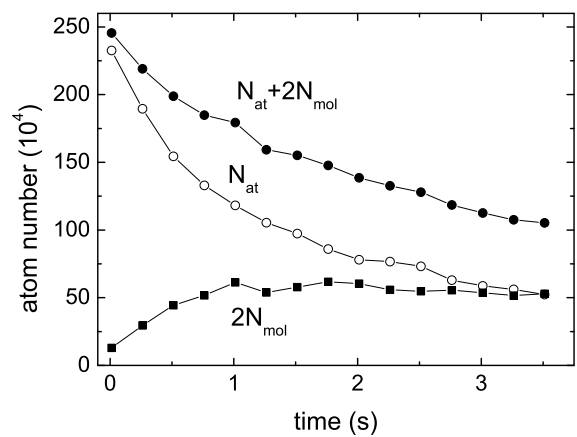}
\caption{An ultracold $^6$Li gas approaches a chemical atom-molecule
equilibrium on the molecular side of the 834-G Feshbach resonance.
The experiment starts with a non-degenerate, purely atomic gas at a
temperature of $2.5\,\mu$K and a peak atomic phase-space density of
$0.04$. The magnetic field is set to 690\,G, where the molecular
binding energy corresponds to $15\,\mu$K. $N_{\rm at}$ and $N_{\rm
mol}$ denote the number of unbound atoms and the number of
molecules, respectively. The total number of unbound and bound atoms
$2 N_{\rm mol} + N_{\rm at}$ slowly decreases because inelastic
loss is not fully suppressed. From \cite{Jochim2003}.} \label{fig:atmolequil}
\end{figure}

The thermal atom-molecule equilibrium was experimentally
investigated by \cite{Jochim2003} in a non-degenerate gas of $^6$Li
atoms. Figure~\ref{fig:atmolequil} illustrates how the initially pure
atomic gas tends to an atom-molecule equilibrium. The observation
that more than 50\% of the atoms form molecules at a phase-space
density of 0.04 highlights the role of the Boltzmann factor. The chemical
atom-molecule equilibrium also played an essential role in the
experiment by \cite{Cubizolles2003}, where a slow
Feshbach ramp, which kept the sample in thermal equilibrium,
led to a conversion efficiency of 85\%.

\subsection{Properties}\label{ssec:mol_prop}
\subsubsection{Dissociation and detection}\label{sssec:mol_diss}

A general way to detect Feshbach molecules is their controlled
dissociation through reverse magnetic-field sweeps, followed by
imaging of the resulting cloud of atoms. If the image is taken
immediately after the forced dissociation, it just reflects the
spatial distribution of the molecules before the onset of
dissociation. However, if the image is taken after a certain time of
flight, it will be strongly affected by a release of kinetic energy.
The image then contains additional information on the dissociation process.

A reverse Feshbach ramp brings the molecule into a quasi-bound state
above the dissociation threshold, from which it decays into two atoms in the
continuum. The decay rate $\Gamma(E)/\hbar$ depends on the energy
$E$ above threshold and can be calculated from Eqs.~(\ref{II.A.13}), (\ref{II.A.16}),
and (\ref{II.A.21b}),
\begin{equation}
\Gamma(E)=2k a_{\rm bg} \delta\mu \Delta = 2k\bar{a}\bar{E} s_{\rm res} \,.\\
\label{dissociation}
\end{equation}
\cite{Mukaiyama2004} gave the energy spectrum of hot atoms by
using this Fermi's golden rule expression for a linear ramp
and showed that it agreed well with measurements with $^{23}$Na$_2$
Feshbach molecules.  \cite{Goral2004} verified the golden rule
theory with a full quantum dynamics calculation of Feshbach molecule
dissociation.

The mean kinetic energy released in a reverse Feshbach ramp
corresponds to the typical energy that the molecules can reach in
the quasi-bound level before the dissociative decay takes place. In
experiments on $^{87}$Rb$_2$, \cite{Durr2004b} studied the
dependence of the energy release on the ramp rate and on the
resonance width. They demonstrated kinetic energy measurements after
reverse Feshbach ramps as a powerful indirect tool to determine the
widths of weak resonances with $\Delta$ in the mG range, where
direct methods are impractical. They also
demonstrated the production of a monoenergetic spherical wave of
atoms by rapidly switching the magnetic field instead of ramping it.

\begin{figure}
\noindent \includegraphics[width=3in]{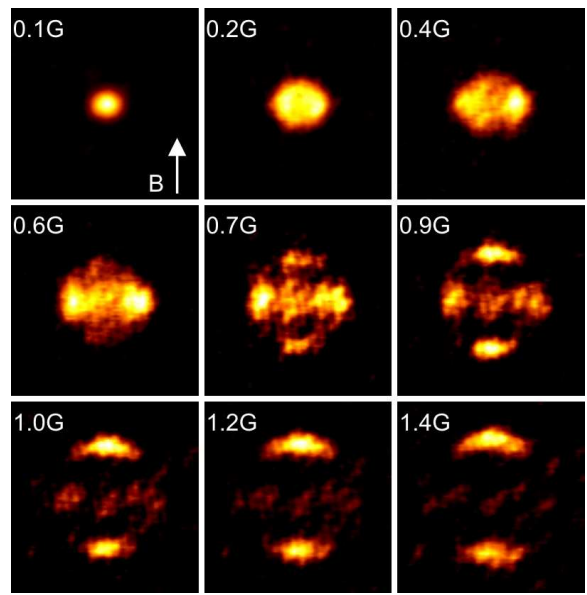}
\caption{Dissociation patterns of $^{87}$Rb$_2$ molecules, showing
the interference of $s$- and $d$- partial waves. At small magnetic
field offsets $B-B_0$ (values given in the upper left corners), the $s$-wave
pattern dominates; at large offsets, $d$-waves are strongly enhanced
due to a $d$-wave shape resonance. From \cite{Volz2005}.}
\label{mol_dissociation}
\end{figure}

The dissociation properties of Feshbach molecules can provide
additional spectroscopic information. \cite{Volz2005} observed
interesting dissociation patterns of $^{87}$Rb$_2$, when the
molecular state was brought high above the threshold with fast jumps
of the magnetic field. The patterns, shown in
Fig.~\ref{mol_dissociation}, reveal a $d$-wave shape resonance. The
dissociation of $^{133}$Cs$_2$ molecules in $l$-wave states was
observed by \cite{Mark2007}. The dissociation pattern showed a
strikingly different behavior from molecules in a $g$-wave state and
allowed to clearly distinguish between these two { types
of molecules}.

We note that direct imaging of Feshbach molecules is not feasible in
most situations because of the absence of cycling optical
transitions. An exception, however, is the direct imaging of atoms
in halo states as demonstrated for $^6$Li
\cite{Zwierlein2003,Bartenstein2004}. In this special case, the
extremely weakly bound dimer absorbs resonant light essentially like
free atoms.

\subsubsection{Halo dimers}\label{sssec:halo}

Broad, entrance-channel dominated $s$-wave resonances feature a
considerable region where Feshbach molecules acquire universal
properties; see Sec.~\ref{sec:theory} and \cite{Kohler2006}.
``Universality'' means that details of the interaction become
irrelevant and that all properties of the dimer are characterized by
a single parameter, the $s$-wave scattering length $a$ or
alternatively the binding energy $E_b = \hbar^2/(ma^2)$. The reason
for this great simplification is the fact that the wavefunction
extends far out of the classical interaction range of the potential.
States of this kind have been coined ``quantum halos''. They have
attracted considerable attention in nuclear physics and, more
recently, in molecular physics and have been extensively reviewed in
\cite{Jensen2004}. An early example is the deuteron, where the
neutron and proton are likely to be found outside of the classically
allowed region. Before the advent of Feshbach molecules, the most
extended halo system experimentally accessible was the helium dimer
($^4$He$_2$), which is about ten times larger than typical diatomic
molecules. An extreme example is given by Bose-condensed
$^6$Li$_2$ Feshbach dimers with a size of $a/2
\simeq 2000\,a_0$, which exceeds the van der Waals length $R_\mathrm{vdw}
\simeq 30\,a_0$ by almost two orders of magnitude.

First experiments on halo dimers have been conducted with bosonic
$^{85}$Rb \cite{Donley2002} and $^{133}$Cs \cite{Mark2007a}, and
fermionic $^{40}$K and $^6$Li; see Sec.~\ref{sssec:molbec}.  For $^{85}$Rb, the dimers are not
formed from atoms in the lowest internal states and thus have open
decay channels. This leads to spontaneous dissociation without the
presence of other atoms or molecules. \cite{Thompson2005} have
observed an $a^{-3}$ scaling of the dissociation rate,
which can be understood as a direct consequence of universality
through wavefunction overlap arguments \cite{Kohler2005}. For
halo dimers of $^{133}$Cs$_2$ created from atoms in their
lowest internal states there are no open dissociation channels.
These molecules cannot decay spontaneously but decay through collisions
with other atoms or molecules \cite{Ferlaino2007,Knoop2008}.

A future promising direction with Feshbach molecules in halo states
is the experimental investigation of universal few-body physics; see
Sec.~\ref{ssec:efimov}.

\subsubsection{Collision properties}\label{sssec:mol_coll}

\begin{figure}
\includegraphics[width=2.5in]{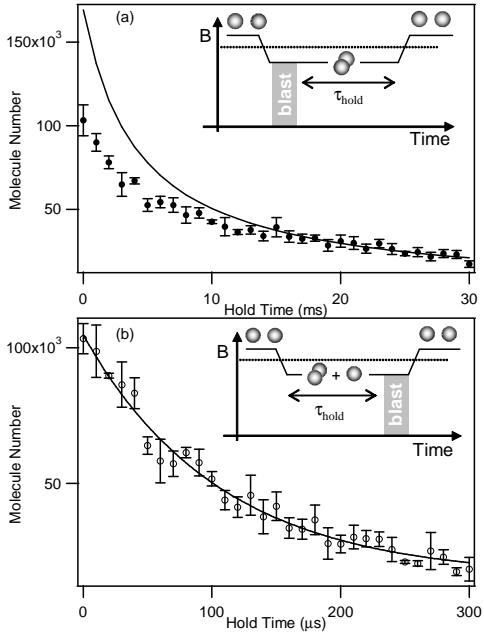}
\caption{Fast collisional decay of $^{23}$Na$_2$ Feshbach molecules.
The molecules are trapped alone (a) or together with atoms (b). From
\cite{Mukaiyama2004}.} \label{fig:moldecay}
\end{figure}

Fast collisional loss is usually observed in trapped samples of
ultracold molecules. This has been seen in experiments with bosonic
$^{23}$Na$_2$ \cite{Mukaiyama2004}, $^{87}$Rb$_2$ \cite{Syassen2006,
Wynar2000} and $^{133}$Cs$_2$ \cite{Chin2005}.
 \cite{Zirbel2008} measured large rate coefficients for fermionic $^{40}$K$^{87}$Rb Feshbach
molecules due to collisions with $^{87}$Rb or $^{40}$K.  Both
atom-dimer and dimer-dimer collisions are generally found to cause
strong inelastic loss, as demonstrated by the example in
Fig.~\ref{fig:moldecay}. Vibrational relaxation is the dominant
mechanism, which leads to large loss rate coefficients of the order
of $10^{-10}$ cm$^3$/s.  Rate coefficients of such magnitude result
in molecular lifetimes on the order of a few ms or less for
densities characteristic of ultracold gases.  Rearrangement
reactions, such as trimer formation, may also play a significant
role in limiting molecular lifetimes.

Inelastic collision rates of Feshbach molecules in the very highest bound state, when they are not universal ``halo'' states, are not significantly different from rates for more deeply bound states, for which fast inelastic loss has been observed~\cite{Staanum2006,Zahzam2006} and predicted~\cite{Quemener2005,Cvitas2007}.  \cite{Hudson2008} measured large inelastic collision rate
coefficients for vibrationally excited triplet $^{87}$Rb$^{133}$Cs molecules
colliding with $^{133}$Cs or $^{87}$Rb atoms.  They also used a simple model to help understand why such large rate constants are typical for atom-molecule vibrational relaxation.  Assuming a probability near unity for inelastic loss when the collision partners approach one another in the short-range region of chemical bonding, the overall collision rate coefficient is then determined from the threshold scattering of the long range van der Waals potential.  If effect, the rate constant is given by Eq.~(\ref{II.A.10}), where the length $b$ turns out to be similar in magnitude to the van der Waals length $R_\mathrm{vdw}$.  Such a simple model gives the typical order of magnitude of $10^{-10}$ cm$^3/$s for the vibrational relaxation rate constant, nearly independent of the vibrational level, found for the $^{87}$Rb$^{133}$Cs system.

Halo molecules comprised of two unlike fermions bound in an $s$-wave state
offer an exception to the rule of fast
inelastic dimer-dimer and atom-dimer collisions.
This has allowed stable molecular samples and even molecular Bose-Einstein
condensation; see Sec.~\ref{ssec:fermi}.
\cite{Petrov2004a} showed that a combination of two effects explains
this stability. The first effect is a small wavefunction overlap of a
halo dimer with more deeply bound dimer states, and the second one
is Pauli suppression in the few-body process. For inelastic
dimer-dimer collisions, \cite{Petrov2004a} predicted the rate
coefficient for inelastic loss to scale as $(a/R_{\rm vdw})^{-2.55}$
whereas, for the elastic part, they obtained a dimer-dimer
scattering length of $0.6\,a$. For the atom-dimer interaction, the
predicted scaling of inelastic loss is $(a/R_{\rm vdw})^{-3.33}$ and
the scattering length is $1.2\,a$.

An interesting case is the observation of stable $^6$Li$_2$ molecules created near the closed-channel dominated resonance at 543~G \cite{Strecker2003}.  These are not halo molecules and would be expected to have a large collisional loss rate coefficient  similar to molecules comprised of bosons.  The collision properties of these molecules still await detailed investigation.

A possible way to overcome harmful inelastic collisional loss is the application of an optical
lattice \cite{Thalhammer2006}. Here a pair of atoms or a single molecule can be trapped in an individual lattice site, which offers shielding from collisions with other molecules or atoms.
Many experiments on ultracold molecules are now being performed with
Feshbach resonances and molecules in an optical lattice; see discussion in Sec.~\ref{ssec:ola}.
Another way to prevent inelastic collisions of ultracold molecules is to transfer them to their lowest energy  ground state, where they do not undergo vibrational, rotational, or spin relaxation.  However, reactive collisions may still be possible.

\subsubsection{Internal state transfer}
\label{sssec:transfer}

A Feshbach resonance can serve as an ``entrance gate'' into the rich variety of molecular states below threshold, allowing preservation of the ultralow temperature of the atomic gas that is used as a starting point.
 The magnetic association technique (Sec.~\ref{sssec:mol_timedep}) produces a molecule in a specific weakly bound molecular state, i.e.\ the particular molecular state
that represents the closed scattering channel of the resonance. This leads to the question how a Feshbach molecule can be transferred to other states with specific
properties of interest or, ultimately, to the absolute
ro-vibrational ground state. Various methods have been developed for
a controlled internal state transfer based on magnetic
field ramps, radio-frequency or microwave radiation, or optical
Raman excitations.

\begin{figure}
\includegraphics[width=2.5in]{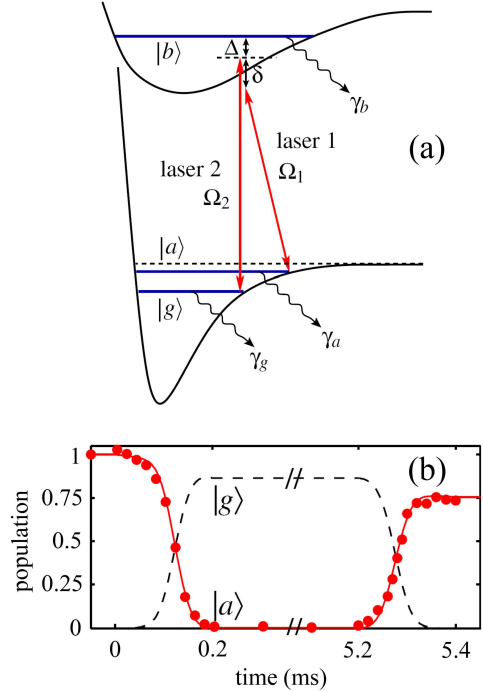}
\caption{STIRAP with Feshbach molecules. (a) After creation of
Feshbach molecules in a weakly bound state $|a\rangle$, a two-photon
transition is used to transfer the molecules into a more deeply
bound state $|g\rangle$. In the counterintuitive pulse sequence of
STIRAP, laser field 2 is switched on before laser field 1, and the
molecule is always kept in a dark superposition state of $|a\rangle$
and $|g\rangle$ during the transfer sequence. This suppresses the
unwanted electronic excitation of state $|b\rangle$ and leads to a
very high transfer efficiency. The measurements shown in (b) refer
to an experiment in $^{87}$Rb$_2$, where $|a\rangle$ and $|g\rangle$
correspond to the last and second-to-last bound vibrational level.
The data show the measured population in $|a\rangle$ during a 0.2-ms
transfer sequence and a reverse transfer sequence after a 5-s hold
time in state $|g\rangle$. The observed efficiency of 75\% for a
double passage corresponds to single-passage transfer efficiency of
87\%. Adapted from \cite{Winkler2007}.} \label{fig:stirap}
\end{figure}

When a Feshbach ramp after initially associating the molecules is
continued over a wider magnetic field range, the molecule will
perform a passage through many level crossings; see e.g.\
Fig.~\ref{fig:Rb87_Marte}. The ramp speed controls whether crossings
are traversed diabatically (fast ramp) or whether they are followed
adiabatically (slow ramp).  \cite{Mark2007a} demonstrated the
controlled transfer of $^{133}$Cs$_2$ molecules into different
states by elaborate magnetic field ramps. In this way, they could
populate various states from $s$- up to $l$-waves with binding
energies of up to $\sim$10\,MHz. In practice, finite ramp speeds
limit this method to rather weak crossings with energy splittings of
up to typically 200\,kHz.  \cite{Lang2007} showed how this problem
can be overcome with the help of radio-frequency excitation.  They
demonstrated the transfer of  $^{87}$Rb$_2$ molecules over nine
level crossings when the magnetic field was ramped down from the
1007-G resonance to zero field in 100\,ms. This produced molecules
having a binding energy $E_b = h \times 3.6$\,GHz with a total
transfer efficiency of about 50\%.

More deeply bound states can be reached by two-photon Raman
transitions, as implemented in a very efficient way by stimulated
Raman adiabatic passage (STIRAP) \cite{Bergmann1998}.
Fig.~\ref{fig:stirap} illustrates STIRAP between the two highest
vibrational levels in $^{87}$Rb$_2$ with binding energies
corresponding to 24\,MHz and 637\,MHz, as demonstrated in a proof-of-principle experiment by \cite{Winkler2007}. The experiment also highlighted the great potential of the approach to combine Feshbach association with stimulated Raman optical transitions, as originally suggested by \cite{Kokkelmans2001} to create deeply bound molecules.

In 2008, enormous experimental progress was made in applications of STIRAP to transfer both homo- and heteronuclear Feshbach molecules into deeply bound states. \cite{Danzl2008} explored $^{133}$Cs$_2$ molecules and demonstrated large binding energies corresponding to 31.8\,THz. \cite{Lang2008} reached the rovibrational ground state in the triplet potential of $^{87}$Rb$_2$, the binding energy of which corresponds to 7.0\,THz.
The heteronuclear case was successfully explored with $^{40}$K$^{87}$Rb. Initial experiments by \cite{Ospelkaus2008} demonstrated the transfer to states with a binding energy corresponding to 10.5\,GHz. Only shortly afterwards, the same group \cite{Ni2008} demonstrated polar molecules in both the triplet and the singlet rovibrational ground state, where the binding energies correspond to 7.2\,THz and 125\,THz, respectively. These experiments opened up a promising new research field related to the exciting interaction properties of ground-state molecular quantum gases.

All the above methods for controlled state transfer rely on coherent
processes. Therefore they can also be applied to produce coherent
superpositions of molecular states. This can, for example, be used
for precise interferometric measurements of the molecular structure.
\cite{Mark2007} and \cite{Lang2007} investigated molecular
level crossings in this way. Using STIRAP, \cite{Winkler2007}
created quantum superpositions between neighboring vibrational
states and tested their coherence interferometrically.

\section{Related topics}
\label{sec:related}

\subsection{Optical Feshbach resonances}
\label{ssec:optfesh}

Magnetic fields have proven to be a powerful
tool to change the interaction strength or scattering length between
ultracold atoms. As discussed at length in this Review this has
been made possible by the presence of a molecular bound state that
is resonantly coupled to the colliding atom pair.  The width of
the resonance ($\Delta$ in Eq.~(\ref{II.A.21})), however, is
governed by the interatomic forces between the two atoms.  Optical Feshbach
resonances promise control of both the resonance location and its
width.

\subsubsection{Analogies}
\label{sssec:of_anlg}

\begin{figure}
\includegraphics[scale=0.3,angle=-90]{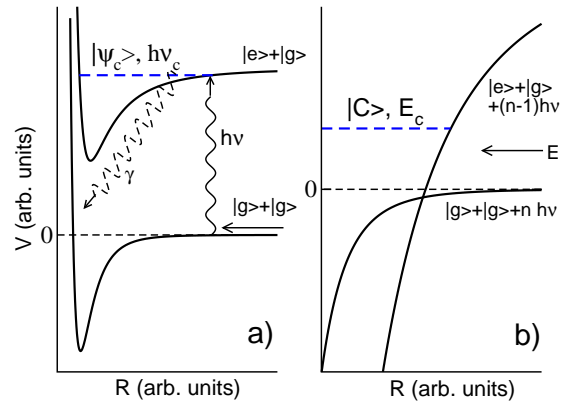}
\caption{Schematic of an optical Feshbach resonance.  Panel (a) refers to
two colliding ground-state atoms, $|g\rangle$ that are coupled through a laser at frequency $\nu$ to a vibrational level $|\psi_c\rangle$ of an excited electronic state that
dissociates to one excited-state $|e\rangle$ and one ground-state
$|g\rangle$ atom. The energy $h\nu_c$ is the energy of the
vibrational level relative to two colliding atoms at zero collision
energy.  The excited bound state $|\psi_c\rangle$
irreversibly decays with an (energy) width $\gamma$.  Panel (b) shows
the dressed state picture corresponding to panel (a).  The
initial state is an atom pair with $n$ photons and the
closed channel state $|C\rangle$ represents $|\psi_c\rangle$
with $n-1$ photons. The atom pair has a relative kinetic
energy $E$ and $E_c=h\nu_c-h\nu$.  In the dressed state
picture the ground and excited interatomic potentials cross at a point called the Condon point.} \label{fig:OFR}
\end{figure}

Figure~\ref{fig:OFR} shows a schematic diagram of an optical
Feshbach resonance.  As first proposed by \cite{Fedichev1996} laser
light nearly resonant with a transition from a colliding atom pair
and a ro-vibrational level of an excited electronic state
induces a Feshbach resonance and modifies the scattering length
of the two atoms.  Excited electronic
states dissociate to one ground- and one electronically-excited atom
for large interatomic separations. For many kinds of atoms the
photon needed to reach such states is in the visible or optical
domain and, hence, the term ``optical Feshbach resonance'' has been
adopted.

The location and strength of an optical Feshbach resonance is
determined by the laser frequency $\nu$ and intensity $I$,
respectively. Both can be controlled experimentally. There is,
however, a crucial difference between magnetic and optical Feshbach
resonances.  For optical Feshbach resonances the resonant state has
a finite energy width $\gamma$ and thus lifetime $\hbar/\gamma$ due
to spontaneous emission.  Hence the scattering length becomes a complex number.

Note that by changing the laser frequency and simultaneously
detecting the population in the excited electronic potentials the
ro-vibrational level structure of these potentials can be
studied.  This is called ultracold photoassociative spectroscopy
and has been reviewed in \cite{Jones2006}.

\cite{Bohn1996,Bohn1999} obtained expressions for the complex scattering length in Eqs.~(\ref{II.A.23})-(\ref{II.A.24}).   Resonances are characterized by a width $\Gamma(E)$ and shift $h\delta \nu_c$, where $E_0=h\nu_c - h\nu -h\delta \nu_c$.  The width is
\begin{equation}
\Gamma(E)=2\pi |\langle C | \vec{d}\cdot \vec{\cal E} | E
\rangle|^2 \,, \label{OFR_width}
\end{equation}
where $\vec{\cal E}$ is the electric field of the laser and
$\vec{d}$ is the molecular electronic transition dipole moment.  Both the
width and the shift of the resonance are proportional to $I$.  As in magnetic resonances
$ | E \rangle$ is the scattering wavefunction at collision energy $E$ in
the entrance channel.

Figure~\ref{fig:opticalscat} in Section~\ref{sssec:resscatt} shows
the real and imaginary part of the scattering length $a-ib$ for
an optical Feshbach resonance. The numbers are based on an analysis
of the strength and lifetime of an experimentally-observed optical
Feshbach resonance in $^{87}$Rb \cite{Theis2004}.  For the intensity used in the figure
the optical length $a_{\rm res}$ defined by Eq.~(\ref{II.A.25}) is 5.47 nm and $\Gamma_0/h=21$ MHz.
That is, $a_{\rm res}\approx a_{\rm bg}$ and $\Gamma_0\approx
\gamma$. Since $k a_\mathrm{bg} \ll 1$, the width $\Gamma(E) \ll \gamma$ so there is negligible power broadening.  Unlike for a magnetic Feshbach resonance, the real part of
the scattering length is now finite for any detuning with a
peak to peak variation of $2a_{\rm res}$.  The
length $b$ peaks at zero detuning.  The maximum
value is $2a_{\rm res}$ and the full-width half-maximum is $\gamma$.

In order for an optical Feshbach resonance to be practical it is
necessary that the change in the real part of the scattering length
$a-a_{\rm bg}$ is large compared to $b$.  This requires that the
detuning $h\nu-h\nu_c$ is large compared to $\gamma$. On the other
hand we would also like to ensure that $a-a_{\rm bg}$ is at least on
the order of $a_{\rm bg}$ at such a detuning.  In order to satisfy
these requirements simultaneously we need $a_{\rm res} \gg a_{\rm
bg}$ or equivalently $\Gamma_0 \gg \gamma$.  For the parameters used
in Fig.~\ref{fig:opticalscat} this is not true. Section
\ref{sssec:earth-alkali} discusses how for alkaline-earth atoms it
is possible to satisfy $a_{\rm res} \gg a_{\rm bg}$.

By adding extra laser fields at different frequencies the scattering
length can be further manipulated.  Of particular interest is the
case where a second field is nearly resonant with the excited bound
state $|\psi_c \rangle$ and a second molecular bound state. If this
second bound state is in the electronic ground
state, then the situation corresponds to a Raman transition.  The analytic expression for the
scattering length
\cite{Bohn1999,Thalhammer2005} is
\begin{eqnarray}
a-ib &=& a_{\rm bg}+ \frac{1}{2k} \frac{\Gamma(E)}
           {h\nu-h\nu_0+\Omega^2/\Delta_2+i (\gamma/2)}
\label{eq:2ph}
\end{eqnarray}
where $a_{\rm bg}$, $\Gamma$, and $\gamma$ are
defined as before, and $h\nu_0 = h \nu_c + h \delta \nu_c$.  The two-photon detuning $\Delta_2$ is zero when the absolute value of the frequency difference of the
two lasers equals the absolute value of binding energy of the ground
bound state relative to two free atoms at rest.  It is positive when
the absolute value of the frequency difference is larger than the
absolute value of the binding energy. The quantity $\Omega$ is the
coupling matrix element between the bound levels in the ground and excited
states and is proportional to the square root of the intensity of the second laser.

\subsubsection{Observations in alkali systems}
\label{sssec:of_alkali}

In a magneto-optical trap filled with { cold} ($<$ 1 mK)
atomic sodium \cite{Fatemi2000} confirmed the predictions of
\cite{Fedichev1996}. They observed the changing scattering length by
detecting the corresponding change in the scattering wavefunction.
Weak detection lasers, with frequencies that differ from those used
for the optical Feshbach resonance, induce a molecular ion signal
that probes the change in the scattering wave function. For the
molecular level used in the experiment \cite{Fatemi2000} were able
to deduce the strength of the resonance as well as the light shift
$\delta v_c$. They found an $a_{\rm res}$ of around 2 nm and
$\Gamma_0/h$ of around 20 MHz for the maximum reported laser
intensity of $I$=100 W/cm$^2$.  For Na the background scattering
length is $a_{\rm bg}=2.8$ nm.

\cite{Theis2004} tuned the scattering length by optical means in a
Bose-Einstein condensate.  In a $^{87}$Rb condensate they were able
to change the scattering length over one order of magnitude from 0.5
nm to 10 nm. The parameters of the optical resonance are given in
the caption of Fig.~\ref{fig:opticalscat}.  The scattering length
was measured using Bragg spectroscopy, where a change in condensate
mean field energy proportional to the scattering length is measured
by a change in the frequency that determines the Bragg condition.

\begin{figure}
\includegraphics[scale=1]{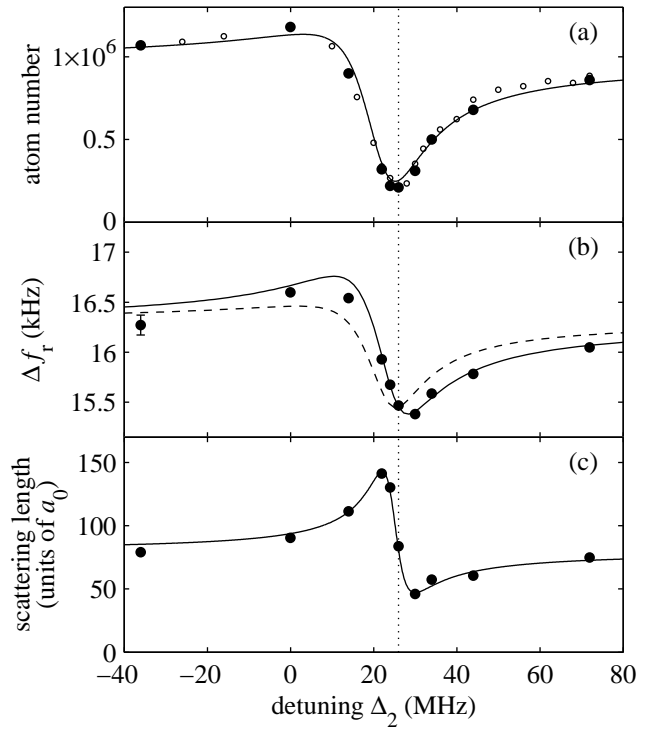}
\caption{Optical Feshbach resonance using a two-color Raman
transition in an $^{87}$Rb BEC.  Panel (a) shows the
measured atom number after a 100 $\mu$s Raman pulse as a function of
the two-photon detuning $\Delta_2$. Panel (b) shows the measured
Bragg resonance frequency and Panel (c) shows the scattering length,
as determined from Panels (a) and (b).  The open and filled
circles correspond to the data.  As indicated by the error bar in Panel (b) the frequency uncertainty in the Bragg spectroscopy is smaller than $100\,\text{Hz}$.  The
solid lines have been obtained from a
fit to Eq.~(\ref{eq:2ph}) adapted to include the finite linewidth of
the ground-bound state.  The dashed line in Panel (b) shows the
expected signal if there was only loss in atom number but no change
in scattering length.  The vertical line indicates the location of
maximal loss in Panel (a) and helps to compare the relative
positions of the three curves.  From \cite{Thalhammer2005}.  }
\label{fig:Raman}
\end{figure}

In \cite{Thalhammer2005} the scattering length was modified by a
two-color Raman transition.  As in \cite{Theis2004} the change in
scattering length was observed in an $^{87}$Rb condensate and
detected by Bragg spectroscopy. In fitting to Eq.~(\ref{eq:2ph}) a
complication arose. \cite{Thalhammer2005} could only explain their
observations if they assumed that the target ground state had a
finite linewidth. It turned out that, even though this state cannot
be lost by spontaneous emission, it can absorb a photon from the
(strong) laser that photoassociates the scattering atom pair.  This
process gave rise to a linewidth of 2 MHz. The results are shown in
Fig.~\ref{fig:Raman}.

\subsubsection{Prospects in alkaline-earth systems}
\label{sssec:earth-alkali}

So far, the scattering length has been experimentally modified by
optical means in ultracold alkali-metal atom collisions. In these
experiments the optical length $a_{\rm res}$ was of the same order
of magnitude as the background scattering length $a_{\rm bg}$ so
that changes in scattering length were accompanied by large atom
losses. \cite{Ciurylo2005} showed that optical Feshbach resonances
in ultracold alkaline-earth atom collisions can have $a_{\rm res}\gg
a_{\rm bg}$.  The presence of intercombination lines in
alkaline-earth systems make this possible.  Atomic intercombination
lines are transitions between the ground $^1$S$_0$ state and the
excited $^3$P$_1$ state. The transition is only weakly allowed by
virtue of relativistic mixing with the $^1$P$_1$ state. For example,
for Sr isotopes $\gamma/h=$ 7.5 kHz is much smaller than for alkali
atoms.

When $\gamma$ is very small, it is possible to use excited bound levels that are very close to the excited state dissociation threshold, while simultaneously maintaining the large detunings that are necessary to suppress losses.    Using levels close to threshold allow a very large value of the ratio $\Gamma_0/\gamma$, and consequently $a_\mathrm{res}$ can be orders of magnitude larger than $a_\mathrm{bg}$.   \cite{Ciurylo2005}
illustrate this by model calculations of optical
lengths of ultracold calcium, for which $\gamma/h=0.7$ kHz.  Using a model that assumes a level with a binding energy on the order of 100 MHz, they predict that $a_{\rm res}$ could be as large as 100 nm at the relatively low intensity of $I=1$ W/cm$^2$.

\cite{Zelevinsky2006} obtained photoassociation spectra near the intercombination line of $^{88}$Sr and measured the strength of various transitions.   They found that the last bound state of the excited potential had an optical length $a_{\rm res}= 24$ $\mu$m at $I=1$ W/cm$^2$.  The very large value of $a_{\rm res}$ implies that practical  changes in the scattering length should be feasible in this species.  Similar photoassociation spectra have been observed for two different isotopes of Ytterbium~\cite{Tojo2006}, which has electronic structure like that of the alkaline earth atoms.  Optical control of both bosonic and fermionic isotopic species may become possible with alkaline earth or Yb atoms. \cite{Enomoto2008} have demonstrated optically induced changes in scattering length for $^{172}$Yb and $^{176}$Yb.

\subsection{Feshbach resonances in optical lattices}
\label{ssec:ola}

{ Ultracold} atoms  in optical lattices are of great
interest because of the exciting prospects to simulate a variety of
condensed matter phenomena, to realize large scale quantum
information processing \cite{Bloch2005}, and to form ultracold
molecules in individual lattice sites to avoid detrimental collision
instability. In all these research directions, Feshbach {
resonances} will provide excellent tools to control the interaction
of the constituent atoms and to explore the transition between
different quantum regime and quantum phases. In the following
sections we review atom-atom scattering in optical lattices and
describe the role of Feshbach resonances therein.

Optical lattices are realized by standing-wave laser fields, which
{ result} in spatially-periodic potentials for the atoms.
The atoms are confined in the individual potential minima or sites
of the lattice potential. One-, two-, and three-dimensional lattices
can be created in this way. In experiments with three-dimensional
configurations the optical lattice can be filled with only one or
two atoms per site.

Section~\ref{sssec:ola_mol} discusses that when two
atoms are held in a single lattice site a Feshbach resonance can be used to very efficiently produce stable molecules.  One advantage of lattice confinement is that such molecules are protected from harmful collisional losses with a third body.  Section~\ref{sssec:ola_lowd} discusses the possibility that confinement in one spatial direction can induce resonant behavior in scattering along the remaining directions.
Finally, Section~\ref{sssec:ola_scat} describes uses of Feshbach resonances in optical lattices
where tunneling between lattices sites is important.

\subsubsection{Atom pairs and molecules}
\label{sssec:ola_mol}

When an atom pair is trapped in a single site of a three-dimensional
optical lattice the motion is fully quantized. For deep optical lattices the
confining potential is harmonic, and the center-of-mass and relative motion of the atom pair separate. In
other words the six-dimensional wavefunction of the two atoms becomes a
product of a center-of-mass and relative  wavefunction. The
center-of-mass motion is harmonic and is solved trivially. The
relative motion is determined by a potential that is the sum of the
atom-atom interaction potentials and a harmonic potential.

The atom-atom interactions between alkali-metal atoms are independent of the relative orientation of the atoms when very weak spin-dependent interactions, $V_{\rm ss}$, are ignored.
Consequently, for a spherically symmetric harmonic trapping potential the
three-dimensional relative motion can be further simplified. The
angular motion can be solved analytically and only a radial
Schr\"odinger equation for the atom-atom interaction potential plus
$\mu\omega^2R^2/2$ needs to be solved. Here, $\mu$ is the reduced
mass and $\omega$ is the oscillation frequency in the trap.

\begin{figure}
\includegraphics[scale=0.5]{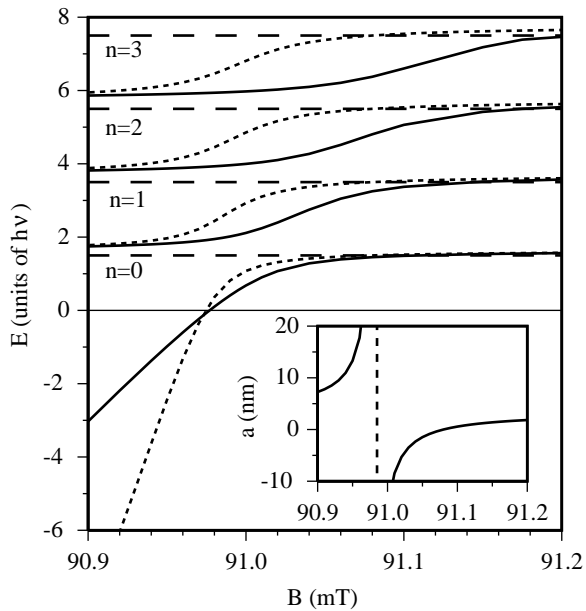}
\caption{The energy in the relative motion of two trapped
interacting Na atoms in their lowest hyperfine state $|a\rangle$, as
a function of magnetic field. The trapping frequency
$\nu=\omega/2\pi$ is 1 MHz. The full lines correspond to energies
obtained from exact numerical calculations. The dotted lines
correspond to { eigenenergies} for trapped Na atoms
interacting via a regularized delta-function potential with a
magnetic field dependent scattering length given by the inset.  This
inset shows the exact scattering length for two freely scattering
$|a\rangle$ states near a Feshbach resonance.  In the theoretical
model $B_0=90.985$ mT, whereas experimentally $B_0=90.7$ mT; see
Section~\ref{sssec:Na}. The long-dashed lines correspond to energies
of the $\ell$=0, $n$=0, 1, 2, and 3 harmonic oscillator states. From
\cite{Tiesinga2000}.  } \label{fig:FeshNa}
\end{figure}

Figure~\ref{fig:FeshNa} shows eigenenergies for two Na atoms with
zero relative orbital angular momentum ($\ell=0$) in a
spherically-symmetric harmonic trap as a function of magnetic field
\cite{Tiesinga2000}.  The atoms are in their lowest hyperfine state
and the energies are obtained from coupled-channels calculations.
For these atomic states there is a Feshbach resonance near 910\,G
{ (91~mT)}. The zero of the vertical axis corresponds to
zero relative kinetic energy in the absence of a trapping potential.
Hence, positive energies correspond to atoms in the trap and
negative energies correspond to molecules bound in the atom-atom
interaction potential.  Avoided crossings between the closed channel
Feshbach state and the trap levels are clearly visible.

An analytic solution for two interacting atoms trapped in a harmonic
trap was found by \cite{Busch1998}. The atom-atom interaction was
modeled by a so-called regularized delta-function potential, $4\pi
\hbar^2/(2\mu) a \delta(\vec{R}) (d/dR) R$, where $a$ is the
scattering length. This approximation of the actual interaction
potential is valid in the Wigner threshold regime. In particular,
they showed that the eigenenergies for the relative motion are
given by
\begin{equation}
         \frac{a}{\sigma} = \frac{ \Gamma( -E/2+1/4 )}{ 2\Gamma( -E/2+3/4 )} \,,
\label{eq:Busch}
\end{equation}
where $\Gamma$ is the Gamma function, the energy $E$ is in units of
$\hbar\omega$, and $\sigma=\sqrt{\hbar/(\mu\omega) }$ is the
harmonic oscillator length for the relative motion.

Near a magnetic Feshbach resonance the scattering length is a
rapidly-changing function of $B$, as for example shown in the inset
of Fig.~\ref{fig:FeshNa}. The dotted lines in Fig.~\ref{fig:FeshNa}
are the eigenenergies found from combining Eq.~(\ref{eq:Busch}) and
the $a(B)$ for the resonance. The exact and model calculations do
not agree where the scattering length is large. As shown by
\cite{Blume2002,Bolda2003} this is due to the breakdown of the
Wigner threshold regime at the finite zero-point energy of the atoms
near the resonance. \cite{Blume2002} introduced an energy-dependent
scattering length, based on the effective range theory.
\cite{Bolda2002} found that an energy-dependent regularized
delta-function potential reproduces well the exact results in
Fig.~\ref{fig:FeshNa}.

 \cite{Dickerscheid2005b,Gubbels2006} developed an analytic
approach extending the theory of \cite{Busch1998}
to the states of two trapped atoms in a single lattice site that interact
strongly through a Feshbach resonance.  They applied their two-body theory
to the broad $^6$Li resonance near 834 G.  They also showed how to
incorporate this theory into a many-body Hubbard model that treats the
tunneling of atoms between lattice sites.

\cite{Mies2000a} have theoretically shown that by varying the
magnetic field in time two atoms in a single lattice site can be
converted into a molecule with near 100\% efficiency. The idea is to
prepare the atoms in the lowest trap level at a magnetic field where
the Feshbach state has a higher energy. In Fig.~\ref{fig:FeshNa}
this corresponds to the nominally $n=0$ state at for example
$B$=912~G.  The magnetic field is then varied. If the ramp is
sufficiently slow the atom pair will be adiabatically converted into
a molecule. \cite{Mies2000a} also showed that this process can be
described by a Landau-Zener curve crossing model. A more recent
derivation is given in \cite{Julienne2004}.

\cite{Widera2004} used a magnetic Feshbach resonance
to entangle two $^{87}$Rb atoms in a site of an optical lattice.
In a Ramsey-type interferometer a sequence of  microwave
pulses manipulate and control the superposition between the
$|0\rangle\equiv|F=1,m_F=1\rangle$ and
$|1\rangle\equiv|F=2,m_F=-1\rangle$ hyperfine states of each
$^{87}$Rb atom.  Initially the atoms are in state $|0\rangle$. The
population in the two hyperfine states after the pulses
will depend on the atom-atom interaction, which entangles the two
atoms.  By controling the scattering length by the magnetic field
and the hold time between the pulses, \cite{Widera2004} were able to
create maximally-entangled Bell states.

\cite{Stoferle2006} spectroscopically mapped the
avoided crossing between the Feshbach and the lowest harmonic
oscillator state as a function a magnetic field. They prepared
fermionic $^{40}$K atoms in a three-dimensional optical lattice
configured such that the bottom of the lattice sites are spherically
symmetric. The two atoms in each lattice site are in different
hyperfine states.  They confirm that the model of trapped atoms
interacting via an energy-dependent delta-function potential agrees
with the experimental observations.

\cite{Thalhammer2006} showed in an experiment with $^{87}$Rb in its
lowest hyperfine state that the Landau-Zener model for a
time-dependent sweep of the magnetic field through the Feshbach
resonance, as developed by \cite{Mies2000a,Julienne2004}, is valid.
The Feshbach resonance near 1007~G was used. In the experiment the
ramp speed $dB/dt$ was varied over four orders of magnitude and
for the slowest ramp speed of $2\times 10^3$ G/s a 95\% conversion
efficiency was observed.  The experiment demonstrated a dramatic increase in the lifetime of the trapped molecules, where the lattice protected them from harmful collisions. Molecular lifetimes up to 700 ms were observed.

\begin{figure}
\includegraphics[scale=0.6]{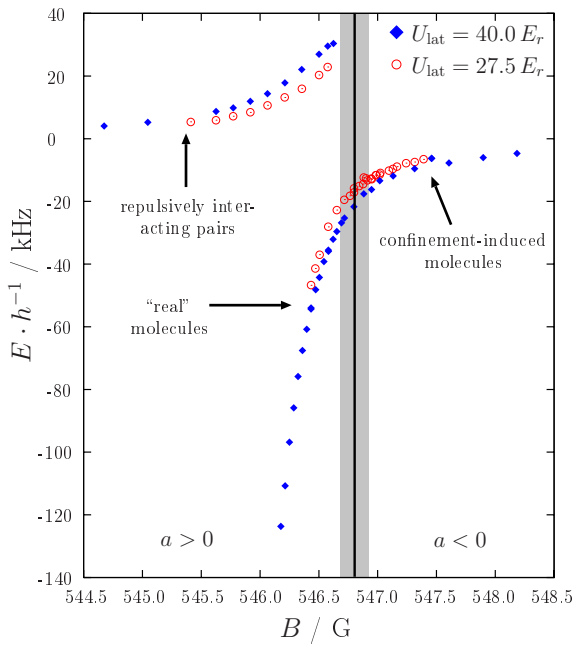}
\caption{Energy of heteronuclear $^{40}$K - $^{87}$Rb molecules versus $B$ in an optical lattice for two different lattice depths $U_{\mathrm{lat}}$ in units of the $^{87}$Rb recoil energy
$E_r=\hbar^2 k^2 / 2m_{\mathrm{Rb}}$ ($E_r/h\approx 3.7$ kHz). The center of the Feshbach resonance is located at $B_0=546.8(1)$ G. The zero of energy corresponds to two atoms in the lattice cell with $B$ far from $B_0$.    Molecules which are stable in free space are observed for $B <B_0$ where $a >0$.   Confinement-induced molecules are observed for $B > B_0$ where $a<0$ and no threshold bound state exists in free space.   In addition, repulsively interacting pairs are observed for $B<B_0$.
 From \cite{Ospelkaus2006b}.  } \label{fig:KRbLat}
\end{figure}

\cite{Ospelkaus2006b} were able to make heteronuclear molecules by
associating atoms of two different species, $^{40}$K and $^{87}$Rb,
trapped on the same lattice site.  They used a rf association
technique both to form the molecule and to measure its binding
energy. Figure~\ref{fig:KRbLat} shows the energy of the
near-threshold states of the atom pair as a function of $B$ and
illustrates an avoided crossing similar to that shown in
Fig.~\ref{fig:FeshNa}. \cite{Deuretzbacher2008} developed a
theoretical model to account for anharmonic corrections, which
couple center-of-mass and relative motion of the atoms in the trap.

\subsubsection{Reduced dimensional scattering}
\label{sssec:ola_lowd}

Optical lattices that confine atoms in only one or two directions in
combination with magnetic Feshbach resonances lead to controllable
quasi-2D or quasi-1D scattering, respectively.  ~\cite{Yurovsky2008} have recently reviewed such reduced dimensional scattering.  By integrating out
the confined spatial direction, effective one- and two-dimensional
atom-atom potentials can be derived.  Their strength is related
to the magnetic-field dependent scattering length for free
scattering.

\cite{Olshanii1998,Bergeman2003} derived the effective atom-atom
potential for quasi one-dimensional scattering.  As in
\cite{Busch1998} the starting point is a regularized
three-dimensional delta-function potential for the atom-atom
interaction potential. The trapping potential along the two confined
dimensions is the same and harmonic with frequency $\omega_\perp$.
They find that for an atom pair in the lowest harmonic oscillator
state of the confined directions the atoms interact via a
one-dimensional delta-function potential $g_{\rm 1D} \delta(z)$,
with coupling constant
\[
  g_{\rm 1D}= 2 \frac{\hbar^2}{\mu} \frac{a}{\sigma^2_\perp}\frac{1}{1-Ca/\sigma_\perp}\,,
\]
where $C=1.4602\dots$ and
$\sigma_\perp=\sqrt{\hbar/(\mu\omega_\perp)}$. The coupling constant
is singular when $a=\sigma_\perp/C$ and approaches the negative
finite value $g_\infty=-2\hbar^2/(\mu C\sigma_\perp)$ for
$a\to\pm\infty$.  \cite{Olshanii1998} has called the singularity a confinement-induced resonance.  In practice,
the resonance condition can be fulfilled by changing $a$ with a
magnetic Feshbach resonance.
For fermionic atoms in quasi one-dimensional confinement the effective
atom-atom potential has been derived by \cite{Granger2004}.

\cite{Petrov2000,Petrov2001} derived a
similar coupling constant for one-dimensional confinement or
two-dimensional scattering. In this case the resonance location not only
depends on $a$ but also logarithmically on the relative wavenumber
between the atoms along the two free spatial directions.  \cite{Naidon2007} describe how these reduced dimensional treatments can be extended to much tighter confinements than previously thought and made more accurate by using the energy-dependent scattering length of  ~\cite{Blume2002} with the effective range expansion for the scattering phase.

\begin{figure}
\epsfxsize=3.3 in \epsfbox{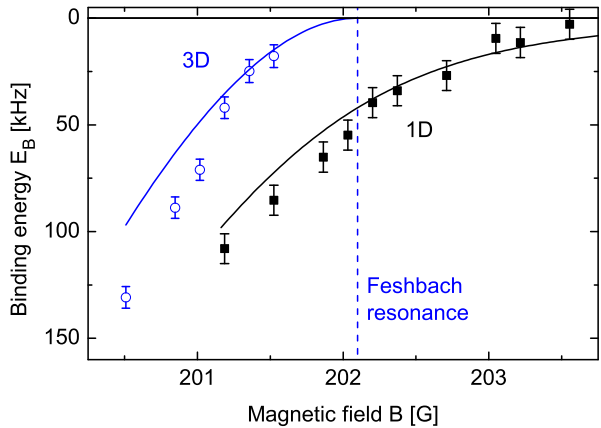} \caption{Measured molecular
bound state binding energy versus magnetic field $B$ near the
202.1-G resonance between the two lowest spin states of $^{40}$K.
The upper curve labeled 3D is for a free space gas, whereas the
lower curve labeled 1D is for a quasi-1D gas with tight confinement
in two dimensions. The points show measured binding energies and the
lines show theoretical predictions.  From \cite{Moritz2005}.}
  \label{fig:Moritz1D}
\end{figure}

\cite{Moritz2005} presented experimental evidence for a
confinement-induced bound state in a one-dimensional system. They
confirmed the existence of the bound state of the one-dimensional
Hamiltonian $H_{\rm 1D}=-\hbar^2/(2\mu) d^2/dz^2+g_{\rm 1D}
\delta(z)$ by changing both $a$ and $\omega_\perp$. { The
experiment was performed by employing an array of 1D tubes, each
containing} about 300 $^{40}$K atoms, equally divided between the
$|f=9/2,\, m=-9/2\rangle$ and $|f=9/2,\, m=-7/2\rangle$ hyperfine
state, were held in a two-dimensional harmonic trap with frequency
$\omega_\perp/(2\pi)= 69$ kHz.  The scattering length for the
collision between these two hyperfine state was varied using the
magnetic Feshbach resonance at $B_0=202$ G.  Figure
\ref{fig:Moritz1D} shows that the measured energy of the tightly
confined atom pair varies as predicted by theory.
Confinement-induced molecules exist in reduced dimension for
$B>B_0$, where they do not exist in free space.
\cite{Dickerscheid2005a} develop an analytical nonperturbative
two-channel theory of the binding energy that is in excellent
agreement with the data.

In \cite{Gunter2005} fermionic $^{40}$K atoms were prepared in a
single hyperfine state and held in either a one- or two-dimensional
optical lattice. By virtue of Fermi statistics the atoms can only
collide via odd partial waves, which for ultracold collision
energies have very small cross sections. One might expect the atomic gas to be a noninteracting Fermi gas. Nevertheless, \cite{Gunter2005} could
observe a Feshbach resonance in the $p$-wave collision by the losses
it induced.  The losses were even sensitive to the orientation of the magnetic field relative to the principal axis of the trap.  Moreover, they showed that the magnetic-field location of the resonance is modified by the confinement.
%In \cite{Stoferle2006} the same group also showed that atom pairs in a %three-dimensional optical lattice
%are stable with respect to collisions.

\cite{Nygaard2008a} investigate the effect of tuning a narrow Feshbach resonance across the Bloch band of a one-dimensional optical lattice for the case when the resonance width is small compared to the width of the band, such as the 414 G $^{87}$Rb resonance studied by ~\cite{Syassen2007}.  They investigate the changes in scattering and bound states due to the band structure in the periodic structure and characterize the time-dependent dynamics of sweeping the resonance across the band.    \cite{Nygaard2008b} extend this work, develop the concept of a generalized scattering length at the band edges, and show the existence of a ''universal'' bound state near the top and bottom band edges at the field strength where the resonance emerges from the band.

\subsubsection{Scattering in shallow lattices}
\label{sssec:ola_scat}

The previous two Sections \ref{sssec:ola_mol} and \ref{sssec:ola_lowd}
discussed deep optical lattices where the tunneling between lattices
sites could safely be neglected. For weaker optical lattices, atoms
tunnel from site to site and then interact with all other atoms.
This leads to many-body systems that can be described by either
a mean-field Gross-Pitaevskii equation or a Bose- or Fermi-Hubbard
Hamiltonian~\cite{Bloch2005}.  The presence of Feshbach resonances has
added and continues to add new twists to these kinds of Hamiltonians.

\begin{figure}
\includegraphics[width=2.8in,angle=270,origin=lB]{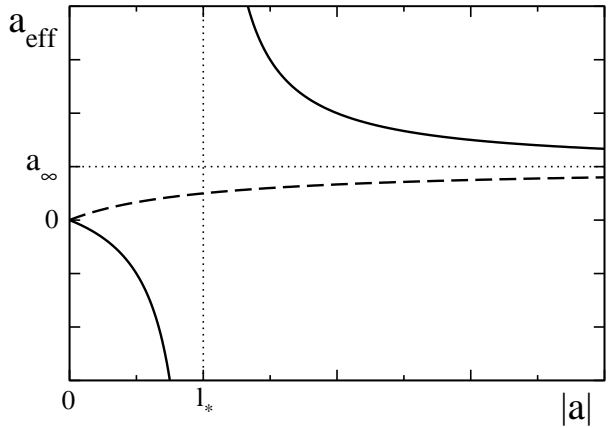}
\caption{The dependence of the effective scattering length
$a_{\mathrm{eff}}$ on the absolute value of the microscopic
scattering length $|a|$. The dashed line corresponds to a repulsive
interaction potential ($a>0$). The solid line corresponds to an
attractive potential ($a<0$) with a geometric resonance at
$|a|=l_{*}$. From \cite{Fedichev2004}. \label{fig:Geo}}
\end{figure}

We will discuss the more simple situation where only two atoms
scatter in an optical lattice. \cite{Fedichev2004} studied the case
of two atoms scattering in a weak three-dimensional optical lattice
of cubic symmetry and interacting with the {regularized}
delta-function potential.  They predict the presence of a
geometrical resonance in the 3D lattice based on a derivation of an
effective atom-atom interaction between the atoms, which by virtue
of the periodic {potential} have an effective mass $m^*$
that is much larger than their atomic mass $m$. Figure~\ref{fig:Geo}
shows the results of their calculation. The resonance occurs at a
scattering length $a=l^*\equiv-(\pi/(2\ln2))(m/m^*)(\sigma/d)^2
\sigma$, where $\sigma$ is the atomic harmonic oscillator length for
motion in a single lattice site and $d$ is the lattice period.  Note
that, in practice, $l^*\ll\sigma$. \cite{Orso2005} presented a
similar analysis for a one-dimensional optical lattice.
\cite{Grupp2007} studied the effect of a very-narrow Feshbach
resonance in scattering in an one-dimensional lattice.

\subsection{Efimov states and universal few-body physics}
\label{ssec:efimov}

Feshbach resonances provide experimental access to systems with very
large values of the scattering length. Such systems are governed by
``universal'' physics, i.e.\ { their low-energy observables are}
independent of details of the interaction \cite{Braaten2006}.
Universality appears as a consequence of the quantum-halo character
of the wave function carrying its dominant part far out of the
classically allowed region. In this case, details of the interaction
potential become irrelevant and the system can be described by
{a few global} parameters. Halo dimers
(Sec.~\ref{sssec:halo}) are the most simple example. For addressing
universal physics with ultracold gases, Feshbach resonances that are
strongly entrance-channel dominated ($s_{\rm res} \gg 1$) are of
particular interest, as they allow a description in terms of a
single-channel model with a large range of universal behavior; see
discussion in Sections~\ref{sssec:resstrength} and
\ref{sssec:AnalMol}.

\subsubsection{Efimov's scenario}
\label{sssec:efimovscenario}

Efimov quantum states in a system of three identical bosons
\cite{Efimov1970, Efimov1971} are a paradigm for universal few-body
physics. These states have attracted considerable interest, fueled
by their bizarre and counter-intuitive properties and by the fact
that they had been elusive to experimentalists for more than 35
years. In 2006, \cite{Kraemer2006} reported on experimental evidence
for Efimov states in an ultracold gas of cesium atoms. By Feshbach
tuning they could identify a pronounced three-body resonance, which
occurs as a fingerprint of an Efimov state at the three-body
scattering threshold. Two years later, \cite{Knoop2008} presented additional evidence for Efimov-like trimer states, reporting on the observation of a decay resonance in atom-dimer scattering.

\begin{figure}
\includegraphics[width=3in]{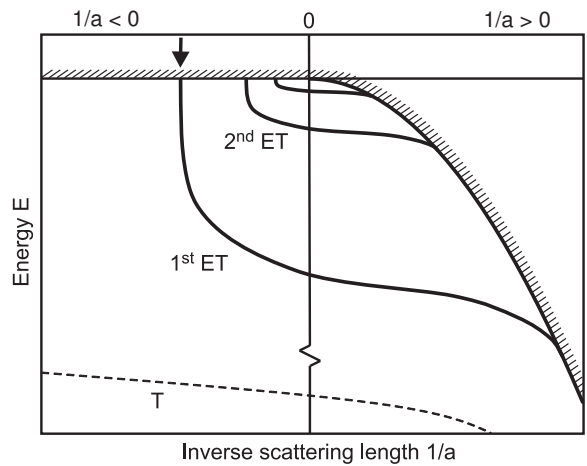}
\caption{Efimov's scenario: Appearance of an infinite series of
weakly bound Efimov trimer states (ET) for resonant two-body
interaction. The binding energy is plotted as a function of the
inverse two-body scattering length $1/a$. The shaded region
indicates the scattering continuum for three atoms ($a<0$) and for
an atom and a dimer ($a>0$). The arrow marks the intersection of the
first Efimov trimer with the three-atom threshold. To illustrate the
series of Efimov states, the universal scaling factor is
artificially reduced from $22.7$ to $2$. For comparison, the dashed
line indicates a more tightly bound non-Efimov trimer (T) which does not
cross the scattering continuum. Adapted by permission from Macmillan Publishers Ltd: Nature \cite{Kraemer2006}, copyright 2006.}
\label{fig:efimovscenario}
\end{figure}

Efimov's scenario is illustrated in Fig.~\ref{fig:efimovscenario},
showing the energy spectrum of the three-body system as a function
of the inverse scattering length $1/a$. For $a<0$, the natural zero
of energy is the three-body dissociation threshold for three atoms
at rest. States below are trimer states and states above are
continuum states of three free atoms. For $a>0$, the dissociation
threshold is given by $-E_b = -\hbar^2/(ma^2)$ where $E_b$ is the
universal binding energy of the weakly bound halo dimer; at this
threshold a trimer dissociates into a dimer and an atom. All states
below threshold are necessarily three-body bound states. Efimov
predicted that in the limit $a \to \pm \infty$ there would be an
infinite sequence of weakly bound trimer states with a universal
scaling behavior. Each successive Efimov state is larger in size by
a universal scaling factor $e^{\pi/s_0} \approx 22.7$ ($s_0 =
1.00624$) and has a weaker binding energy by a factor of  $(22.7)^2
\approx 515$.

Efimov states exist on both sides of a resonance, and
Fig.~\ref{fig:efimovscenario} shows the adiabatic connection between
both sides. For $a>0$, an Efimov state near the atom-dimer
dissociation threshold can be regarded as a weakly bound state of an
atom and a dimer with a size set not by $a$ but by the even larger
atom-dimer scattering length \cite{Braaten2006}. For $a<0$, Efimov
states are ``Borromean'' states \cite{Jensen2004}, which means that
a weakly bound three-body state exists in the absence of a weakly
bound two-body state. This property that three quantum objects stay
together without pairwise binding is part of the bizarre nature of
Efimov states.

Resonant scattering phenomena arise as a natural consequence of this
scenario \cite{Efimov1979}, and they are closely related to the
basic idea of a Feshbach resonance. When an Efimov state intersects
with the continuum threshold for $a<0$, three free atoms resonantly
couple to a trimer. This results in a ``triatomic Efimov
resonance''. When an Efimov state intersects with the atom-dimer threshold for $a>0$, the result is an ``atom-dimer Efimov resonance" \cite{Nielsen2002}.

\subsubsection{Observations in ultracold cesium}
\label{sssec:efimov_Cs}

In an ultracold atomic gas with resonant interactions, Efimov physics manifests itself in three-body decay properties \cite{Nielsen1999, Esry1999, Bedaque2000, Braaten2001, Braaten2006}. The three-body loss coefficient $L_3$ (Sec.~\ref{sssec:lossspect}) can be conveniently
expressed in the form $L_3 = 3 \, C(a) \hbar a^4 /m$, which
separates an overall $a^4$-scaling from an additional dependence
$C(a)$. Efimov physics is reflected in a logarithmically periodic
behavior $C(22.7a) = C(a)$, corresponding to the scaling of the
infinite series of weakly bound trimer states. A triatomic Efimov
resonance leads to giant recombination loss \cite{Esry1999, Braaten2001},
as the resonant coupling of three atoms to an Efimov state opens up fast decay
channels into deeply bound dimer states plus a free atom.

\begin{figure}
\includegraphics[width=3.4in]{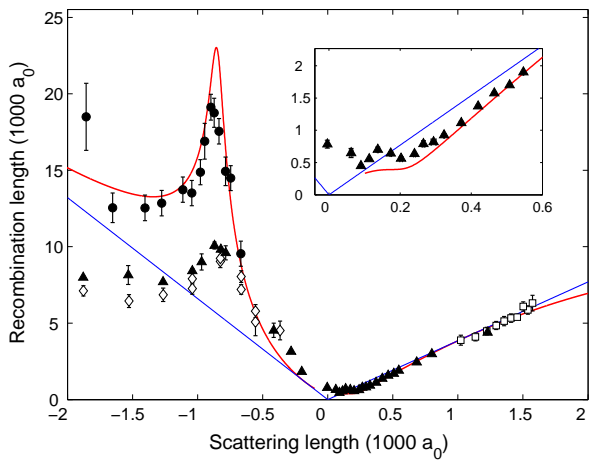}
\caption{Observation of an Efimov resonance in three-body decay of
an ultracold gas of cesium atoms. The data are presented in terms of
a recombination length $\rho_3 = [2m/(\sqrt{3}\hbar)\,L_3]^{1/4}$
\cite{Esry1999}. The general $a^4$-scaling of $L_3$ corresponds to a
linear behavior in $\rho_3(a)$ (straight lines). The filled circles represent
measurements taken at temperatures around 10\,nK, whereas the filled
triangles and open diamonds refer to measurements in the range of
200\,nK - 250\,nK. The solid line is a fit to the low-temperature data
based on effective-field theory \cite{Braaten2006}. The inset shows an expanded view of the region of positive scattering lengths up to 600\,$a_0$. Reprinted by permission from Macmillan Publishers Ltd: Nature \cite{Kraemer2006}, copyright 2006.} \label{fig:efimovres}
\end{figure}

\cite{Kraemer2006} observed a triatomic Efimov resonance in an
ultracold, thermal gas of Cs atoms. They made use of the strong
variation in the low-field region (Fig.~\ref{fig:cs_aa_slength}).
This tunability results from a strongly entrance-channel dominated
resonance at $-12$\,G with $s_{\rm res}=566$
(Table~\ref{tab:resonances}), which provides a broad range of
universal behavior. By applying magnetic fields between $0$ G and
$150$\,G, \cite{Kraemer2006} varied the $s$-wave scattering length
$a$ between $-2500\,a_0$ and $1600\,a_0$, large enough to study the
universal regime, which requires $|a| \gg R_{\rm vdw} \simeq
100\,a_0$. The occurrence of one triatomic Efimov resonance could be
expected in the accessible negative-$a$ region. The position,
however, could not be predicted from knowledge of the scattering
length alone as, for a three-body process, a second parameter is
required to characterize the universal properties
\cite{Braaten2006}.

Figure~\ref{fig:efimovres} shows the results of \cite{Kraemer2006}. The three-body loss resonance was found at a magnetic field of 7.5\,G, corresponding to a scattering length of $-850\,a_0$. The behavior of loss at
temperatures around 10\,nK  closely resembles the theoretical predictions of \cite{Esry1999}, who numerically solved the three-body Schr\"odinger equation for a generic two-body model potential. The observed behavior is also
well fit with a universal analytic expression obtained in the
framework of effective-field theory \cite{Braaten2006}. Experimental
data taken at higher temperatures demonstrated the unitarity
limitation of three-body loss \cite{Dincao2004} and showed how the
Efimov resonance evolved into a triatomic continuum resonance \cite{Bringas2004}.

For positive scattering lengths, theory predicts a variation of
$C(a)$ between very small values and a maximum of about $70$
\cite{Nielsen1999, Esry1999, Bedaque2000}. The results in
Fig.~\ref{fig:efimovres} are consistent with the upper loss limit,
represented by the straight line for $a>0$. For $a$ below
$600\,a_0$, the experimentally determined recombination length
significantly drops below this limit, as seen in the inset of the
figure. Further measurements of \cite{Kraemer2006} revealed the
existence of a loss minimum at $B=21\,G$, where $a=+210\,a_0$. It is
interesting to note that earlier experiments by the same group
\cite{Kraemer2004} had identified 21\,G as an optimum magnetic field
for evaporative cooling of cesium and attainment of BEC; see
Sec.~\ref{sssec:attainment}. The nature of the minimum may be
interpreted in the framework of universal physics, following
theoretical predictions of an interference effect between two
different recombination pathways \cite{Nielsen1999, Esry1999}.
However, as the minimum occurs at a scattering length which is only
a factor of two larger than $R_{\rm vdw} \simeq 100\,a_0$ (Table
\ref{tab:vdw}) the application of universal theory to describe this
feature is questionable. \cite{Massignan2008} presented an
alternative theoretical approach, which reproduced both this minimum
and the maximum observed for negative $a$
%without invoking the basic
%assumptions of universal theory. Their approach is
{ on the basis of the two-body physics of the particular Feshbach resonance.}
%without
%the need to introduce any free parameter for
%three-body phenomena.

In a pure sample of trapped atoms, as discussed so far, three-body recombination is the only probe for Efimov physics. Mixtures of atoms and dimers can provide complementary information on Efimov states through resonances in inelastic atom-dimer collisions \cite{Nielsen2002,Braaten2007}. In a recent experiment, \cite{Knoop2008}
prepared an optically trapped mixture of Cs atoms and Cs$_2$ halo
dimers. Their measurements revealed an atom-dimer scattering
resonance, which is centered at a large value of the two-body
scattering length, $a \simeq +390\,a_0$ at a magnetic field of 25\,G. This observation provides strong evidence for a trimer state approaching the atom-dimer threshold. The situation is close to the atom-dimer resonance in Efimov's scenario, but it probably remains a semantic question whether, at $a \simeq 4 R_{\rm vdw}$, the underlying trimer state may be called an Efimov state.

In the Cs experiments described in this Section, the Efimov state
that causes the observed triatomic resonance at 7.5\,G does not
connect to the state that causes the atom-dimer resonance at 25\,G
when the magnetic field is varied. This is because these two cases
are separated by a zero crossing in the scattering length
(Fig.~\ref{fig:cs_aa_slength}) and not by the pole as in Efimov's
scenario in Fig.~\ref{fig:efimovscenario}. A universal relation
between these regimes may nevertheless exist \cite{Kraemer2006}.
\cite{Lee2007} provided a further interpretation of these
observations in terms of the underlying Cs$_3$ states and point out
the analogies to trimer states of helium \cite{Schollkopf1994}.

\subsubsection{Prospects in few-body physics}
\label{sssec:efimovprospects}

Ultracold gases with resonantly tuned interactions offer many
opportunities to study universal, Efimov-related few-body physics.
Cesium alone has much more to offer than the experiments could
explore so far. Moreover, several other systems with broad Feshbach
resonances promise new insight into this field.

In cesium, a predicted broad Feshbach resonance near 800\,G
\cite{Lee2007} in the $aa$ channel offers similar properties as the
low-field region explored in previous experiments, but overcomes the
disadvantage that only the tail of the resonance is accessible at
low fields. The broad 155-G resonance in $^{85}$Rb
(Sec.~\ref{sssec:Rb}) might be another interesting candidate, but
experiments may suffer from strong two-body decay which is absent
for the discussed cesium resonances. A further interesting candidate
is the 402-G resonance in $^{39}$K (Sec.~\ref{sssec:K}); \cite{Zaccanti2008} have studied three-body decay near this resonance and found features strongly indicative of Efimov physics.

Many more opportunities for studying Efimov-related physics in
ultracold gases with resonant interactions are offered by mixtures
of different spin states or different species. In $^6$Li, all three combinations of the lowest three spin states
(channels $ab$, $ac$, and $bc$) have broad Feshbach resonances \cite{Bartenstein2005} that overlap in a magnetic-field range between 650\,G and 850\,G. For such a three-component fermionic spin mixture \cite{Luu2007} predicted a novel Borromean three-body state. \cite{Ottenstein2008} and \cite{Huckans2008} experimentally investigated the stability of a $^6$Li three-component spin mixture and found evidence for a three-body resonance at 130\,G. { Its interpretation in terms of coupling to a three-body bound state is supported by several theoretical studies \cite{Braaten2008, Schmidt2008, Naidon2008}.}
A particularly interesting situation arises in mixtures of atoms with different masses. With increasing mass ratio the Efimov factor substantially decreases from its value of $22.7$ at equal masses to values as low as $4.9$ for the mass ratio of $^{133}$Cs combined with $^{6}$Li. \cite{Dincao2006}
pointed out that this will substantially enhance the observability
of the Efimov effect in terms of the logarithmically periodic
variation of the three-body loss coefficient with increasing
two-body scattering length.

{ Four-body processes at large values of the $s$-wave scattering length $a$ represent a logical next step in understanding universal few-body physics. Theoretical studies \cite{Hammer2007,Vonstecher2008,Wang2008} predict the existence of universal four-body states, and consider the process of atomic four-body recombination. A first experimental step into this field was made by \cite{Ferlaino2007}, who studied collisions of Cs$_2$ halo dimers at large positive $a$. They observed a loss minimum in the same region where atom-dimer scattering shows a maximum ($a \simeq 500\,a_0$ at 30\,G), which may be related to a universal connection between four- and three-body physics \cite{Hammer2007,Vonstecher2008}.}

Optical lattices have proven a powerful tool for the manipulation of
ultracold Feshbach molecules, see Sec.~\ref{sssec:ola_mol} and may also open up new possibilities for the creation of Efimov trimers and, more generally, for the controlled production of few-body quantum states \cite{Stoll2005,Luu2007}.

\subsection{Molecular resonances and cold chemistry}
\label{ssec:molchem}

While we have reviewed the formation of cold molecules from
ultracold atoms, parallel progress has been made by other techniques
for preparing samples of cold molecules that extend the range far
beyond alkali-metal species.  These advances have been made possible
by Stark deceleration of molecules  such as ND$_3$, OH, and
formaldehyde~\cite{Meerakker2006} or by buffer gas cooling with
liquid helium \cite{DeCarvalho1999}.  In contrast to the association
of cold atoms to make Feshbach molecules, which have a high level of
vibrational excitation,  these methods can produce cold molecules in
the rotational and vibrational ground state.   See
\cite{Hutson2007,Polar2004,Krems2005} for overviews of the issues
involved in trapping, cooling, and colliding such molecules.

Resonances will play a very prominent role in atom-molecule and molecule-molecule collisions.  The complexity of these systems will increase the number of closed channels and lead to numerous resonances with diverse properties.   Their presence will
make it possible to change molecular scattering properties as well as to
create more complex molecules.  Both static magnetic and
electric fields can be used to tune the molecule-molecule resonance
and provide control over collisions, as many molecules not
only have a magnetic moment but an electric dipole moment as well.
Here we will briefly review some of this work.

\cite{Forrey1998} pointed out that Feshbach resonances occur in ultracold atom-diatom scattering, giving an example from collisions of H$_2$ with He. The effect of resonance states in chemical reactions has been studied by ~\cite{Balakrishnan2001} for F+H$_2$$\to$FH+H and~\cite{Weck2005} for
Li$+$HF$\to$H$+$LiF reactions.   Recent coupled channels models of Rb$+$OH~\cite{Lara2007} and He$+$NH~\cite{Gonzalez2007} collisions have been developed to give a realistic assessment of atom-molecule scattering.  The latter study demonstrated the effect of magnetic tuning of a decaying resonance across a threshold, using the resonance length formalism described in Sec.~\ref{sssec:resscatt}.

In 2002 \cite{Bohn2002} realized that, unlike for atomic systems
where Feshbach resonances originate from the hyperfine structure of
the atoms, for molecules the resonances can also be due to
rotational states.  For many molecules the rotational spacing for
low-lying rotational levels is of the same order of magnitude as
hyperfine interactions in atoms.  Based on the rotational splittings
and a potential energy surface \cite{Bohn2002} estimated the mean
spacing and widths of the resonances and found for collisions
between oxygen molecules as many as 30 resonances for collision
energies below $E/k_B=1$ K.

\begin{figure}
\includegraphics[width=2.9in]{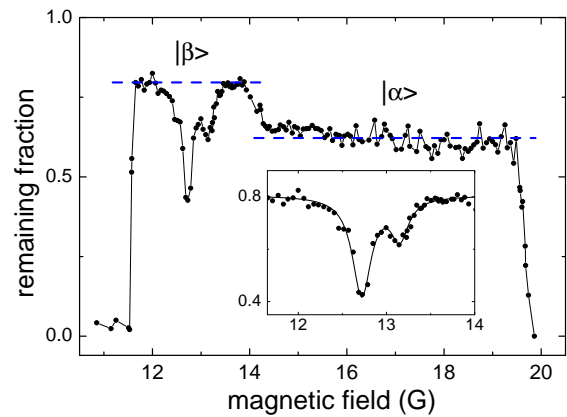}
\caption{Observation of Feshbach-like resonances in the collisions
of $^{133}$Cs$_2$ molecules. The figure shows the number of
remaining $^{133}$Cs molecules after fixed storage time in an optical trap
as a function of external magnetic field. The two features near 13 G shown
magnified in the inset have been
interpreted as tetramer resonances. From \cite{Chin2005}.} \label{fig:Cs4}
\end{figure}

\cite{Chin2005} observed magnetic Feshbach-like resonances  between
two weakly bound $^{133}$Cs$_2$ molecules that temporarily form a
tetramer during a collision. Their data are shown in
Fig.~\ref{fig:Cs4}. The $^{133}$Cs$_2$ molecules in this experiment
are bound by no more than $E/h$=5 MHz and have a temperature and
peak density of 250 nK and $5\times10^{10}$ cm$^{-3}$, respectively.
As the magnetic field is varied near $B=$13 G the lifetime of the
molecules rapidly changes, indicating two resonances.

Heteronuclear molecules can be manipulated by static electric fields
in addition to magnetic fields. The electric field Stark shifts the
rotational levels of the molecule. These level shifts can then give
rise to electric Feshbach resonances~\cite{Avdeenkov2002}.  In atoms
the levels can also be sufficiently Stark shifted to induce
collisional resonances but rather large fields are
required~\cite{Marinescu1998}.

\begin{figure}
\includegraphics[scale=.85]{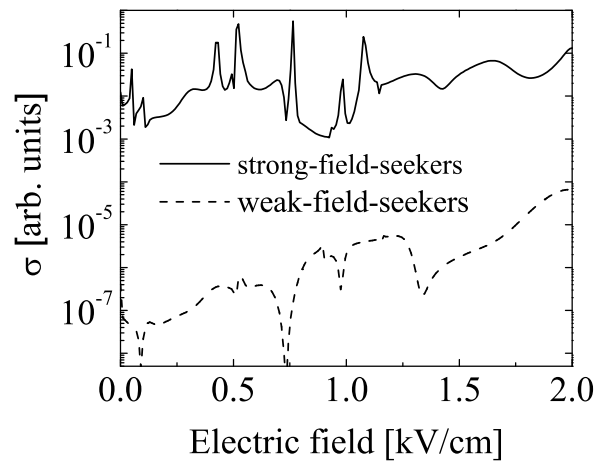}
\caption{Calculated chemical reaction cross section of
H$_2$CO+OH$\to$ HCO+H$_2$O at a collision energy of $E/k_B=$1 mK as
a function of applied electric field. Strong-/weak-field seekers
correspond to states of H$_2$CO and OH that can be held in an
electric trap where the center of the trap has the larger/smallest
electric field strength, respectively. Figure and model are from
\cite{Hudson2006}.} \label{fig:formaldehyde}
\end{figure}

Figure~\ref{fig:formaldehyde} shows the results of a calculation on
reaction rate coefficients of formaldehyde H$_2$CO reacting with
 OH to yield HCO and H$_2$O \cite{Hudson2006}.  Both H$_2$CO
and OH are in their lowest vibrational state of their ground
electronic configuration. Multiple resonances occur for electric
fields up to 2 kV/cm. More recently, \cite{Tscherbul2008} have studied
reaction rates of LiF with H in the presence of electric fields.

We can expect atom-molecule and molecule-molecule collisions to exhibit a
rich variety of resonance phenomena in elastic, inelastic, and reactive
collisions.  Such phenomena are likely to become progressively more
important to understand as sources of trap loss and for coherent control
of molecular ensembles.

\section*{Acknowledgments}
Over the past decade, many people have contributed to advancing our
knowledge on Feshbach resonances in ultracold gases. For valuable
discussions and insights related to this exciting field, we
particularly acknowledge B.\ Esry, F.\ Ferlaino, C. Greene, J.\
Hecker Denschlag, J.\ Hutson, S.\ Knoop, H.-C.\ N\"agerl, W.\
Phillips, F.\ Schreck, and G.\ Shlyapnikov. We thank E.\ Braaten, S.\
D\"urr, T.\ Esslinger, M.\ Gustavsson, L.\ Khaykovich, H.\ Moritz, W.\
Ketterle, S.\ Kokkelmans, T.\ Pfau, H.\ Stoof, B.\
Verhaar, J.\ Walraven, M.\ Zaccanti, C.\ Zimmermann, and in particular I.\ Spielman and S.\ Jochim for helpful comments on the manuscript.

C.C.\ acknowledges support from NSF No. PHY-0747907, NSF-MRSEC
DMR-0213745, and ARO No. W911NF0710576 with funds from the DARPA OLE
Program. R.G.\ thanks the Austrian Science Fund FWF and the Austrian
Science ministry BMWF for support. P.J.\ and E.T.\ acknowledge
support by the Office of Naval Research (ONR).

\appendix \section*{Tables of selected resonances}

\begin{table*}
\caption{Properties of selected Feshbach resonances. The first
column describes the atomic species and isotope. The next three
columns characterize the scattering and resonance state, which
includes the incoming scattering channel (ch.), partial wave $\ell$,
and the angular momentum of the resonance state $\ell_{\rm c}$. This
is followed by the resonance location $B_0$, the width $\Delta$, the
background scattering length $a_{\rm bg}$, the differential magnetic
moment $\delta\mu$, the dimensionless resonance strength
$s_\mathrm{res}$, the background scattering length in van der Waals
units $r_\mathrm{bg}=a_{\rm bg}/\bar{a}$, and the bound state
parameter $\zeta$ from Eq.~(\ref{zeta}). Here $a_0$ is the Bohr
radius and $\mu_B$ is the Bohr magneton. Definitions
 are given in Sec.~\ref{sec:theory}. The last column gives the source. A string ``na'' indicates that the corresponding property is not defined. For
example $a_\mathrm{bg}$ is not defined for $p$-wave scattering.}
%\begin{center}

\begin{ruledtabular}
\begin{tabular}{lcccrrrrrrrl}
  atom&ch.& $\ell$ & $\ell_{\rm c}$ & $B_0$(G) & $\Delta$(G) & $a_\mathrm{bg}/a_0 $ & $\delta\mu/\mu_{B}$ & $s_\mathrm{res}$ & $r_\mathrm{bg}$ & $\zeta$ & reference\\ \hline
  $^6$Li       &$ab$ & $s$ & $s$ & 834.1 & -300    & -1405 & 2.0   & 59 & -47 & 1400  & \cite{Bartenstein2005}\\
               &$ac$ & $s$ & $s$ & 690.4 & -122.3  & -1727 & 2.0   & 29 & -58 & 850   & \cite{Bartenstein2005}\\
               &$bc$ & $s$ & $s$ & 811.2 & -222.3  & -1490 & 2.0   & 46 & -50 & 1200  & \cite{Bartenstein2005}\\
               &$ab$ & $s$ & $s$ & 543.25 & 0.1     & 60    & 2.0   & 0.001 & 2.0 & 0.001 & \cite{Strecker2003}  \\
               &$aa$ & $p$ & $p$ & 159.14 & na      &  na   & 2.0   & na & na & na  & \cite{Schunck2005, Zhang2004}\\
               &$ab$ & $p$ & $p$ & 185.09 & na      &  na   & 2.0   & na & na & na  & \cite{Schunck2005, Zhang2004}\\
               &$bb$ & $p$ & $p$ & 214.94 & na      &  na   & 2.0   & na & na & na  & \cite{Schunck2005, Zhang2004}\\

  $^7$Li       &$aa$ & $s$ & $s$ & 736.8  & -192.3      & -25   & 2.0   & 0.80 & -0.79 & 0.31 & \cite{Pollack2008, Strecker2002} $^a$ \\    \hline

  $^{23}$Na    &$cc$ & $s$ & $s$ & 1195   & 1.4    & 62    & -0.15 & 0.0050 & 1.4 & 0.004 & \cite{Stenger1999, Inouye1998} $^a$\\
               &$aa$ & $s$ & $s$ & 907    & 1       & 63    & 3.8   & 0.09   & 1.5 & 0.07   & \cite{Stenger1999, Inouye1998} $^a$\\
               &$aa$ & $s$ & $s$ & 853    & 0.0025  & 63    & 3.8   & 0.0002 & 1.5 & 0.0002 & \cite{Stenger1999, Inouye1998} $^a$\\    \hline

  $^{39}$K     &$aa$ & $s$ & $s$ & 402.4  & -52     & -29   & 1.5   & 2.1    & -0.47 & 0.49 &\cite{Derrico2007} \\
  $^{40}$K     &$bb$ & $p$ & $p$ & 198.4  &  na    & na     & 0.134 & na  & na  &  na & \cite{Regal2003,Ticknor2004}$^a$ \\
               &$bb$ & $p$ & $p$ & 198.8  &  na    & na     & 0.134 & na  & na  &  na & \cite{Regal2003,Ticknor2004}$^a$ \\
               &$ab$ & $s$ & $s$ & 202.1  & 8.0    & 174    & 1.68  & 2.2 & 2.8 & 3.1 & \cite{Regal2004} $^a$\\
               &$ac$ & $s$ & $s$ & 224.2  & 9.7    & 174    & 1.68  & 2.7 & 2.8 & 3.8 & \cite{Regal2003b} $^a$\ \\    \hline

  $^{85}$Rb    &$ee$ & $s$ & $s$ & 155.04 & 10.7   & -443   & -2.33 & 28  & -5.6 & 80 & \cite{Claussen2003} \\

  $^{87}$Rb    &$aa$ & $s$ & $s$ & 1007.4 & 0.21   & 100   & 2.79 & 0.13   & 1.27 & 0.08   & \cite{Durr2004b, Volz2003} $^a$\\
               &$aa$ & $s$ & $s$ & 911.7  & 0.0013 & 100   & 2.71 & 0.001  & 1.27 & 0.0006 & \cite{Marte2002} $^a$\\
               &$aa$ & $s$ & $s$ & 685.4  & 0.006  & 100   & 1.34 & 0.006  & 1.27 & 0.004 & \cite{Durr2004b, Marte2002} $^a$\\
               &$aa$ & $s$ & $s$ & 406.2  & 0.0004 & 100   & 2.01 & 0.0002 & 1.27 & 0.0001 & \cite{Marte2002} $^a$\\
               &$ae$ & $s$ & $s$ & 9.13   & 0.015  & 99.8  & 2.00 & 0.008  & 1.27 & 0.005  & \cite{Widera2004} \\        \hline

  $^{133}$Cs   &$aa$ & $s$ & $s$ & -11.7 & 28.7  & 1720  & 2.30  & 560 & 17.8 & 5030  & \cite{Chin2004, Lange2008} $^a$\\
               &$aa$ & $s$ & $d$ & 47.97  & 0.12   & 926   & 1.21  & 0.67 & 9.60& 3.2   & \cite{Chin2004, Lange2008} $^a$\\
%               &$aa$ & $s$ & $g$ & 11.02  & -0.005  & -450  & 1.05 & 0.012 & -4.66 & -0.027 & \cite{Chin2004} $^a$\\
%               &$aa$ & $s$ & $g$ & 14.4   & -0.01   & -174  & 0.92 & 0.0080 & -1.80 & -0.0072 & \cite{Chin2004}  $^a$\\
%               &$aa$ & $s$ & $g$ & 15.1   & -0.003  & -121  & 0.26 & 0.00047 & -1.25& -0.0003 & \cite{Chin2004} $^a$\\
               &$aa$ & $s$ & $g$ & 19.84  & 0.005  & 160   & 0.57 & 0.002 & 1.66 & 0.002 & \cite{Chin2004} $^a$\\
               &$aa$ & $s$ & $g$ & 53.5   & 0.0025 & 995   & 1.52 & 0.019 & 10.3 & 0.1  & \cite{Chin2004, Lange2008} $^a$\\
               &$aa$ & $s$ & $s$ & 547    & 7.5    & 2500  & 1.79 & 170& 26 & 2200  & $^a$\\
               &$aa$ & $s$ & $s$ & 800    & 87.5   & 1940  & 1.75 & 1470& 20 & 15000  & $^a$\\    \hline
  $^{52}$Cr    &$aa$ & $s$ & $d$ & 589.1  & 1.7    & 105   & 2.00 & 0.31  & 2.45 & 0.38 & \cite{Werner2005} $^a$ \\
               &$aa$ & $s$ & $d$ & 499.9  & 0.08   & 107   & 4.00 & 0.03  & 2.49 & 0.04 & \cite{Werner2005} $^a$\\
%               &$aa$ & $s$ & $d$ & 290.3  & 0.05   & 105   & 4.00 & 0.02  & 2.45 & 0.02 & \cite{Werner2005} $^a$\\
\hline
  $^6$Li$^{23}$Na    & $aa$ & $s$ & $s$ & 746   &  0.44 & 14.0 &   &  &  &  & \cite{Stan2004}, \cite{Gacesa2008}\\
%                     & $aa$ & $s$ & $s$ & 759.6 &  0.31 & 13.9 &   &  &  &  & \cite{Stan2004}, \cite{Gacesa2008}\\
                     & $aa$ & $s$ & $s$ & 795.6 &  2.177& 13.0 &   &  &  &  & \cite{Stan2004}, \cite{Gacesa2008}\\  \hline
  $^6$Li$^{40}$K     & $aa$ & $s$ & $s$ & 157.6 & 0.25  &      &   &  &  &  & \cite{Wille2007} \\
                     & $aa$ & $s$ & $s$ & 168.2 & 0.15  &      &   &  &  &  & \cite{Wille2007} \\
%                     & $aa$ & $p$ & $p$ & 249   & 0.1  &      &   &  &  &  & \cite{Wille2007}\\
\hline
  $^6$Li$^{87}$Rb    & $aa$ & $p$ & $p$ & 882   & na  &  na    &   & na  & na & na & \cite{Deh2008, Li2008}\\
                     & $aa$ & $s$ & $s$ & 1067  & 10.62 &      &   &  &  &  & \cite{Deh2008, Li2008}\\
  $^7$Li$^{87}$Rb    & $aa$ & $s$ & $s$ & 649  &  -70  &  -36    &   &  &  &  & \cite{Marzok2008}\\  \hline
$^{39}$K$^{87}$Rb    & $aa$ & $s$ & $s$ & 317.9 & 7.6& 34  &  2.0  & 0.74 & 0.50 & 0.18 &\cite{Simoni2008}\\
$^{40}$K$^{87}$Rb    & $aa$ & $s$ & $s$ & 546.9 & -3.10 & -189 & 2.30  & -1.96 & -2.75 & 2.70 & \cite{Pashov2007,Simoni2008}\\
$^{41}$K$^{87}$Rb    & $aa$ & $s$ & $s$ & 39   & 37   & 284  & 1.65 & 25.8 & 4.11 & 53.0 &\cite{Thalhammer2008,Simoni2008}\\
$^{41}$K$^{87}$Rb    & $aa$ & $s$ & $s$ & 79 & 1.2 & 284  & 1.59  &
0.81 & 4.11 & 1.66 &\cite{Thalhammer2008,Simoni2008}\\ \hline
$^{85}$Rb$^{87}$Rb   & $ec$ & $s$ & $s$ &265.4 & 5.8 & 213  &  & & & &\cite{Papp2006}\\
$^{85}$Rb$^{87}$Rb   & $ec$ & $s$ & $s$ & 372  & 1   & 213  &  & & &
&\cite{Papp2006} \\  \hline
%\multicolumn{9}{l}{$^j$ \cite{durr2004, Marte2002}} \\
%\multicolumn{9}{l}{$^n$ \cite{Widera2004}} \\
%\multicolumn{9}{l}{$^l$ \cite{Werner2005}, $^k$ \cite{Chin2004}} \\
%\multicolumn{10}{l}{$^a$ Table entries partially based on unpublished calculations by the authors.} \\
%\multicolumn{8}{l}{$^n$ \cite{Stan2004}} \\
%\multicolumn{8}{l}{$^o$ \cite{Inouye2004}} \\
%\multicolumn{8}{l}{$^p$ \\
\end{tabular}
$^a$ Table entries partially based on unpublished calculations by
the authors.
\end{ruledtabular}

%\end{center}
\label{tab:resonances}
\end{table*}

Table \ref{tab:resonances} lists positions and properties of
resonances for various species. The data is a combination of
experimentally determined as well as theoretically derived values.
Most of the magnetic field locations are experimentally determined.
Most of the widths $\Delta$ and background scattering lengths
$a_{\rm bg}$ are determined from theoretical calculations. Where
unavailable from the literature we calculated values based on the
best Born-Oppenheimer potentials obtained from the literature. For a
complete list as well as error bars on the resonance locations the
reader is referred to the literature.  The notation defined in
Secs.~\ref{sssec:resscatt} and \ref{sssec:feshcoupl} has been
used in the table.

The table shows a richness in the kinds of resonances available for
magnetic field values that are relatively easily created in
laboratories. Some of the resonances are very narrow with $\Delta$
on the order of a mG. Others are very broad with $\Delta$ larger
than 100 G. The background scattering length can be either negative
or positive, its absolute value ranging from a few tens to several
thousands Bohr radii. The magnetic moment of the resonance state is
always on the order of the Bohr magneton, which reflects the form of
the Zeeman interaction. The partial wave of the resonance states
ranges from zero to four ($\ell_c = 0...4$). Finally, the resonances are characterized in terms of their background
scattering lengths $a_{\rm bg}$, their strengths $s_{\rm res}$
(Sec.~\ref{sssec:resstrength}), and the parameter $\zeta$
(Sec.~\ref{sssec:AnalMol}).

For atomic cesium, a resonance location  is given with a negative
magnetic field value. This is not an experimental value. Here $B_0$
is determined from a fit of Eq.~(\ref{II.A.21}) to the slowly
varying scattering length as shown in Fig.~\ref{fig:cs_aa_slength}.
\cite{Vogels1998} give the physical interpretation of a negative
$B_0$, namely, taking $B<0$ corresponds to the case for $B>0$ with
the spin projections of each atom reversed in sign. For the case of
cesium, a negative magnetic field in the $aa$ channel corresponds to
a positive field in the $gg$ channel.

\begin{figure}
\includegraphics[angle=-90,width=4in]{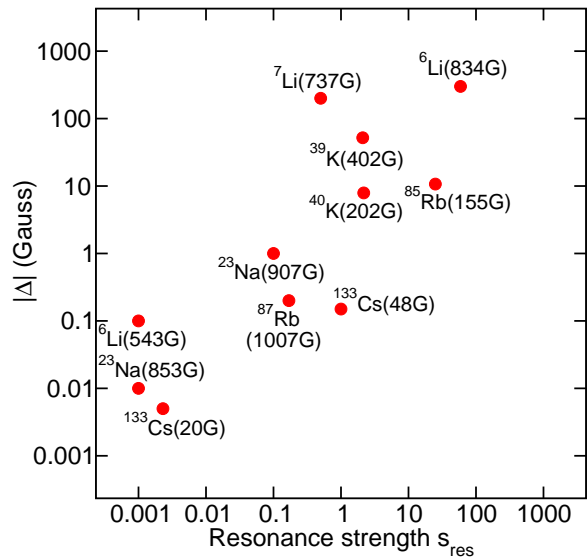}
\caption{Overview of selected Feshbach resonances in terms of widths $\Delta$ and strengths $s_{\rm res}$.}
\label{fig:ResStrength}
\end{figure}

Figure~\ref{fig:ResStrength} illustrates the rich variety of
Feshbach resonances in terms of their widths $\Delta$ and strengths
$s_{\rm res}$. Both parameters change over six orders of magnitude.
Resonances with $\Delta > 1\,$G tend to be entrance-channel
dominated ($s_{\rm res} > 1$). A notable exception is the $^7$Li
{ 737}-G resonance mentioned in Sec.~\ref{sssec:Li6res}

%\bibliographystyle{apsrmplong}
%\bibliography{feshbach_allrefs}

\end{document}